\documentclass[a4paper,11pt]{article}
\usepackage{graphicx}
\pdfoutput=0 

\usepackage{jheppub} 

\usepackage[T1]{fontenc} 

\title{\boldmath Dynamics of wave fluctuations\\
in the homogeneous Yang-Mills condensate}

\author[a]{Roman Pasechnik\note{Corresponding author.}}
\author[b,1]{George Prokhorov}
\author[b,2]{Grigory Vereshkov\note{Also at
Institute for Nuclear Research of Russian Academy of Sciences,
117312 Moscow, Russian Federation.}}

\affiliation[a]{Theoretical High Energy Physics, Department of
Astronomy and Theoretical Physics, Lund University, S\"olvegatan
14A, SE 223-62 Lund, Sweden}

\affiliation[b]{Research Institute of Physics, Southern Federal
University, 344090 Rostov-on-Don, Russian Federation}

\emailAdd{Roman.Pasechnik@thep.lu.se}

\abstract{In the present work, the Yang-Mills (YM) quantum-wave excitations 
of the classical homogeneous YM condensate have been studied in 
quasi-classical approximation. The formalism is initially formulated in the Hamilton 
gauge and is based upon canonical quantisation in the Heisenberg representation. 
This canonical framework is then extended and related to YM dynamics in 
arbitrary gauge and symmetry group containing at least one $SU(2)$ subgroup.
Such generic properties of the interacting YM system as excitation of longitudinal
wave modes and energy balance between the evolving YM condensate and waves
have been established. In order to prove these findings, the canonical quasi-classical 
YM system "waves + condensate" in the pure simplest $SU(2)$ gauge theory has 
been thoroughly analysed numerically in the linear and next-to-linear approximations 
in the limit of small wave amplitudes. The effective gluon mass dynamically generated 
by wave self-interactions in the gluon plasma has been derived. A complete set of equations 
of motion for the YM ``condensate + waves'' system accounting for second- and 
third-order interactions between the waves has been obtained. In the next-to-linear 
approximation in waves we have found that due to interactions between 
the YM waves and the YM condensate, the latter looses its energy leading to the growth 
of amplitudes of the YM wave modes. A similar effect has been found in the 
maximally-supersymmetric ${\cal N}=4$ Yang-Mills theory as well as in two-condensate 
$SU(4)$ model. Possible implications of these findings to Cosmology and gluon plasma physics 
have been discussed.}

\begin{document}
\maketitle
\flushbottom

\section{Introduction}
\label{Sec:Intro}

A consistent dynamical theory of the non-perturbative Yang-Mills (YM) vacuum,
responsible e.g. for the spontaneous chiral symmetry breaking and color confinement 
phenomena in quantum chromodynamics (QCD)
\cite{Belavin:1975fg,instantons,Gogohia:1999ip}, has not yet been
created. This is indeed one of the biggest theoretical challenges of 
modern quantum field theory which has a large variety of important 
implications. For example, self-interacting YM fields play an important role in yet
poorly known quark-gluon plasma dynamics at high and low
temperatures, including the problem of QCD phase transition and 
QCD dynamics at large distances. The YM fields are expected to play 
an important role in pre-equilibrium evolution at early times of relativistic heavy-ion
collisions when a Bose condensation of gluons could develop 
\cite{Blaizot:2011xf,Blaizot:2013lga,Huang:2013lia}.
Also, the non-Abelian gauge fields dynamics have important applications in early Universe which 
are being intensively studied in many different aspects, in particular, 
in the context of the Dark Energy \cite{Galtsov1,Galtsov2,YM-DE} 
and the non-Abelian fields driven inflation (``gauge-flation'') \cite{gauge-flation}. 
Moreover, the modern Dark Energy can be in principle generated by the quasi-classical 
gravity corrections to the QCD vacuum energy \cite{Urban:2009yg,Pasechnik:2013poa}. 
Very recently, it was understood that the unknown non-perturbative dynamics of the
quantum-topological and quantum-wave modes of the YM vacuum could
also be responsible for (partial or complete) compensation of the
QCD instanton vacuum energy to the ground state energy of the
Universe at macroscopic length scales and thus may be tightly related 
the colour confinement phenomenon \cite{Pasechnik:2013poa,Pasechnik}.

The major goal of the present paper is to study dynamical properties of the
{\it spatially-inhomogeneous wave modes propagating in the spatially-homogeneous 
YM condensate} incorporating interactions between these two components 
in the simplest pure YM $SU(2)$ theory. Note, the YM wave modes are interpreted 
as particles after a quantisation procedure giving rise to the {\it ultra-relativistic YM 
plasma}, and our purpose is to study the plasma properties consistently taking into
account its interactions with the YM condensate considered as a non-trivial background.

We consider the simplest situation when $SU(2)$ is the complete gauge symmetry
group. Note, the equations of motion of such a theory are form-invariant w.r.t.
$SU(2)$ gauge transformations whereas their solutions are not required to
have exactly same symmetry as the equations. This situation is called spontaneous
symmetry breaking (SSB). The appearance of the homogeneous and
isotropic YM condensate is a specific case of the SSB. Namely, the condensate
by itself is not gauge invariant, therefore, breaks the initial symmetry of 
the Lagrangian $SU(2)\otimes SO(3)$ down to the global $SO(3)$ group 
of spatial 3-rotations.

The basic new result of our study is observation that the interactions between 
the small YM waves and large condensate in a simple $SU(2)$ YM theory trigger a
significant energy transfer in one particular direction, namely,
from the condensate to the wave modes. Such a specific {\it energy swap 
effect} between the two YM subsystems may have serious consequences 
to the theory of non-perturbative YM vacuum and, in particular, may have 
important phenomenological implications e.g. in the theory of QCD phase
transition in early Universe and in particle production mechanisms
in the hot cosmological plasma. One may further extend this first 
simple analysis incorporating effects of mutual interactions
between different wave modes. Certainly, the latter should be taken
into account in a complete theory of ultra-relativistic gluon plasma.

To start with, we have constructed the exact quasi-classical
equations for the wave modes and the condensate in the Hamilton gauge, 
and investigated them in the linear approximation (in small wave amplitudes limit).
In this case, the equations of motion have a characteristic form of
Mathieu equations having certain regions of parametric resonance
instability which leads to an increase of amplitude of the waves. As
was argued above, the constraints written for the system of
interacting homogeneous YM condensate and inhomogeneous waves do not allow to
exclude the longitudinal modes, such that these extra d.o.f.'s acquire
their own dynamical properties due to interactions between the two
subsystems.

As a consequence of energy conservation, an increase of the YM waves
energy reflected in a corresponding increase of their amplitudes has
to be accompanied by a corresponding decrease of the YM condensate energy. In
particular, this fact must be taken into account in derivation of
the next-to-linear condensate equation of motion where the ``back reaction''
effect of the wave modes to the condensate is consistently incorporated.
The numerical analysis of the resulting system of equations has
indeed revealed the energy swap effect satisfying the energy
conservation: a decrease of the condensate energy is exactly compensated by
an increase of energy attributed to the wave modes, which is an
important test of our calculations. 

In addition, equations of motion for pure gluon plasma (without the 
YM condensate) in the $SU(3)$ gauge theory have been derived. It turned
out that the third-order interaction terms between wave modes alone
lead to excitation of a local dynamical (constituent) gluon mass whose expression
has been obtained in a simple renormalized form and is shown to be consistent 
with the corresponding thermodynamic prediction. An inclusion of the second-order 
interactions leads to dynamical mass terms in a time non-local form. Also, 
the exact integro-differential equations for the condensate and waves 
evolution incorporating both second- and third-order interaction terms in waves 
have been obtained in the $SU(2)$ theory and can further be studied numerically 
which is a subject of future studies. 

Note, a generalisation of these studies to the $SU(3)$ theory appears to be conceptually 
straightforward but technically much more involved since it incorporates (i) three 
overlapping condensates which are thus expected to interact strongly with each other and 
(ii) a larger variety of the wave modes compared to the $SU(2)$ case, and this analysis 
should be eventually performed. In this work, we have considered the gluon plasma 
dynamics in the $SU(3)$ neglecting the contributions from the overlapping condensates. 
The next non-trivial example is the $SU(4)$ theory which contains two non-overlapping 
gauge $SU(2)$ groups and hence gives rise to two non-interacting condensates 
to the leading order in waves. Such a model has been discussed in some major 
details in this work. Note, that in the $SU(4)$ toy-model we have disregarded 
all possible overlapping condensates in order to consider some basic
features of heterogenic multi-condensate systems although a more realistic 
study involving overlapping condensates is necessary.

We have looked also into the question of gauge dependence of 
the YM condensate and waves. In fact, the YM condensate can be consistently extracted 
only in the Hamilton gauge where $A_0(x)=0$. Departing from this statement, we have proposed 
a new method which allows to extract the condensate and waves in arbitrary
gauge and connect them to the condensate and waves evaluated in the Hamilton gauge.
In short, for non-degenerate YM ``condensate + waves'' systems it is possible to establish a firm 
connection between the canonical results in the Hamilton gauge and those in arbitrary 
gauge by absorbing the gauge dependence (an infinitesimal depart from 
the Hamilton gauge is encoded by the gauge-fixing function $\delta\theta(x^\mu)$) into 
a coordinate transformation from $x^\mu$ to new abstract coordinates 
$z^\mu=z^\mu(x^\mu,\theta(x^\mu))$. In new coordinates, the YM field $A_\mu(z)$ in 
arbitrary gauge is constrained in the same way as is typically done in the Hamilton 
gauge, i.e. $A_0(z)=0$, whereas $z^\mu$ absorbs the information about an underlined 
gauge choice. Then, the YM condensate and waves are consistently defined in such $z$-space 
and then transformed back to usual space-time. This approach has allowed to define the 
condensate in arbitrary gauge as a functional of the YM solution in the Hamilton gauge and 
the gauge-fixing function. This therefore justifies the rest of our studies of YM dynamics 
performed in the Hamilton gauge which is the most convenient in the present setting.

Further, we have generalised our study to the maximally (${\cal
N}=4$) supersymmetric YM (SYM) theory (see e.g.
Ref.~\cite{Gliozzi}). As one of the specific features of this theory
is its conformality such that its $\beta$-function disappears (i.e.
the coupling constant does not acquire radiative corrections and
therefore does not run), which significantly simplifies our
calculations. The ${\cal N}=4$ SYM theory includes four different
fermion fields, three scalar and pseudo-scalar fields. We have shown,
both numerically and analytically, that interactions of
supersymmetric wave modes with the YM condensate lead to a similar energy swap
effect from the YM condensate to the (pseudo)scalar wave modes as it was
earlier observed for the vector wave modes in $SU(2)$ gauge theory. 
These findings strongly suggest that the observed dynamics in energy 
balance of the interacting YM system (``wave + condensate'') is 
a general phenomenon and specific property inherent to YM dynamics. 
Inclusion of coloured fermion modes into our quasi-classical analysis 
is relevant for particle production mechanisms in early Universe and 
will be done elsewhere.

Finally, we have looked into conformal time dynamics of the free YM condensate 
(without waves) on expanding background of the Friedmann Universe. A non-linear 
oscillatory behaviour of the condensate fluctuations contributing to the radiation-type 
matter has turned out to be unstable w.r.t. quantum radiative corrections.
So we have incorporated the leading vacuum polarisation effects at one-loop
and considered such a log-corrected quasi-free condensate dynamics. As a result, 
the condensate contribution to the energy density in this case becomes constant in 
time instead (and positive!) which is in accordance with the vacuum 
equation of state and can be used to eliminate large negative topological 
vacua contributions possibly providing a smallness of the observed cosmological constant.
In order to investigate the latter possibility one should further incorporate the wave 
contributions together with the vacuum polarisation effects.

One should notice the following important point. A specific character of the considered SSB 
mechanism by means of the YM condensate compared to the traditional versions of the SSB 
(e.g. the quark-gluon condensate in spontaneous chiral symmetry breaking of the Higgs condensate 
in spontaneous electroweak symmetry breaking) is as follows. In other known versions one deals 
with the SSB of the vacuum which remains Lorentz-invariant, i.e. stationary 
in time and spatially-homogeneous and isotropic in 3-space with energy-momentum 
tensor invariant w.r.t. Lorentz transformations (the well-known $\Lambda$-term). 
In our case, the YM condensate in its present form (i.e. without including vacuum 
polarisation effects) is not Lorentz-invariant due to a non-stationary character of 
the corresponding energy momentum tensor (it behaves as a radiation-type matter 
as the Universe expands). 

As was mentioned above, it turned out that an inclusion of the vacuum polarisation 
effects dramatically change the time dependence 
of the YM condensate and its energy density -- the latter becomes stationary in time 
and emerges as an extra constant (positive) contribution to the Lorentz-invariant 
$\Lambda$-term (i.e. satisfies to the vacuum equation of state $p=-\varepsilon$) although 
the condensate $U$ by itself is non-stationary which is therefore an instanton-type solution 
(This is due to the fact that the equations of motion for the free condensate are not 
form-invariant under small quantum fluctuations.) So, the Lorentz-invariance of 
the energy-momentum tensor of such a quasi-free condensate $U(t)$ is restored 
by the vacuum polarisation effects. For this very first study, the latter argument 
motivates our simplest choice of initially free Lorentz non-invariant YM condensate 
(with no vacuum polarisation incorporated) which is spatially-homogeneous
and isotropic. At the next step, it would be important to incorporate 
vacuum polarisation and the wave dynamics on the same footing as it may give 
an access to topological effects, e.g. to a dynamical compensation of the 
topological and condensate excitations. Note, we are focused now on the simplest 
version of the SSB, but the YM condensate can, in principle, be spatially-inhomogeneous 
and at the same time not possess a wave structure (i.e. not be a superposition of waves). 
An analysis of non-homogeneous condensates could also be a good point for future studies.

Note also that if the ground state does not have a space-time symmetry of the initial 
theory (likewise, our free YM condensate is not Lorentz-invariant) then the conditions 
of the Goldstone theorem do not apply to this case of SSB. Therefore, there are no arguments 
to expect an appearance of the massless Goldstone modes after the SSB in this case. However, 
instead of the Goldstone effect we observe another interesting effect: the SSB in our case leads 
to a reconstruction of the excitations spectrum such that the underlined degrees of freedom 
of the initial YM theory (YM potentials) get reclassified into a new set of physical 3-tensor, 
3-vector and 3-scalar d.o.f.'s. whose dynamics has been studied here.

The paper is organised as follows. In Sect.~\ref{Sec:YM-Ham}, the generic 
properties of the YM dynamics in both degenerate (without the condensate) and non-degenerate 
interacting YM system ``condensate + waves'' in the Hamilton gauge have been discussed. 
Sect.~\ref{Sec:arb-gauge} is devoted to a formulation of the interacting YM theory 
in arbitrary gauge and its important relation to the canonical framework developed 
in subsequent sections. Sect.~\ref{Sec:YM-Ham-analysis} contains a thorough 
analytical and numerical analysis of the $SU(2)$ ``condensate + waves'' system 
in the Hamilton gauge in the limit of small wave amplitudes compared to the condensate 
in zeroth, quasi-linear and next-to-linear approximations in wave modes. 
In Sect.~\ref{Sec:YM-plasma} we have discussed physics of the gluon 
plasma (without the condensate) and dynamical gluon mass generation due 
to wave self-interactions in different orders as well as the exact YM equations 
of motion of the ``condensate + waves'' system accounting for all 
the higher-order interaction terms. In Sect.~\ref{Sec:SYM}
we apply the quasi-classical approach to the ${\cal N}=4$ conformal
Super-Yang-Mills theory with the condensate. Sect.~\ref{Sec:YM-cosm} contains a thorough
discussion of cosmological implications of the YM condensate with and without
the vacuum polarisation effects as well as discusses typical condensate 
decay time scales due to condensate-wave interactions in quasi-linear 
approximation. Sect.~\ref{Sec:Two-YMC} provides an overview of the simplest
two-condensate model based upon $SU(4)$ gauge theory. A few concluding 
remarks were made in Sect.~\ref{Sec:concl}. Appendix \ref{A1} is devoted to 
details of the Hamilton formulation of a degenerate YM theory where the 
YM propagator has been derived by using the method of infinitesimal parameter.
Finally, canonical quantisation of the YM wave modes in the
classical YM condensate has been performed in Appendix \ref{A2} 
as a consistency check of the quasi-classical framework 
developed in this work.

\section{Overview of Yang-Mills theory in the Hamilton gauge}
\label{Sec:YM-Ham}

\subsection{Degenerate Yang-Mills system}
\label{Sec:YM-Ham:deg}

Let us remind a few important basics of the classical degenerate 
YM theory useful for the forthcoming analysis below. Consider first 
the simplest non-Abelian $SU(2)$ gauge symmetry group.
The gauge invariant Lagrangian of the corresponding YM field is
\begin{equation}
\mathcal{L}=-\frac{1}{4}F^a_{\mu\nu}F_a^{\mu\nu}\,, \label{L}
\end{equation}
where
\begin{equation*}
F^a_{\mu\nu}=\partial_\mu A^a_\nu - \partial_\nu A^a_\mu +
g\,e^{abc} A^b_\mu A^c_\nu
\end{equation*}
is the YM stress tensor with isotopic $a=1,2,3$ and Lorentz $\mu,\nu=0,1,2,3$
indices. So in the degenerate case there are twelve equations of motion
\begin{equation}
\partial^\mu F_{\mu\nu}^{a} - gF_{\mu\nu}^{b}e_{abc}A^{\mu}_{c} =
0\,.
\end{equation}

Consider first the simplest (ghost-free) Hamilton gauge. Note, the Hamilton gauge 
is the only known gauge which enables us to formulate the YM theory in the Heisenberg 
representation consistently beyond the perturbation theory (for more details, see 
e.g. Ref.~\cite{Faddeev,Bogolubov}). Indeed, the Heisenberg representation 
is the most useful and appropriate one in the analysis of real-time evolution 
of the homogeneous YM condensate. Imposing the Hamilton gauge condition
\begin{equation}
A^a_0=0\,, \label{gauge}
\end{equation}
one ends up with nine equations of motion
\begin{equation}
\partial_0 F_{0k}^{a}=\partial_i F_{ik}^{a} -
gF_{ik}^{b}e_{abc}A_{i}^{c}\,, \label{eq}
\end{equation}
and three additional constraint equations in the form of 
first integrals of motion
\begin{equation}
 \partial^0(\partial^i F_{i0}^{a} - gF_{i0}^{b}e_{abc}A^{i}_{c}) = 0\,,
 \label{motionint}
\end{equation}
such that the total time derivative can be eliminated 
in the non-gauged (before the gauge is fixed) case, i.e.
\begin{equation}
 \partial^i F_{i0}^{a} - gF_{i0}^{b}e_{abc}A^{i}_{c} = 0\,.
\label{coupleq}
\end{equation}

It is worth to mention shortly a few important aspects of the YM quantum theory in 
the Hamilton gauge. Typically, such a theory is formulated in terms of the 
functional (or path) integral \cite{Faddeev}. After introducing the Hamilton 
gauge into the path integral for the YM field, one defines the $S$-matrix 
and the YM propagator. As one of the attractive features of the
Hamilton gauge, the asymptotic states of such $S$-matrix
automatically contain transverse modes only. Then, the YM propagator 
is defined as a Green function of the equations of motion 
in the Hamilton gauge. The longitudinal mode gives 
a certain contribution to the YM propagator while it naturally disappears 
in the asymptotic states.

Below we adopt the following theoretically consistent and pragmatic approach 
which has certain methodological advantages in the analysis of interacting YM
``waves + condensates'' systems compared to the standard path integral formulation.
\begin{itemize}
\item Firstly, following to the Bohr's correspondence principle one starts with 
the pure YM Lagrangian in the Hamilton gauge and writes down the Lagrange 
equations of motion in the operator form;
\item Secondly, since in the Hamilton gauge the zeroth component of 
the YM field $A_0$ is absent in the YM Lagrangian, there are no
explicit constraint equations in the corresponding system of
Lagrange equations. Such equations of constraint can be instead 
obtained as integrals of motion of the Lagrange equations;
\item Thirdly, starting from the Lagrange formulation of the YM theory 
one easily turns to the Hamilton (or canonical) formulation and the 
canonical quantisation procedure.
\end{itemize}

Let us discuss the third point of the above scheme in more detail.
In the case of free YM field (without its interactions with the
condensate), the free (non-interacting) longitudinal YM mode 
does not have any proper frequency and dispersion, thus this 
mode is aperiodic in Minkowski space without the presence of 
YM condensate. Such a property has the following three important 
consequences: (1) the free longitudinal mode does not contribute to 
the Hamiltonian; (2) it is impossible to calculate its contribution 
to the YM propagator as a vacuum expectation value of the time-ordered 
YM field operator product; (3) there is a related issue with canonical 
quantisation. The first consequence is one of the advantages 
of the Hamilton gauge which automatically excludes any non-physical 
degrees of freedom from the YM Hamiltonian. The other two consequences 
point out to the fact that the YM theory in the Hamilton gauge cannot be 
constructed according to standard algorithms of a non-degenerate field theory. 

One could advise a simple alternative and methodologically elegant way to resolve 
the latter issue: for a YM system without the presence of a homogeneous 
YM condensate one can introduce extra ``virtual'' infinitesimal terms (proportional to 
a single infinitesimal parameter $\xi$) into the YM Lagrangian in the Hamilton 
gauge. These terms can be chosen in such a way that the longitudinal mode acquires 
a small dispersion proportional to the small parameter $\xi$. Further, this enables 
to perform the canonical quantisation procedure in a standard way and to calculate 
a contribution of this mode to the YM propagator as a vacuum expectation value of 
the time-ordered YM field operator product. In the end, the infinitesimal parameter 
can be safely turned to zero $\xi\to0$ leading to exactly the same $S$-matrix as 
the one defined in the framework of standard path integral formulation.
Such an approach therefore leads to theoretically consistent results 
and enables one to work in the scheme described above. The method
of infinitesimal parameter is described in Appendix \ref{A1} in more detail.

\subsection{Non-degenerate Yang-Mills system: ``condensate + waves''}
\label{Sec:YM-Ham:nondeg}

Let us now consider an interacting non-degenerate YM system ``condensate + wave'' 
in the pure $SU(2)$ gauge theory. Here, the situation is different from the degenerate case 
described above in Sect.~\ref{Sec:YM-Ham:deg}. Namely, as will be explicitly shown below 
in Sect.~\ref{Sec:YM-Ham-analysis:lin:long} there is a new effect of dynamical generation 
of the longitudinal plasma waves as collective excitations of macroscopic medium (e.g. condensate)
also known as plasmons. The latter effect is well-known in physics of ordinary plasma 
\cite{plasma} and the quark-gluon plasma \cite{Blaizot:1994vs,Bannur:2002dk} (see also
Ref.~\cite{Schenke:2008hw} and references therein). The longitudinal YM waves 
acquire both proper frequency (proportional to density of the medium) and 
dispersion (proportional to thermal wave velocity squared) and thus should 
be properly taken into consideration.

Now consider how the homogeneous YM condensate can be extracted from the YM field 
in the Hamilton gauge. It has been demonstrated in Refs.~\cite{Galtsov1,Galtsov2} 
that due to isomorphism of the isotopic $SU(2)$ group and the $SO(3)$ group of spatial 
3-rotations, the unique (up to re-scaling) $SU(2)$ YM configuration can be
parameterized in terms of a scalar time-dependent spatially-homogeneous 
field. Indeed, such an isomorphism enables one to introduce a mixed 
space-isotopic orthonormal basis $e^a_i$ in which the YM vector field 
$A_\mu^a$ transforms into a tensor field $A_{ik}$ with two spatial 
indices\footnote{For the spacial components, one does not distinguish 
upper and lower indices while repeated indices are summed up by default.} 
$i$ and $k$ as follows
\begin{equation} \label{e-basis}
e^a_i A^a_k=A_{ik}\,, \qquad e^a_i e^a_k=\delta_{ik}\,,\qquad e^a_i e^b_i=\delta_{ab}\,.
\end{equation}
Then, the resulting spacial tensor $A_{ik}$ can be separated into two parts
\begin{equation}
A_{ik}(t,\vec x)=\delta_{ik}U(t)+\widetilde{A}_{ik}(t,\vec x)\,,
\label{TheFirst}
\end{equation}
Here, the first spatially-homogeneous time-dependent scalar field $U(t)$
corresponds to the YM condensate and can be found as an average over a 
substantially large 3-space domain $\Omega\to \infty$
\begin{equation}
U(t) \equiv \frac13\delta_{ik} \langle A_{ik} (t,\vec x) \rangle_{\vec x}\,,  \qquad 
\langle A_{ik} (t,\vec x) \rangle_{\vec x} = \frac{\int_{\Omega} d^3x A_{ik} (t,\vec x)}
{\int_{\Omega} d^3x} \,,
\end{equation}
In what follows, we consistently relate such an average procedure over 3-spacial 
domain $\langle \dots \rangle_{\vec x}$ with the average over the state 
vector $\langle \dots \rangle$
\begin{eqnarray}
\langle \dots \rangle\equiv \langle \Psi | \dots | \Psi \rangle \sim \langle \dots \rangle_{\vec x} \,.
\end{eqnarray}
It is important to notice that the condensate 
effectively removes the degeneracy of the underlined YM theory reducing the initial 
$SU(2)\otimes SO(3)$ symmetry of 
the YM Lagrangian to the global $SO(3)$ symmetry of spacial rotations.
The second quantum-wave part 
\begin{equation}
\widetilde{A}_{ik}=\widetilde{A}_{ik}(t,\vec x)\equiv A_{ik}(t,\vec x)-\langle A_{ik} (t,\vec x) \rangle_{\vec x}
\end{equation}
is spatially-inhomogeneous and describes motion of YM quanta, namely, 
physical particles after quantisation. It turns out that due to interactions 
between the YM condensate and waves, the longitudinal d.o.f.'s which is 
unphysical in the degenerate theory are parametrically excited and must 
be treated as physical ones.

Connections between the YM condensate and classical electric $E$ and magnetic $H$ 
fields can be easily derived, by definition of the YM potentials, e.g.
\begin{equation}
\partial_0 U= -\frac{1}{3}\langle E_{kk}\rangle \,, \qquad U^2 = \frac{1}{3g}\langle H_{kk} \rangle \,,
\end{equation}
such that the condensate emerges as a mixture of $E$ and $H$ components. 
Analogous connections between the wave modes and the fields $E$ and $H$ 
can be obtained in a similar way.

Note, the above procedure remains valid also for the $SU(3)$ gauge theory 
which however is much more complicated and contains three different 
but overlapping $SU(2)$ subgroups isomorphic to $SO(3)$. This situation 
potentially gives rise to three strongly interacting condensates, and needs 
to be analyzed separately. In a generic $SU(N)$ model with $N>3$ the 
number of non-overlapping and overlapping $SU(2)$ subgroups 
can be even larger which leads to formation of the heterogenic system 
of a few YM condensates. The simplest case of such a heterogenic 
system with two non-overlapping condensates is the $SU(4)$ gauge theory. 
The latter has been discussed below in Sect.~\ref{Sec:Two-YMC} without an
account for overlapping condensates.

\section{Yang-Mills theory in an arbitrary gauge}
\label{Sec:arb-gauge}

In the previous Section, we have discussed generic properties of the YM condensate theory in the Hamilton 
gauge and in isotopic $SU(2)$ symmetry. Now we turn to an algorithm which allow us to describe YMC 
dynamics and YM waves in arbitrary gauge in connection to that known in the Hamilton gauge. 
Here, we consider the quasi-classical limit of the YM theory so at this point we do not discuss
issues with ghost fields which appear via radiative corrections in a proper 
quantum formulation.

\subsection{Degenerate Yang-Mills system}
\label{Sec:arb-gauge:deg}

Consider for simplicity a non-Abelian gauge group $SU(N),\, N=2,3$ with a set of generators $T_a$,
which contains one isotopic $SU(2)$ subgroup. In a degenerate theory, 
the corresponding YM field operator $A_\mu\equiv T^aA^a_\mu$, $\langle A_\mu \rangle_{\vec x}=0$ 
with four Lorentz components $\mu=0,1,2,3$ contains only two physical components 
which can be identified transverse polarizations of the respective quanta.
Recall, the local gauge transformation condition and the covariant derivative 
of the degenerate theory are given by
\begin{eqnarray}
&& A^{(\theta)}_\mu=GA^{(\theta_0)}_\mu G^{-1}+\frac{i}{g}(\partial_\mu G) G^{-1}\,, 
\quad  D_\mu=\hat{\partial}_\mu + ig A^{(\theta_0)}_\mu \,, \\
&& G\equiv G(x)=\exp\Big[ig\delta\theta(x)\Big]\simeq {\rm I} + ig\delta\theta(x) + \, \dots \,, 
\qquad \delta\theta(x)=T^a \delta\theta^a(x)\,,
\end{eqnarray}
The initial gauge $SU(N)$ symmetric YM Lagrangian 
\begin{equation}
\mathcal{L}=-\frac{1}{2}{\rm Tr}[F_{\mu\nu}F^{\mu\nu}]\,, \qquad
F_{\mu\nu}\equiv -\frac{1}{ig}[D_\mu,D_\nu]=
\partial_\mu A_\nu-\partial_\nu A_\mu+g[A_\mu,A_\nu]\,,
\end{equation}
is manifestly invariant under the above gauge transformation condition which takes the following 
infinitesimal form
\begin{eqnarray}
A^{(\theta)}_\mu=A^{(\theta_0)}_\mu+ig[\delta\theta,A_{\mu}^{\theta_0}] - 
\partial_\mu \delta\theta \,.
\end{eqnarray}
Above, $A^{(\theta_0)}_\mu$ is the YM field in a given {\it reference gauge}.
In what follows, it will be convenient to choose a simple reference gauge 
where the YM solutions of the non-degenerate theory are assumed to be known. 
So whenever it is reasonable we will use the simplest Hamilton gauge $\theta_0\equiv\theta_H$ 
as our reference gauge assuming that the corresponding solutions are known.

As was advocated above in the Hamilton gauge, due to isomorphism of the local isotopic $SU(2)$ (sub)group and 
the global $SO(3)$ group of spatial 3-rotations, the YM quanta condensation can effectively induce 
a homogeneous classical background field, the YM condensate, whose vacuum expectation value is non-trivial. 
As was discussed above, the condensate removes the degeneracy of the underlined YM theory 
and effectively breaks the symmetry of initial YM Lagrangian down to the $SO(3)$ rotation symmetry. 
In this case, besides two physical transversely polarised modes one extra physical wave d.o.f. gets 
parametrically excited in the YM condensate and can be associated with the longitudinally polarised plasma mode. 
Therefore, we effectively end up with three independent YM wave operators $\widetilde A_k(t,\vec x)$
where index $k$ was associated with three spacial components in the Hamilton gauge. In what follows,
we consider a non-degenerate theory and rely on the fact that the number of physical modes $\widetilde A^a_k(t,\vec x)$ 
is equal to $3(N^2-1)$ and gauge-independent which allows us to determine the YM condensate dynamics in an analogical way 
to that in the Hamilton gauge.

In order to construct a consistent theory of the interacting system ``condensate + waves'' in arbitrary 
gauge let us introduce an abstract Minkowski 4-space with coordinates depending on 
the gauge-fixing function $\delta\theta$, i.e.
\[
z_\alpha=z_\alpha(t,\vec x;\delta\theta(t,\vec x))\equiv \{\tau,\vec z\} \,, \quad \alpha=0,1,2,3\,,
\]
which are related to usual space-time coordinates $x_\mu=\{t,\,\vec x\}$ as
\begin{equation}
dz_\alpha=\varepsilon^\mu_\alpha dx_\mu\,, \qquad  \varepsilon^\mu_\alpha=
\frac{\partial z_\alpha}{\partial x_\mu}\,, \qquad 
\varepsilon^{-1}_{\mu \alpha}=\frac{\partial x_\mu}{\partial z^\alpha}\,.
\end{equation}
The new Lorentz-covariant coordinate transformation matrix $\varepsilon^\mu_\alpha$ is normalized as
\begin{equation}
\varepsilon^\mu_\alpha=\varepsilon^\mu_\alpha\big(\tau(t,\vec x;\delta\theta),\vec z(t,\vec x;\delta\theta)\big)\,, \qquad 
\varepsilon^\mu_\alpha \varepsilon^{-1}_{\mu \beta}=g_{\alpha\beta}\,, 
\qquad  \varepsilon^\alpha_\mu (\varepsilon_{\nu \alpha})^{-1}=g_{\mu\nu}\,.
\end{equation}
In what follows, we call such an abstract 4D space as the {\it $z$-space}. In general, the basis 
transformation matrix $\varepsilon_{\mu \alpha}$ depends on the gauge-fixing function 
$\delta\theta=\delta\theta(t,\vec x)$ and can therefore effectively absorb all the information 
about the underlined gauge choice. In particular, in the Hamilton gauge corresponding to 
$\theta\to \theta_H$ it takes a simple form
\begin{equation}
\varepsilon_{\mu \alpha}=\delta_{\mu \alpha}\,,
\end{equation}
which means that $z$-space simply coincides with usual coordinate 3-space in the Hamilton gauge
\[
\lim_{\delta\theta\to 0} z_\alpha(t,\vec x;\delta\theta) = x_\alpha \,.
\]

Let us now define the $z$-space projections of the YM field $A^{(\theta)}_{\mu}$ and 
the stress tensor $F^{(\theta)}_{\mu\nu}$ in a given gauge as follows
\begin{eqnarray} \nonumber
&& A^{(\theta)}_{\alpha}(\tau,\vec z) \equiv (\varepsilon^{\mu}_\alpha)^{-1} A^{(\theta)}_{\mu} \,, \;\,
F^{(\theta)}_{\alpha\beta}(\tau,\vec z)\equiv (\varepsilon^{\mu}_\alpha)^{-1}(\varepsilon^{\nu}_\beta)^{-1}
F^{(\theta)}_{\mu\nu}=\delta_\alpha A^{(\theta)}_\beta-\delta_\beta A^{(\theta)}_\alpha+g[A^{(\theta)}_\alpha,A^{(\theta)}_\beta]\,, 
\end{eqnarray}
where 
\[
\delta_\alpha\equiv \frac{\partial}{\partial z^\alpha}=\varepsilon^{-1}_{\mu \alpha}\partial^\mu\,,
\]
and the relation $\delta_\alpha \varepsilon_{\mu \beta}=\delta_\alpha \varepsilon_{\mu \beta}$ has been used.
The equations of motion in $z$-space
\begin{eqnarray}
\delta^\alpha F^{(\theta)}_{\alpha\beta}(\tau,\vec z) + ig [A^{\alpha(\theta)}(\tau,\vec z),F^{(\theta)}_{\alpha\beta}(\tau,\vec z)]=0\,,
\label{eom}
\end{eqnarray}
are covariant w.r.t. local gauge transformations of the YM field from a chosen reference
gauge $\theta'$ to a given gauge $\theta$ in infinitesimal form
\begin{eqnarray} \label{GT}
A^{(\theta)}_\alpha(\tau,\vec z)=A^{(\theta')}_\alpha(\tau,\vec z)+ig[\delta\theta,A_\alpha^{(\theta')}(\tau,\vec z)]-\delta_\alpha \delta\theta\,.
\end{eqnarray}
The particular gauge null-transform to/from the Hamilton gauge $\theta'=\theta_H$
 \[
 \delta\theta_{[0]}=C\,\exp\Big[ig\int A_k(t,\vec x) dx_k\Big] \,.
 \]
corresponds to a closed loop in the space of gauge configurations such that the gauge field transforms to itself.
In what follows, we assume that $\delta\theta\not=\delta\theta_{[0]}$ unless noted otherwise.
For finite transformation, one writes consequently
\begin{eqnarray}
A^{(\theta)}_\alpha(\tau,\vec z)=\frac{1}{ig}G(\hat{\delta}_\alpha + igA^{(\theta')}_\alpha(\tau,\vec z)) G^{-1}\,.
\end{eqnarray}

At the next step, in a complete analogy with the Hamilton gauge the zeroth component 
of the gauge field in $z$-space is required to be zeroth in {\it any gauge}, i.e.
\begin{eqnarray}
A^{(\theta)}_0\big(\tau(t,\vec x;\delta\theta),\vec z(t,\vec x;\delta\theta)\big)=0\,, 
\qquad \delta\theta=\delta\theta(t,\vec x)\,,
\end{eqnarray}
such that the other three components $A^{(\theta)}_k(\tau,\vec z),\,k=1,2,3$ should be identified with 
physical YM operators. Starting from the Hamilton gauge as our reference gauge, the latter identification 
allows to consider a coordinate transformation of usual 
4D $x$-space to a new 4D $z$-space as a gauge transformation, i.e.
\[
A^{(\theta)}_k\big(\tau(t,\vec x;\delta\theta),\vec z(t,\vec x;\delta\theta)\big)=
A^{(\theta_H)}_k(t,\vec x)\equiv A_k(t,\vec x)\,, \qquad \delta\theta=\delta\theta(t,\vec x)\,,
\]
which provides a connection between the coordinate transformation matrix $\varepsilon_{\mu \alpha}$ 
and the gauge fixing function $\delta\theta$. Indeed, using this relation in Eq.~(\ref{GT}) one obtains 
the following constraint equations
\begin{eqnarray} \label{GInv-an}
\varepsilon^{-1}_{\mu 0}\partial^\mu\delta\theta=0\,, \qquad 
\varepsilon^{-1}_{\mu k}\partial^\mu\delta\theta(t,\vec x)=ig[\delta\theta(t,\vec x),A_k(t,\vec x)]\,,
\end{eqnarray}
in terms of the YM solution in the Hamilton gauge $A_k(t,\vec x)$ in the Hamilton gauge. 
The relations (\ref{GInv-an}) then determine the coordinate transformation 
matrix $\varepsilon_{\mu \alpha}$ and hence the complete YM solution in arbitrary 
gauge $A^{(\theta)}_\mu$ as functionals of the gauge 
fixing function $\delta\theta(t,\vec x)$ and the YM solution 
in the Hamilton gauge $A_k(t,\vec x)$, namely
\begin{eqnarray}\label{eps-sol}
 \bar{\varepsilon}_{\mu \alpha}(t,\vec x)=\varepsilon_{\mu \alpha}\big[A_k(t,\vec x); 
 \delta\theta(t,\vec x)\big]\,, \qquad A^{(\theta)}_\mu(t,\vec x)=-A_k(t,\vec x) 
 \bar{\varepsilon}_{\mu k}(t,\vec x)\,.
\end{eqnarray}
Therefore, the YM solution in arbitrary gauge is expressed in terms of the canonical solution 
in the Hamilton gauge and the transformation matrix $\bar{\varepsilon}_{\mu \alpha}(t,\vec x)$ which
absorbs all the information about the underlined gauge transformation.

\subsection{Non-degenerate Yang-Mills system: ``condensate + waves''}
\label{Sec:arb-gauge:nondeg}

Based upon a direct analogy of the YM field consideration in ordinary space-time and 
in the Hamilton gauge made in the previous Section, isomorphism of the global $SO(3)$ 
symmetry group of 3-rotations in $z$-space and local $SU(2)$ subgroup of $SU(N)$, $N=2,3$ 
gauge group enables us to define the YM condensate and YM waves in arbitrary gauge.
Indeed, by employing the mixed space-isotopic orthonormal basis $e^a_i$ defined 
in Eq.~(\ref{e-basis}) one writes for the physical YM operators in $z$-space 
\begin{eqnarray}
A^{(\theta)}_k(\tau,\vec z)\equiv T^a A_k^{a(\theta)}(\tau,\vec z)=
{\cal T}_i A^{(\theta)}_{ik}(\tau,\vec z)\,,
\end{eqnarray}
where
\begin{eqnarray}
{\cal T}_i \equiv T^a e^a_i \,, \quad A^{(\theta)}_{ik}(\tau,\vec z)=A_{ik}(t,\vec x)=
\delta_{ik}W^{(\theta)}(t)+\widetilde{A}^{(\theta)}_{ik}(t,\vec x)\,,
\end{eqnarray}
in terms of the homogeneous YM condensate $W^{(\theta)}(t)$ and non-homogeneous 
YM waves $\widetilde{A}^{(\theta)}_{ik}(t,\vec x)$ which coincide with $U(t)$ and 
$\widetilde{A}_{ik}(t,\vec x)$ in the Hamilton gauge limit, respectively, 
corresponding to $\delta\theta\to \delta\theta_{[0]}$. The time-dependent scalar field $W^{(\theta)}(t)$ 
is defined normal way as an average over a large $z$-spacial domain $\Theta\to \infty$ 
at whose boundary all the fields (more precisely, wave excitations) are vanishingly small
\begin{equation}
W^{(\theta)}(t) \equiv \frac13\delta_{ik} \langle A^{(\theta)}_{ik} (\tau,\vec z) \rangle_{\vec z}\,,  \quad 
\langle A^{(\theta)}_{ik} (\tau,\vec z) \rangle_{\vec z} = \frac{\int_{\Theta} d^3z A^{(\theta)}_{ik} (\tau,\vec z)}
{\int_{\Theta} d^3z}=\frac{\int_{\Omega} d^3x\, |\bar{\varepsilon}| A_{ik}(t,\vec x)}
{\int_{\Omega} d^3x\, |\bar{\varepsilon}|}\,,
\end{equation}
where $A_{ik}(t,\vec x)$ is the known YM solution in the Hamilton gauge 
\begin{equation}
A_{ik}(t,\vec x)=\delta_{ik}U(t)+\widetilde{A}_{ik}(t,\vec x)\,,
\end{equation}
and the Jacobian of the 3-volume element transformation is
\begin{equation}
 |\bar{\varepsilon}| \equiv \Big |\det \Big\{\bar{\varepsilon}^l_m(t,\vec x)\Big\}\Big|=
 \Big| \frac{\partial(z_1,z_2,z_3)}{\partial(x_1,x_2,x_3)} \Big|\,.
\end{equation}
The YM condensate in arbitrary gauge finally reads
\begin{equation} \label{W}
W^{(\theta)}(t)=\frac13\delta_{ik} \langle A_{ik} (t,\vec x) \rangle_{\vec z}=
U(t)+\frac13\delta_{ik} \langle \widetilde{A}_{ik} (t,\vec x) \rangle_{\vec z}\not=U(t)\,, 
\quad \delta\theta\not=\delta\theta_{[0]}\,,
\end{equation}
where the last term vanishes in the Hamilton gauge $\delta\theta\to \delta\theta_{[0]}$ 
corresponding to $|\bar{\varepsilon}|\to1$.
The physical quantum-wave components are then given by
\begin{eqnarray}
\widetilde A^{(\theta)}_{ik}(t,\vec x)\equiv A_{ik}(t,\vec x) - \delta_{ik}W^{(\theta)}(t)\not=
\widetilde A_{ik}(t,\vec x)\,, \quad \delta\theta\not=\delta\theta_{[0]}\,.
\end{eqnarray}
Therefore, in this formulation the quasi-classical YM theory in an arbitrary gauge can be constructed in analogous 
way to that in the Hamilton gauge. The latter analogy allows to extract the YM condensate and to study its 
properties from the universal perspectives. The suggested formulation establishes a tight connection between 
the YM condensate and waves in the Hamilton gauge and those in an arbitrary gauge. So in the coming sections 
we will be focused primarily on analysis of the YM dynamics in the Hamilton gauge and its extension to an arbitrary 
gauge is now conceptually straightforward.

\subsection{Non-local gauge transformations}
\label{Sec:arb-gauge:nonloc}

As a result of the above discussion, we observe that despite the initial field $A^{(\theta)}_{ik} (t,\vec z)$ 
in the $z$-space coincides with the one in the Hamilton gauge in $A_{ik} (t,\vec x)$ in 
the $x$-space by construction, the corresponding separate YM condensates and wave contributions 
do not coincide. Indeed, an extra weighting factor $|\bar{\varepsilon} (t,\vec x)|$ appears in the average (\ref{W}) such
that it may lead to the appearance of additional contributions to the condensate $U(t)$ from non-vanishing 
averages of higher powers of the waves $\widetilde{A}_{ik}(t,\vec x)$. This is a direct consequence of the fact 
that the gauge symmetry is dynamically broken by the condensation of YM quanta. Since the total energy of 
the interacting system ``condensate + waves'' is conserved, this means that the gauge fixing operator $\delta\theta$ 
acquires a dynamical role and describes possible energy exchanges between the YM condensate and waves, 
and in this case may play a role of the evolution operator.

Interestingly enough, this effect, has a close analogy to the real time evolution of the YM system in 
the fixed Hamilton gauge where energy of the YM condensate is dynamically transferred into 
the waves in the quasi-classical limit of small wave amplitudes as will be shown below.
Let us pay a closer attention to a possible connection between the 
``gauge evolution'' in abstract $z$-space and real time evolution of the non-degenerate 
``condensate + waves'' system in usual space-time.

Consider a YM field in $z$-space fixed by constraints (\ref{GInv-an}) in a given gauge $\theta\not=\theta_H$, i.e.
\begin{eqnarray}
A^{(\theta)}_0\big(z[\delta\theta(x)]\big)\equiv 0\,, \quad A^{(\theta)}_k\big(z[\delta\theta(x)]\big)=A_k(x)\,, 
\quad \delta\theta\not=\delta\theta_{[0]} \,,
\end{eqnarray}
where $z=\{\tau,\vec z\}$ and $x=\{t,\vec x\}$. Then, define another gauge $\theta'\not=\theta$, $\theta'\not=\theta_H$ 
such that the infinitesimal gauge transformation $\theta\to\theta'$ takes the following generic non-local form
\begin{eqnarray}
A^{(\theta)}_k\big(z[\delta\theta]\big)=A^{(\theta')}_k\big(z[\delta\theta']\big)+
ig\big[\gamma[\delta\theta,\delta\theta'],A^{(\theta')}_k(z[\delta\theta'])\big]-
\delta_k\gamma[\delta\theta,\delta\theta']\,,
\end{eqnarray}
where $\delta\theta$ and $\delta\theta'$ denote the gauge transforms corresponding 
to $\theta_H\to \theta$ and $\theta_H\to \theta'$, respectively, and  the non-local operator 
$\gamma[\delta\theta,\delta\theta']$ describes the gauge transform $\theta\to \theta'$ 
with boundary condition 
 \[
 \gamma[\delta\theta,\delta\theta]\equiv \delta\theta=C\,\exp\Big[ig\int A^{(\theta)}_k\big(z[\delta\theta]\big) dz_k\Big] \,.
 \]
Following to the scheme above, one notices that the gauge transformations $\delta\theta$ and $\delta\theta'$
are equivalent to coordinate transforms 
\[ 
x\to z\equiv z[\delta\theta]\,, \qquad x\to z'\equiv z[\delta\theta'] \,,
\]
respectively. Absorbing the gauge dependence into the $z$-coordinate transformation as usual
\begin{eqnarray}
A^{(\theta)}_k\big(z[\delta\theta(x)]\big)=A^{(\theta')}_k\big(z[\delta\theta'(x)]\big)=A_k(x)\,,
\end{eqnarray}
one observes that
\begin{eqnarray}
\label{non-loc}
\varepsilon^{-1}_{\mu k}[\delta\theta'(x)]\partial^\mu
\gamma[\delta\theta(x),\delta\theta'(x)]=ig\big[\gamma[\delta\theta(x),
\delta\theta'(x)],A_k(x)\big]\,, \quad \varepsilon^{-1}_{\mu k}[\delta\theta'(x)]=
\frac{\partial x_\mu}{\partial z'_k}\,,
\end{eqnarray}
which is valid for any $\delta\theta$ and $\delta\theta'$. At last, for a given pair of 
non-intersecting non-trivial gauges
\[
\delta\theta'(x)\not=\delta\theta(x)
\]
one could always find two different space-time points $x$ and $x'$ for which
\begin{eqnarray} \label{xx'}
\delta\theta'(x)=\delta\theta(x')
\end{eqnarray}
is satisfied. In this case, the functional relation (\ref{non-loc}) finally transforms to
\begin{eqnarray}
\varepsilon^{-1}_{\mu k}\big[A_k(x');\delta\theta(x')\big]\partial^\mu
\gamma[\delta\theta(x),\delta\theta(x')]=ig\big[\gamma[\delta\theta(x),
\delta\theta(x')],A_k(x)\big]\,,
\end{eqnarray}
where $\varepsilon^{-1}_{\mu k}$ has been found earlier in Eq.~(\ref{eps-sol}).
We therefore notice that the gauge evolution between two points $\theta$ and $\theta'$
in space of non-equivalent gauge configurations driven by the non-local operator 
$\gamma[\delta\theta,\delta\theta']$ can be unambiguously mapped onto real 
physical motion between two space-time points $x$ and $x'$ related via (\ref{xx'}) 
in a fixed gauge $\theta$ whose evolution operator has the same form 
$\gamma[\delta\theta(x),\delta\theta(x')]$. Indeed, in the considered 
non-degenerate theory the values of the YM field in the Hamilton gauge 
defined in two different space-time points $A_k(x)$ and $A_k(x')$ have 
turned out to be related via the operator $\gamma$ which therefore acquires 
a dynamical role. A detailed development of this concept goes significantly
beyond the scope of the present paper and will be done elsewhere. Instead,
we turn to analysis of the YM condensate and waves in the Hamilton gauge.

\section{Yang-Mills dynamics in the Hamilton gauge: one condensate model}
\label{Sec:YM-Ham-analysis}

Below in this section as the very first step we consider the simplest 
one-condensate model in $SU(2)$ gauge theory analysing consequently 
its dynamics in different limiting cases and approximations.

\subsection{Free condensate case}
\label{Sec:YM-Ham-analysis:freeYMC}

\begin{figure*}[!h]
\begin{minipage}{0.45\textwidth}
 \centerline{\includegraphics[width=1.0\textwidth]{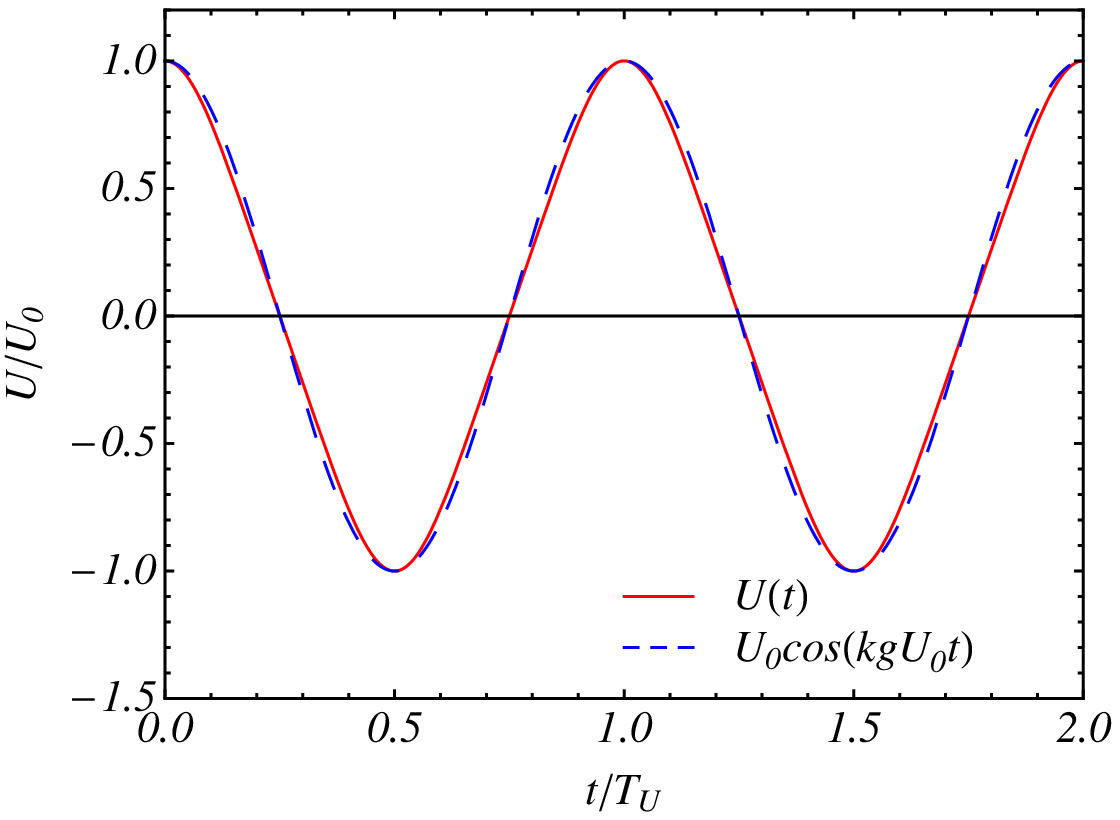}}
\end{minipage}
\hspace{1cm}
\begin{minipage}{0.45\textwidth}
 \centerline{\includegraphics[width=1.0\textwidth]{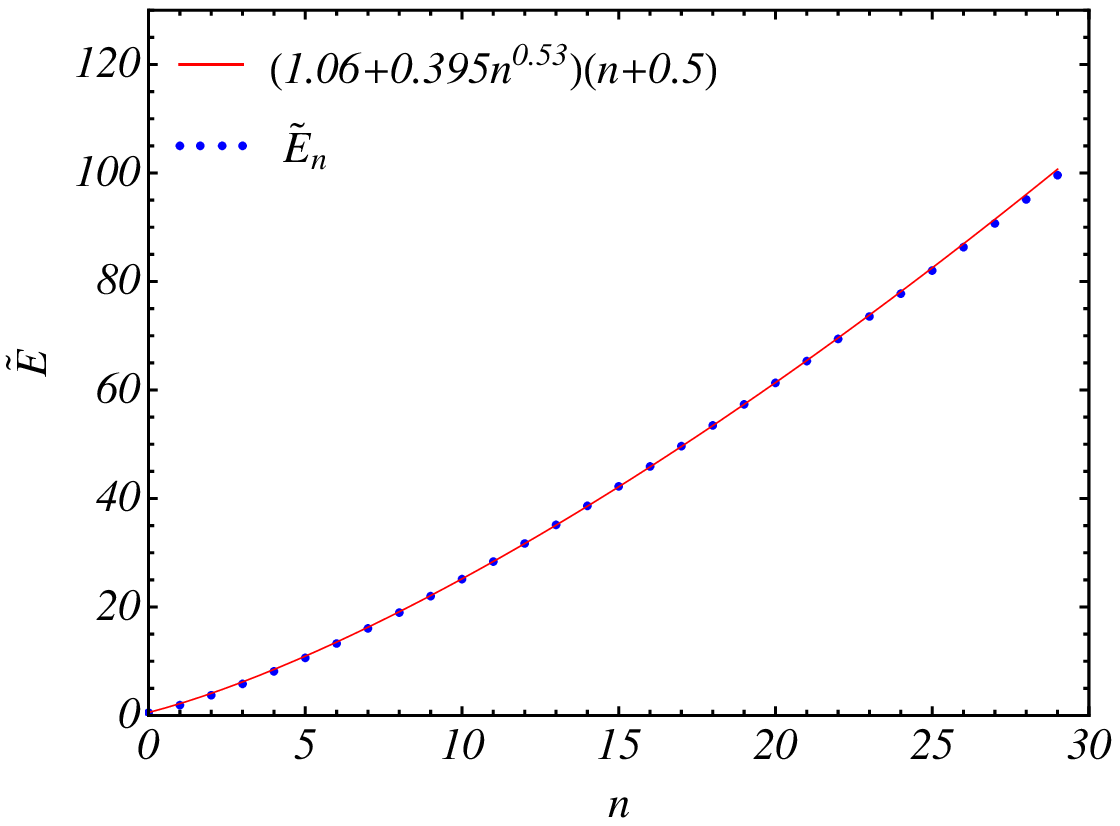}}
\end{minipage}
   \caption{Numerical solution of the YM equation of motion (\ref{eqU})
   for the time dependence of free YM condensate potential, $U=U(t)$, and its
   analytical approximation (\ref{appU}) for a fixed initial phase
   $\phi_0=0$ (left) and numerical result for the energy
   spectrum of free YM condensate, $E_n$, and its continuous analytical
   approximation (\ref{appEU}) (right). This plot is given in terms of
   dimensionless quantities and thus does not depend on initial 
   value of the condensate $U_0$.}
 \label{fig:free-YMC}
\end{figure*}

To the zeroth order in small waves $|\widetilde{A}_{ik}|\ll |U|$, 
the representation (\ref{TheFirst}) enables us to rewrite 
the Hamiltonian and YM equation of motion (\ref{eq}) 
for the real-time evolution of the YM condensate alone, $U=U(t)$. Namely,
\begin{equation}
\mathcal{H}_{\rm YM}\simeq \mathcal{H}_{\rm YMC}=
\frac{3}{2}\Big[(\partial_0U)^2 + g^2U^4\Big]\,,\qquad  
\partial_0\partial_0U+2g^2\,U^3=0\,.
\label{eqU}
\end{equation}
Its numerical solution is shown in Fig.~\ref{fig:free-YMC}(left). 
The exact solution of the equation of motion for the ``free'' YM condensate is given by
\begin{equation}
t=-\int_{U_0}^U \frac{dU}{\sqrt{g^2U_0^4-g^2U^4}}\,, \qquad  U(0)=U_0\,, \qquad U'(0)=0\,.
\end{equation}
According to this result the free YM condensate can 
exist only in a free non-stationary slowly-oscillating state. 
Indeed, to a good accuracy, the latter exhibits a non-linear oscillation pattern and can be
approximated by a quasi-harmonic function with frequency of
oscillations depending on their amplitude, e.g.
\begin{eqnarray}
U\simeq U_0 \cos(\omega t+\phi_0)\,, \quad \omega\equiv \frac{2\pi}{T_U}=kgU_0\,, \quad
k=\frac{2\pi}{B(\frac{1}{4},\frac{1}{2})}\simeq 1.2 \label{appU}
\end{eqnarray}
where $T_U$ is the period of YM condensate oscillations, and $B(x,y)$ is the 
Euler beta function. The maximal error of this approximation is limited by $\Delta U/U\lesssim 0.07$.

The energy spectrum of quasi-harmonic YM condensate fluctuations can be found
in standard way from the Schr\"odinger steady-state equation and is
shown in Fig.~\ref{fig:free-YMC}(right). Starting from the
Hamiltonian density for the free (non-interacting) YM condensate (\ref{eqU})
one arrives at the Schr\"odinger equation
\begin{equation}
\Big[\frac{1}{6}\frac{d^{\,2}}{dU^2}+\Big(E-\frac{3}{2}\,g^2
U^4\Big)\Big]\Psi=0\,.
 \label{E4}
\end{equation}
It straightforward to show that the free YM condensate spectrum corresponds to
a potential well of the fourth power. Numerical calculation provides us 
with the first few energy levels in the spectrum (see also, Ref.~\cite{Hautot}), e.g.
\begin{equation}
E_n=\widetilde{E}_n\;\frac{g^{2/3}}{3^{1/3}}\,,\qquad
\widetilde{E}_0\simeq 0.5\,,\quad \widetilde{E}_1\simeq 1.9\,,\quad
\widetilde{E}_2\simeq 3.7\,,\quad \widetilde{E}_3=5.8\,,\quad
\widetilde{E}_4=8.13\,,\quad \dots
\end{equation}
For practical use, it is convenient to come up with an approximate
analytic formula for the lower end of this spectrum, e.g. in the
following form
\begin{equation}
\widetilde{E}_n=(1.06+0.395\,n^{0.53})(n+0.5)\,,\qquad
n=0,1,2,\,\dots \label{appEU}
\end{equation}
The maximal error of this formula for the first thirty energy levels
does not exceed 4\%.

\subsection{Condensate and wave dynamics in the linear approximation}
\label{Sec:YM-Ham-analysis:lin}

Consider now the physically interesting and realistic case of the YM 
wave modes interacting with the YM condensate. We start with the linear (in waves) 
approximation where only interactions between YM wave modes and YM condensate
are taken into account while interactions between different
wave modes are not included. In practice, this situation corresponds
to a YM system in the beginning of its real time evolution with very few
wave modes such that the interactions between waves are negligibly 
small compared to ``wave-condensate'' interactions. 
Let us consider dynamics of such a linearized system in detail.

\subsubsection{First-order Yang-Mills equations of motion}
\label{Sec:YM-Ham-analysis:lin:eom}

The all-order YM equation of motion (\ref{eq}) for the YM condensate, 
$U=U(t)$, interacting with the wave modes, $\widetilde{A}_{ik}$, 
can be written as follows
\begin{eqnarray} \nonumber
&& -\,\delta_{lk}(\partial_0\partial_0 U+2g^2 U^3)
+(-\partial_0\partial_0 \widetilde{A}_{lk} +\partial_i\partial_i
\widetilde{A}_{lk} -\partial_i\partial_k \widetilde{A}_{li}- g
e_{lmk}\partial_i\widetilde{A}_{mi}U
-2ge_{lip}\partial_i\widetilde{A}_{pk}U \\
&& -\,ge_{lmi}\partial_k\widetilde{A}_{mi}U
+g^2\widetilde{A}_{kl}U^2- g^2\widetilde{A}_{lk}U^2
-2g^2\delta_{lk}\widetilde{A}_{ii}U^2)+
(-ge_{lmp}\partial_i\widetilde{A}_{mi}\widetilde{A}_{pk} \nonumber \\
&& -\,2ge_{lmp}\widetilde{A}_{mi}\partial_i\widetilde{A}_{pk}
-ge_{lmp}\partial_k\widetilde{A}_{mi}\widetilde{A}_{pi}
+g^2\widetilde{A}_{li}\widetilde{A}_{ik}U+
g^2\widetilde{A}_{li}\widetilde{A}_{ki}U
+g^2\widetilde{A}_{ik}\widetilde{A}_{il}U \nonumber \\
&& -\,2g^2\widetilde{A}_{ii}\widetilde{A}_{lk}U-
g^2\delta_{lk}\widetilde{A}_{pi}\widetilde{A}_{pi}U)+
g^2(\widetilde{A}_{li}\widetilde{A}_{pk}\widetilde{A}_{pi}-
\widetilde{A}_{pi}\widetilde{A}_{pi}\widetilde{A}_{lk})=0\,.
\label{main}
\end{eqnarray}
The constraint equation (\ref{coupleq}) linearized in waves reads
\begin{equation}
\partial_i\partial_0\widetilde{A}_{li}
-ge_{lmi}\partial_0U\widetilde{A}_{mi}
+ge_{lmi}\partial_0\widetilde{A}_{mi}U
+ge_{lmp}\partial_0\widetilde{A}_{mi}\widetilde{A}_{pi}=0\,.
\label{inmcoupeq}
\end{equation}
Note, the equation (\ref{main}) is separable by averaging over the
Heisenberg state vector. Also, as the matter of perturbation theory approach 
the zeroth-order YM condensate equation (\ref{eqU}) has to be fulfilled in order to find 
the equations of motion for the YM wave modes to the first (linear) approximation.

It is convenient to turn to Fourier transforms for $\widetilde{A}_{ik}$
modes and expand them over the tensor basis \cite{Landau}. In terms
of symmetric and antisymmetric parts, the tensor field
$\widetilde{A}_{ik}$ reads
\begin{equation}
\widetilde{A}_{ik}=\psi_{ik}+e_{ikl}\chi_{l}\,.
\end{equation}
Then, we expand the Fourier transforms of antisymmetric $\chi_{l}$
and symmetric $\psi_{ik}$ modes into the tensor basis as
\begin{equation}
\chi^{\vec p}_l=s_l^\sigma \eta^{\vec p}_\sigma + n_l \lambda^{\vec
p}\,, \label{newAntiSymmModes}
\end{equation}
and
\begin{equation}
\psi^{\vec p}_{ik}=\psi^{\vec
p}_{\lambda}Q^\lambda_{ik}+\varphi^{\vec p}_\sigma(n_i
s^\sigma_k+n_k s^\sigma_i)+(\delta_{ik}-n_i n_k)\Phi^{\vec p} +n_i
n_k \Lambda^{\vec p}\,, \label{newModes}
\end{equation}
respectively, where the coefficients satisfy the following
conditions
\begin{equation}
Q^{\lambda}_{ik}=Q^{\lambda}_{ki}\,, \qquad Q^{\lambda}_{ii}=0\,,
\qquad p_i Q^\lambda_{ik}=0\,, \qquad p_k s_k=0\,.
\end{equation}
In Eqs.~(\ref{newAntiSymmModes}) and (\ref{newModes}), the 3-vectors 
$n_i$ and $s_k$ are the longitudinal and transverse unit vectors, respectively, 
and $\vec p$ is the corresponding Fourier 3-momentum. In what follows, we omit the
Fourier momentum index $\vec p$. 

Next, let us rewrite the general equation of motion (\ref{main}) 
(after a proper subtraction of the YM condensate equation (\ref{eqU})) 
through the new d.o.f.'s to the linear (in waves) approximation as follows
\begin{eqnarray}
 && \partial_0\partial_0\psi_\lambda + p^2\,\psi_\lambda +
 igp\,U\,Q^{\lambda\gamma}\,\psi_{\gamma}=0\,,
 \label{eq1}
\end{eqnarray}
and
\begin{eqnarray}
 && \partial_0\partial_0\Lambda + 2igp\,\lambda\, U + 2g^2U^2\,
 (2\Phi+\Lambda)=0\,, \nonumber \\
 && \partial_0\partial_0\Phi + p^2\,\Phi + 2igp\,\lambda\, U
 + 2g^2\,U^2\,(2\Phi+\Lambda)=0\,, \nonumber \\
 && \partial_0\partial_0\lambda + p^2\,\lambda -
 igp\,(2\Phi+\Lambda)\,U + 2g^2\,U^2\,\lambda=0\,,
 \nonumber \\
 &&
 \partial_0\partial_0\phi_\sigma + \frac{p^2}{2}\,\phi_\sigma -
 \frac{p^2}{2}\,e^{\sigma\gamma}\,\eta_\gamma-
 igp\,U\,e^{\gamma\sigma}\,\phi_\gamma=0\,, \nonumber \\
 && \partial_0\partial_0\eta_\sigma + \frac{p^2}{2}\,\eta_\sigma -
 \frac{p^2}{2}\,e^{\gamma\sigma}\,\phi_\gamma+
 igp\,U\,e^{\gamma\sigma}\,\eta_\gamma +
 2g^2\,U^2\,\eta_\sigma=0\,, \label{eq2}
\end{eqnarray}
where the following shorthand notations
\begin{equation}
e^{\alpha\beta}\equiv e_{ikm}s_i^{\alpha}n_k s^{\beta}_m\,, \qquad
Q^{\lambda\gamma}\equiv e_{lip}n_iQ^{\lambda}_{lk}Q^{\gamma}_{kp}\,,
\end{equation}
are adopted. Therefore, one arrives at the system of nine equations of motion for
nine d.o.f.'s. Finally, the equations of constraint (\ref{inmcoupeq})
can be conveniently transformed to the following explicit form in
terms of new d.o.f.'s
\begin{eqnarray}
 &&\partial_0(ip\partial_0\,\Lambda-2g\,\lambda\,\partial_0
 U+2g\,\partial_0\lambda\,
 U)=0\,,\label{mainCeq1} \\
 &&\partial_0(ip\partial_0\,\phi_\gamma + ipe^{\gamma\sigma}\,
 \partial_0 \eta_\sigma - 2g\,\eta_\gamma\,\partial_0 U +
 2g\,\partial_0\eta_\gamma\,U)=0\,. \label{mainCeq2}
\end{eqnarray}
It is straightforward to check that these two constraints are
automatically satisfied for a solution of the system of YM equations
(\ref{eq1}) and (\ref{eq2}) (see Appendix \ref{A2} for consistency 
checks of the corresponding equations of motion). Note, 
in a non-gauged YM theory, these constraints do not explicitly 
contain time derivatives, i.e.
\begin{eqnarray}
&& ip\partial_0\Lambda-2g\lambda\partial_0 U+2g\partial_0\lambda
U=0\,, \label{1CeqNOd0} \\
&& ip\partial_0\phi_\gamma+ipe^{\gamma\sigma}\partial_0
\eta_\sigma-2g\eta_\gamma\partial_0 U+2g\partial_0\eta_\gamma
U=0\,,\label{2CeqNOd0}
\end{eqnarray}
which are thus the first integrals of motion.

For further considerations, it is instructive
to represent quadratic Lagrangian and Hamiltonian densities of the
$SU(2)$ YM wave modes interacting with the YM condensate $U=U(t)$ in
terms of the new d.o.f. as follows
\begin{eqnarray} \nonumber
 \mathcal{L}^{\rm waves}_{\rm YM}&=&\,\frac12\Big\{\partial_0\psi_\lambda\,\partial_0\psi_\lambda^\dagger +
 \partial_0\phi_\sigma\,\partial_0\phi_\sigma^\dagger + \partial_0\Phi\,\partial_0\Phi^\dagger +
 \frac{1}{2}\,\partial_0\Lambda\,\partial_0\Lambda^\dagger +
 \partial_0\eta_\sigma\,\partial_0\eta_\sigma^\dagger \\
 &+&
 \partial_0\lambda\partial_0\lambda^\dagger - p^2\,\psi_\lambda\psi_\lambda^\dagger -
 \frac{p^2}{2}\,\phi_\sigma\phi_\sigma^\dagger - p^2\,\Phi\Phi^\dagger -
 \frac{p^2}{2}\,\eta_\sigma\eta_\sigma^\dagger - p^2\,\lambda\lambda^\dagger  \nonumber \\
 &+&
 \frac{p^2}{2}\,e^{\gamma\sigma}(\eta_\sigma\phi_\gamma^\dagger +
 \phi_\gamma\eta_\sigma^\dagger) - igp\,U\,e^{\sigma\gamma}\eta_\sigma\eta_\gamma^\dagger +
 igp\,U\,Q^{\lambda\gamma}\psi_\lambda\psi_\gamma^\dagger   \label{Lmain} \\
 &+&
 igp\,U\,e^{\sigma\gamma}\phi_\sigma\phi_\gamma^\dagger +
 igp\,U\,(2\Phi\lambda^\dagger - 2\lambda\Phi^\dagger +
 \Lambda\lambda^\dagger - \lambda\Lambda^\dagger)  \nonumber \\
 &-&
 2g^2\,U^2\,\eta_\sigma\eta_\sigma^\dagger - 2g^2\,U^2\,\lambda\lambda^\dagger -
 g^2\,U^2\,(4\Phi\Phi^\dagger + 2\Phi\Lambda^\dagger + 2\Lambda\Phi^\dagger +
 \Lambda\Lambda^\dagger)\Big\}\,,\nonumber
\end{eqnarray}
\begin{eqnarray} \nonumber
 \mathcal{H}^{\rm waves}_{\rm YM}&=&\,\frac12\Big\{\partial_0\psi_\lambda\,\partial_0\psi_\lambda^\dagger +
 \partial_0\phi_\sigma\,\partial_0\phi_\sigma^\dagger + \partial_0\Phi\,\partial_0\Phi^\dagger +
 \frac{1}{2}\,\partial_0\Lambda\,\partial_0\Lambda^\dagger +
 \partial_0\eta_\sigma\,\partial_0\eta_\sigma^\dagger \nonumber \\
 &+& \partial_0\lambda\,\partial_0\lambda^\dagger +
 p^2\,\psi_\lambda\psi_\lambda^\dagger +
 \frac{p^2}{2}\,\phi_\sigma\phi_\sigma^\dagger + p^2\,\Phi\Phi^\dagger +
 \frac{p^2}{2}\,\eta_\sigma\eta_\sigma^\dagger + p^2\,\lambda\lambda^\dagger \nonumber \\
 &-& \frac{p^2}{2}\,e^{\gamma\sigma}(\eta_\sigma\phi_\gamma^\dagger +
 \phi_\gamma\eta_\sigma^\dagger) + igp\,U\,e^{\sigma\gamma}\eta_\sigma\eta_\gamma^\dagger -
 igp\,U\,Q^{\lambda\gamma}\psi_\lambda\psi_\gamma^\dagger \label{H} \\
 &-&  igpUe^{\sigma\gamma}\phi_\sigma\phi_\gamma^\dagger -
 igp\,U\,(2\Phi\lambda^\dagger-2\lambda\Phi^\dagger+\Lambda\lambda^\dagger -
 \lambda\Lambda^\dagger) \nonumber \\
 &+& 2g^2\,U^2\,\eta_\sigma\eta_\sigma^\dagger + 2g^2\,U^2\,\lambda\lambda^\dagger +
 g^2\,U^2\,(4\Phi\Phi^\dagger + 2\Phi\Lambda^\dagger +
 2\Lambda\Phi^\dagger + \Lambda\Lambda^\dagger)\Big\}\,.\nonumber
\end{eqnarray}
The terms like $U^3$, $U^3\Phi$, $U^2\Phi$ etc do not appear in the 
``condensate + waves'' Hamiltonian above since $U(t)$ is a spatially-homogeneous
function such that any spacial derivatives of $U(t)$ which could give rise 
to these terms simply disappear. It is straightforward to check that the system of equations
(\ref{eq1}) and (\ref{eq2}) can be obtained directly from
Eq.~(\ref{Lmain}) or (\ref{H}) in the usual way consistent with canonical quantisation (see Appendix \ref{A2}). 
Finally, the complete effective $SU(2)$ YM Hamiltonian density properly including the YM condensate
dynamics can be represented in terms of $\mathcal{H}^{\rm waves}_{\rm YM}$ (\ref{H}) as follows
 \begin{eqnarray}
 \mathcal{H}_{\rm YM}=\mathcal{H}_{\rm YMC} +
 \sum_{\vec{p}}\mathcal{H}^{\rm waves}_{\rm YM}\,,
 \label{HYM}
\end{eqnarray}
which will be used below in studies of the dynamical properties of
the ``waves + condensate'' system below.

\subsubsection{Longitudinal Yang-Mills wave modes: free vs interacting case}
\label{Sec:YM-Ham-analysis:lin:long}

To start with, let us consider the limiting case of free YM field without taking
into account its interactions with the YM condensate, i.e. setting $U=0$ in
Eqs.~(\ref{eq1}) and (\ref{eq2}). In this case we have the following
reduced system
\begin{eqnarray*}
&& \partial_0\partial_0\Lambda=0\,,\quad
   \partial_0\partial_0\Phi+p^2\Phi=0\,,\quad
   \partial_0\partial_0\lambda+p^2\lambda=0\,,\quad
   \partial_0\partial_0\psi_\lambda+p^2\psi_\lambda=0\,,\\
&&
\partial_0\partial_0\phi_\sigma+\frac{p^2}{2}\phi_\sigma
- \frac{p^2}{2}e^{\sigma\gamma}\eta_\gamma=0\,,\quad
\partial_0\partial_0\eta_\sigma+\frac{p^2}{2}\eta_\sigma-
\frac{p^2}{2}e^{\gamma\sigma}\phi_\gamma=0\,,
\end{eqnarray*}
and two constraints
\begin{equation*}
ip\partial_0\Lambda=0\,,\qquad
ip\partial_0\phi_\gamma+ipe^{\gamma\sigma}\partial_0
\eta_\sigma=0\,,
\end{equation*}
since the considered case with $U=0$ corresponds to a degenerate YM
theory. Notably, now the constraints allow to eliminate three
d.o.f., namely, $\Lambda$ and $\eta_\alpha$, which can be associated
with three unphysical longitudinal polarisations of free gauge field
$A_\mu^a$ such that
\begin{equation}
\Lambda=0\,, \qquad \eta_\alpha=e^{\alpha\beta}\phi_\beta\,.
\end{equation}

Thus, in the considering limiting case the constraints reduce the
number of physical d.o.f.'s from nine down to six transverse ones.
Note, such a reduction is not possible for non-zeroth interactions
with the YM condensate, e.g. when $U\not=0$ in Eqs.~(\ref{eq1}) and
(\ref{eq2}). This is because in the general case the constraints
(\ref{1CeqNOd0}) and (\ref{2CeqNOd0}) cannot be represented in the
form of motion integrals as it used to take place in the standard
case without the YM condensate such that the longitudinal d.o.f.'s cannot be
eliminated anymore. This fact essentially means that interactions of
the wave modes with the homogeneous YM condensate dynamically generate three
additional d.o.f.'s $\Lambda$ and $\eta_\alpha$, such that both the
longitudinal and transverse polarisations of interacting YM field
have the status of physical d.o.f.'s and therefore must be treated on
the same footing. The latter statement is in a good agreement with
the canonical quantisation procedure (see Appendices \ref{A1} and \ref{A2}) and is
confirmed by numerical analysis of the complete system (\ref{eq1})
and (\ref{eq2}).

The effect of dynamical generation of the longitudinal modes in a YM medium and their 
dynamical role has been previously discussed in the literature (see e.g. Ref.~\cite{Schenke:2008hw}).
Noticeably enough, it turns out that even in the simple linearized model considered above 
the longitudinal modes acquire proper frequencies due to YM condensate-wave interactions. As long as 
one incorporates the homogeneous condensate in the initial YM Lagrangian, the canonical 
quantisation of longitudinal modes appears to be natural and theoretically consistent without 
introducing any ``virtual'' infinitesimal terms as in the degenerate case discussed 
in Sect.~\ref{Sec:YM-Ham:deg} (see also Appendix \ref{A1}). Physically, it means that 
the longitudinal (plasma) waves in the condensate get excited together with transverse 
ones and the both contribute to observable quantities and should be studied on the same footing.

\subsubsection{Evolution of the Yang-Mills wave modes}
\label{Sec:YM-Ham-analysis:lin:waves}

Now, consider dynamics of the wave modes in the linear approximation
given by the system of YM equations (\ref{eqU}),
(\ref{eq1}) and (\ref{eq2}) without taking into account ``back reaction'' effects 
of the wave modes to the YM condensate. In fact, Eq.~(\ref{eq1}) is a closed
system of two equations for two functions $\psi_k,\;k=1,2$, and thus
can be analysed separately. By an appropriate choice of the frame of
reference, the matrices $Q^{\lambda}_{ij}$ and vector $n_i$ can be
represented in the following simple form
\begin{equation*}
Q^{\lambda=1}_{ij} = \left( \begin{array}{ccc}
1 & 0 & 0 \\
0 & -1 & 0 \\
0 & 0 & 0
\end{array} \right)\,, \qquad
Q^{\lambda=2}_{ij} = \left( \begin{array}{ccc}
0 & 1 & 0 \\
1 & 0 & 0 \\
0 & 0 & 0
\end{array} \right)\,, \qquad n_i=(0,\;0,\;1)\,.
\end{equation*}
Further, introducing superpositions
\begin{equation}
\psi_1\equiv \psi_1' + \psi_2'\,, \qquad \psi_2\equiv i(\psi_1' -
\psi_2') \label{circpol}
\end{equation}
equation (\ref{eq1}) in the above basis falls apart into two
independent equations for $\psi'_{1,2}$
\begin{eqnarray*}
&&\partial_0\partial_0\psi'_1 + (p^2 + 2gp\,U)\psi'_1=0\,, \\
&&\partial_0\partial_0\psi'_2 + (p^2 - 2gp\,U)\psi'_2=0\,,
\end{eqnarray*}
which are recognized as Mathieu equations. The above transformation (\ref{circpol}) has 
a direct analogy with transition from linearly polarised modes to circularly polarised ones,
and is very convenient in the current setting. Above, $U=U(t)$ is a solution of YM 
condensate equation (\ref{appU}). Notably, the parametric
resonance (or instability) domains are well known for this type of
equations. In particular, for the tensor $\psi_k$ mode the first
such parametric resonance instability domain can be found
approximately as $0.15 g\,U_0 \lesssim p \lesssim 4.55 g\, U_0$,
where $U_0\equiv U(t=t_0)$ is an initial value of YM condensate. Other wave
modes have different {\it resonance-like instability domains} which
can be found numerically.
\begin{figure*}[!h]
\begin{minipage}{0.45\textwidth}
 \centerline{\includegraphics[width=0.8\textwidth]{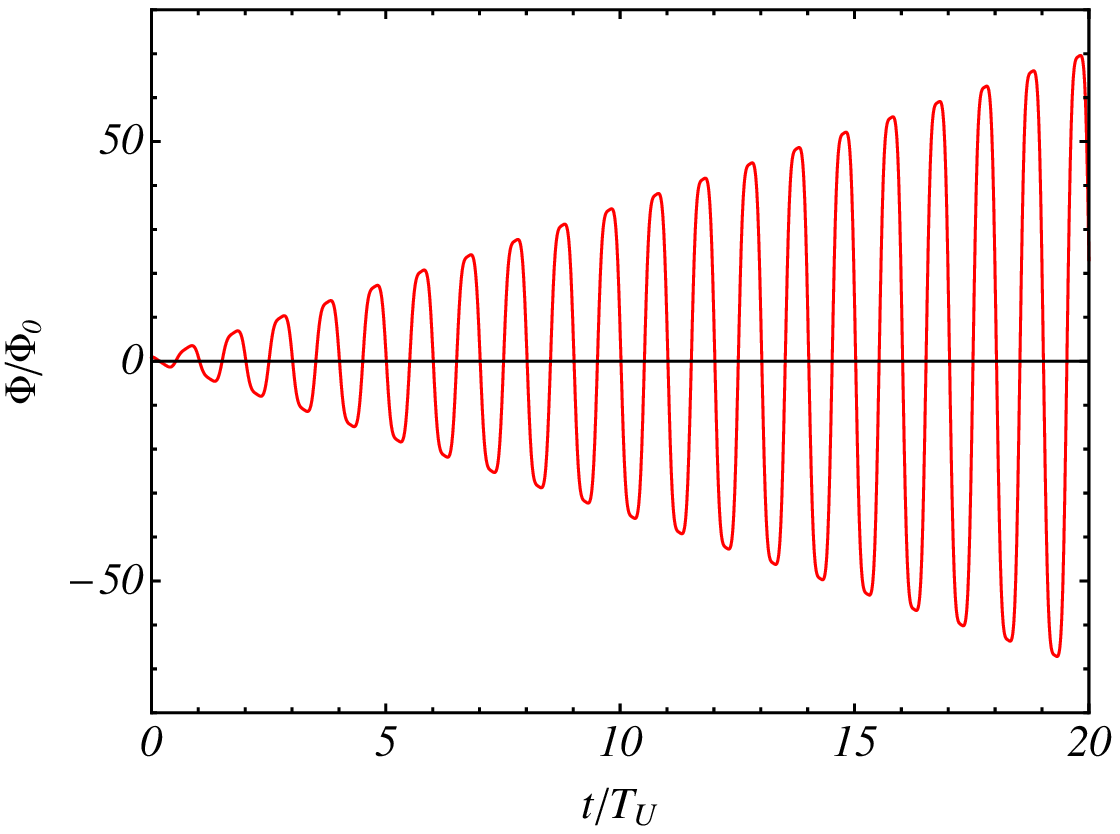}}
\end{minipage}
\hspace{1cm}
\begin{minipage}{0.45\textwidth}
 \centerline{\includegraphics[width=0.8\textwidth]{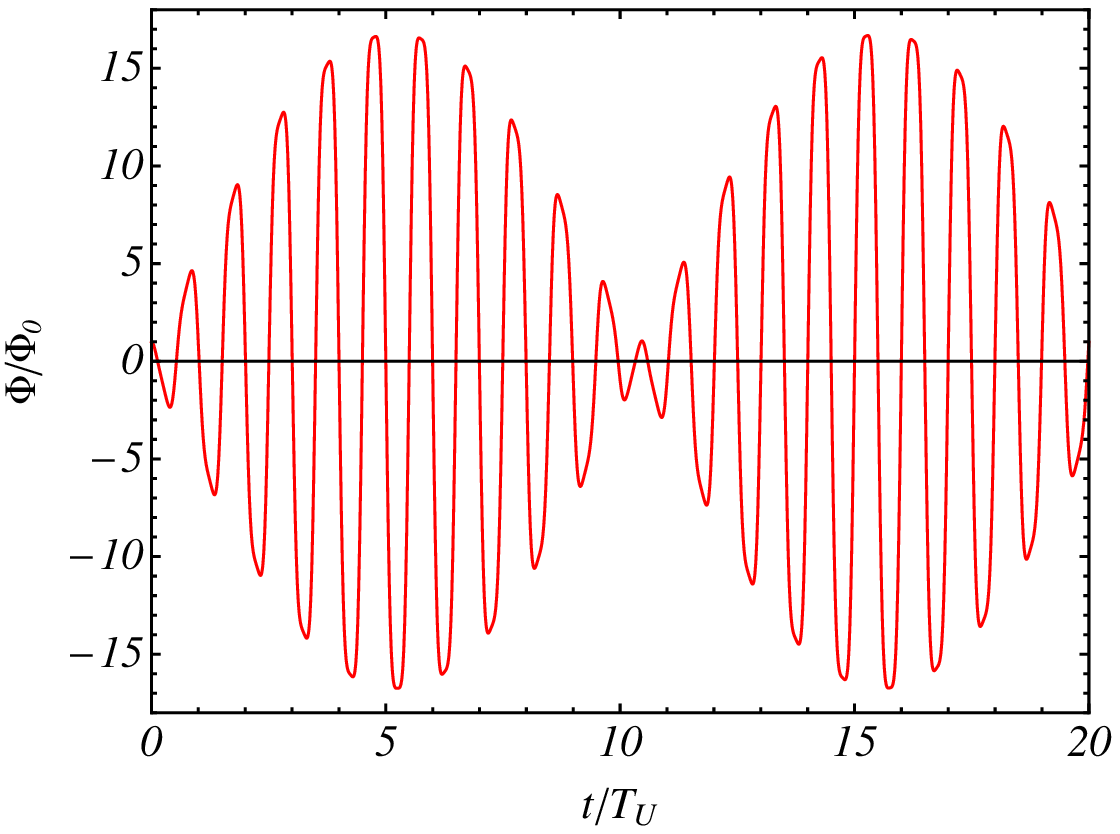}}
\end{minipage}
   \caption{An example of normalized numerical solution of the system
   of YM equations (\ref{eqU}), (\ref{eq1}) and (\ref{eq2}) for one
   of the wave modes, $\Phi=\Phi(t)$, in the case of monotonic growth
   of oscillations for $p=0$ (left), and in the resonance-like
   instability domain with harmonic impulses for a particular
   value $p=0.1g\,U_0$ (right). Here and below, the initial values of all 
   the wave modes ($\Phi_0,\,\lambda_0,\, \dots$) are 
   fixed to be small in comparison to the amplitude of YM 
   condensate oscillations.}
 \label{fig:Phi}
\end{figure*}

An analytical analysis of remaining equations (\ref{eq2}) is less
feasible due to the presence of quadratic term in YM condensate, $\sim U^2$.
In addition, our numerical study has shown that one should not use
the approximation (\ref{appU}) in this case such that dynamics of
the wave modes becomes very different from the well-known picture of
parametric resonance in the case of Mathieu equations. Nevertheless,
exact numerical analysis of the complete system of equations reveals
the existence of the resonance-like instability domains for all of
the wave d.o.f. analogical to the rigorous parametric (Mathieu)
resonance one of the $\psi_k$ mode.

As an example, in Fig.~\ref{fig:Phi} we represent the normalised
numerical solution for one of the wave modes, $\Phi=\Phi(t)$, for
two distinct cases: a solution with monotonically growing amplitude
for $p=0$ (left) and a solution from the parametric resonance-type
instability domain with harmonic impulses for a particular value of
momentum $p=0.1\,g\,U_0$ (right). The observed growth of $\Phi$
amplitude or, equivalently, its energy is triggered by its
interactions with the YM condensate. The same effect has been observed for all
other modes as well. We thus conclude that the particles energy
dynamically increases in the course of time evolution of ``particles
$+$ condensate'' system due to parametric resonance-like instability
of numerical solutions of the non-linear YM equations.

Note, in all the numerical calculations the initial values of all the wave modes 
($\Phi_0,\,\lambda_0,\, \dots$) are fixed to be small in 
comparison to the amplitude of YM condensate oscillations.
This is in order to be consistent with the quasi-classical approximation.
Under this key assumption the characteristic behaviour of all 
the numerical solutions in terms of the dimensionless quantities 
does not strongly depend on actual numerical values for the 
initial fields used in the calculations.

\subsection{``Back reaction'' of the waves to the condensate}
\label{Sec:YM-Ham-analysis:BR}

Due to energy conservation the growth of energy of the wave modes
observed in the previous Section has to be followed by a certain
redistribution of energy between YM condensate and wave modes. In order to
take into account this effect consistently it is necessary to
incorporate second-order contributions to the YM condensate equation for
$U=U(t)$ (\ref{eqU}) which account for interactions between
condensate and particles as follows
\begin{eqnarray} \nonumber
&& \partial_0\partial_0 U + 2g^2\,U^3 +
\frac{g^2}{6}\,U\,\sum_{\vec{p}}\; \Big\langle
2\eta_\sigma\eta_\sigma^\dagger + 2\lambda\lambda^\dagger +
4\Phi\Phi^\dagger + \Lambda\Lambda^\dagger + 2\Phi\Lambda^\dagger \\
&& +\; 2\Lambda\Phi^\dagger + 2\eta^\dagger_\sigma\eta_\sigma +
2\lambda^\dagger\lambda + 4\Phi^\dagger\Phi + \Lambda^\dagger\Lambda +
2\Phi^\dagger\Lambda + 2\Lambda^\dagger\Phi \Big\rangle  \nonumber \\
&& +\; \frac{ig}{12}\,\sum_{\vec{p}} p\; \Big\langle 2\Lambda^\dagger\lambda
- 2\Lambda\lambda^\dagger + 2\Phi^\dagger\lambda -
2\Phi\lambda^\dagger + 2\lambda\Phi^\dagger - 2\lambda^\dagger\Phi \nonumber \\
&& -\; Q^{\lambda\sigma}(\psi_\lambda\psi^\dagger_\sigma -
\psi^\dagger_\lambda\psi_\sigma) -
e^{\sigma\gamma}(\phi_\sigma\phi^\dagger_\gamma -
\phi^\dagger_\sigma\phi_\gamma) + \phi^\dagger_\sigma\eta_\sigma -
\phi_\sigma\eta^\dagger_\sigma \nonumber \\
&& +\; \eta_\sigma^\dagger\phi_\sigma - \eta_\sigma\phi^\dagger_\sigma +
e^{\sigma\gamma}(\eta_\sigma\eta^\dagger_\gamma -
\eta^\dagger_\sigma\eta_\gamma) \Big\rangle = 0 \,. \label{BackReact}
\end{eqnarray}
Here, the averaging $\langle\dots\rangle$ is performed over the
Heisenberg vector state. Since equal wave modes with different
momenta do not interact with each other then all their products
disappear upon the averaging, so only products of different
(interacting) modes remain. Besides, all linear and cubic terms in
waves also disappear in the considering next-to-linear approximation
as well.

The effective second-order Hamiltonian density incorporating the
``back reaction'' effect of the wave modes to the YM condensate can be
represented as a sum of three components corresponding to the free
YM condensate, $\mathcal{H}_{\rm U}$, free wave modes (or particles),
$\mathcal{H}_{\rm particles}$, and the term accounting for
interactions between these two subsystems, $\mathcal{H}_{\rm int}$,
respectively. The corresponding terms in the total effective Hamiltonian 
density $\mathcal{H}_{\rm YM}$ given by Eq.~(\ref{HYM}) are as follows
\begin{eqnarray*}
\mathcal{H}_{\rm U} &=& \frac{3}{2}\big(\partial_0U\partial_0U +
g^2\,U^4\big) \,, \\
\mathcal{H}_{\rm particles} &=& \frac{1}{2}\sum_{\vec{p}}\,\Big( \partial_0\psi_\lambda\,\partial_0\psi_\lambda^\dagger +
 \partial_0\phi_\sigma\,\partial_0\phi_\sigma^\dagger + \partial_0\Phi\,\partial_0\Phi^\dagger +
 \frac{1}{2}\,\partial_0\Lambda\,\partial_0\Lambda^\dagger +
 \partial_0\eta_\sigma\,\partial_0\eta_\sigma^\dagger \nonumber \\
 &+& \partial_0\lambda\,\partial_0\lambda^\dagger +
 p^2\,\psi_\lambda\psi_\lambda^\dagger +
 \frac{p^2}{2}\,\phi_\sigma\phi_\sigma^\dagger + p^2\,\Phi\Phi^\dagger +
 \frac{p^2}{2}\,\eta_\sigma\eta_\sigma^\dagger + p^2\,\lambda\lambda^\dagger \nonumber \\
 &-& \frac{p^2}{2}\,e^{\gamma\sigma}(\eta_\sigma\phi_\gamma^\dagger +
 \phi_\gamma\eta_\sigma^\dagger) \Big) \,, \\
\mathcal{H}_{\rm int} &=& \frac{1}{2}\sum_{\vec{p}}\,\Big[  igp\,U\,e^{\sigma\gamma}\eta_\sigma\eta_\gamma^\dagger -
 igp\,U\,Q^{\lambda\gamma}\psi_\lambda\psi_\gamma^\dagger \\
 &-&  igpUe^{\sigma\gamma}\phi_\sigma\phi_\gamma^\dagger -
 igp\,U\,(2\Phi\lambda^\dagger-2\lambda\Phi^\dagger+\Lambda\lambda^\dagger -
 \lambda\Lambda^\dagger) \nonumber \\
 &+& 2g^2\,U^2\,\eta_\sigma\eta_\sigma^\dagger + 2g^2\,U^2\,\lambda\lambda^\dagger +
 g^2\,U^2\,(4\Phi\Phi^\dagger + 2\Phi\Lambda^\dagger +
 2\Lambda\Phi^\dagger + \Lambda\Lambda^\dagger) \Big] \,.
\end{eqnarray*}
It can be seen from these expressions that interaction term
$\mathcal{H}_{\rm int}$ is not sign-definite in distinction to
positively-definite condensate $\mathcal{H}_{\rm U}$ and waves
$\mathcal{H}_{\rm particles}$ contributions.
\begin{figure*}[!h]
\begin{minipage}{0.45\textwidth}
 \centerline{\includegraphics[width=0.9\textwidth]{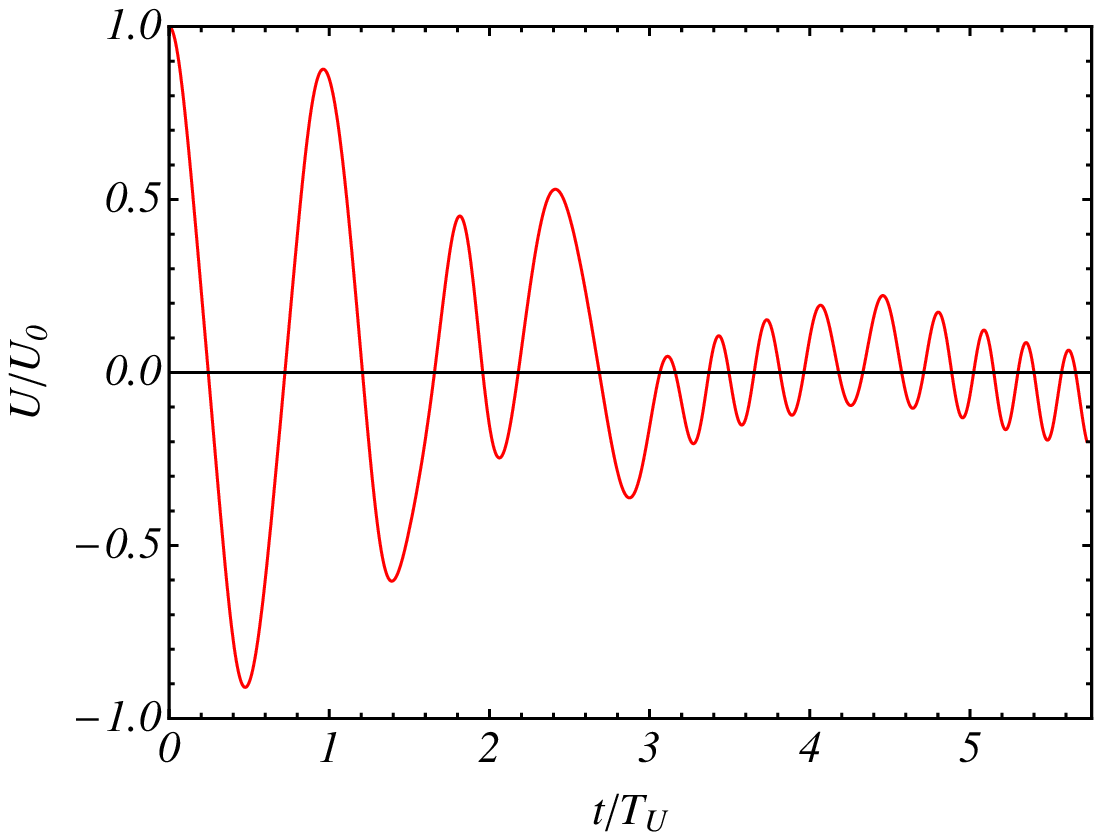}}
\end{minipage}
\hspace{1cm}
\begin{minipage}{0.45\textwidth}
 \centerline{\includegraphics[width=0.9\textwidth]{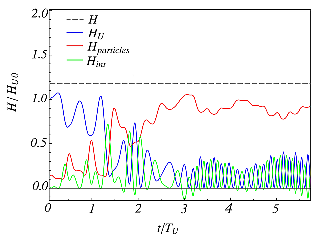}}
\end{minipage}
   \caption{Time dependence of the YM condensate
   in the quasilinear approximation in the complete system of $SU(2)$
   wave modes (left), and the evolution of condensate $\mathcal{H}_{\rm U}$, 
   YM wave modes (or particles)
   $\mathcal{H}_{\rm particles}$, and interaction term
   $\mathcal{H}_{\rm int}$ contributions to the total energy
   in the complete ``condensate + waves'' system (right). The illustrated
   numerical solutions are realistic in the quasi-classical limit of relatively small wave
   amplitudes, i.e. when $|\widetilde{A}_{ik}|\ll |U|$ corresponding
   to $t/T_{\rm U}\lesssim 1$. On the right panel, $\mathcal{H}_{U_0}$ is the 
   initial energy of the YM condensate which is taken to be strongly 
   dominating over the initial values of $\mathcal{H}_{\rm particles}$ and
   $\mathcal{H}_{\rm int}$ terms.}
 \label{fig:YMC-int}
\end{figure*}

In our numerical analysis and in all the plots in this paper we
consider the complete system of all nine wave d.o.f. and YM condensate
including interactions between them. We found that wave-condensate
interactions lead to a decrease of amplitude of the YM condensate oscillations
in time as is seen in Fig.~\ref{fig:YMC-int} (left). An analogical
picture of damping of the condensate oscillations is observed in the
reduced (closed) system of $\Phi$, $\Lambda$ and $\lambda$ wave
modes and the YM condensate. We have also calculated the energy evolution of
particles and condensate shown in Fig.~\ref{fig:YMC-int} (right).
These plots clearly illustrate the energy transfer (swap) effect
from the YM condensate to particles due to interactions between them.
\begin{figure*}[!h]
 \centerline{\includegraphics[width=0.4\textwidth]{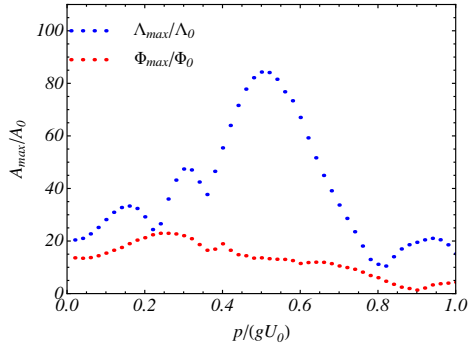}}
   \caption{The particle momentum dependence of the ratio of 
   wave amplitude (maximum over the period of oscillations) at a fixed final $t>t_0$
   (when most of the energy of the system is concentrated in waves)
   to the corresponding initial amplitude at $t=t_0$ (when all energy of the system
   is concentrated in the YM condensate). The results are shown for transverse
   $\Phi$ and longitudinal $\Lambda$ modes.}
 \label{fig:Phi-energy}
\end{figure*}

One of the important issues in the YM theory relates to investigation of 
energy redistribution is the momentum spectrum of YM waves as a result 
of plasma self-interactions \cite{Kurkela:2012hp,Berges:2012ev,Berges:2013fga}.
Interactions between the YM condensate and wave modes (particles)
lead to a redistribution of energy between the modes with different
impulses which rather strongly depends on particles momentum due to
parametric resonance-like instability of YM solutions. The latter
happens because the interaction strength, and hence the energy
transfer rate, depends on the amplitude of impulses which is
different for different modes and particles momenta (for a given
mode). 

As an illustration of this effect, in Fig.~\ref{fig:Phi-energy} we show the 
particle momentum dependence (in dimensionless units) of the ratio of wave amplitude 
maximal over the period of oscillations at a fixed final $t>t_0$ to the corresponding 
initial amplitude at $t=t_0$ (when all energy of the system is concentrated 
in the YM condensate). The results are shown for transverse $\Phi$ and longitudinal
$\Lambda$ modes and are qualitatively the same for all the other
modes. We notice that an increase in wave amplitude due to
parametric instability may be rather strong close to the peak
regions in corresponding energy spectra. One therefore observes the
ultra-relativistic particles production effect with momenta close to
the resonant momenta $p\sim g U_0$ in a vicinity of maxima points in
Fig.~\ref{fig:Phi-energy}.

As the main physical result to be emphasised here, we have found the
significant energy swap effect between the YM condensate and particle-like
modes of the ultra-relativistic YM plasma due to their interactions
in quasilinear approximation. Due to energy conservation it is clear
that the parametric resonance for the $\psi_k$ mode or a
resonance-like instability for other modes in general is accompanied
by an energy flow from the YM condensate to the waves. So the resonance-like
instability of the quantum-wave solutions in the classical
condensate is the physical reason of the energy swap effect. Note,
that in Fig.~\ref{fig:YMC-int} (right) we should restrict ourselves
to maximal time scales of about one period of YM condensate fluctuations,
$t/T_{\rm U}\lesssim 1$. At larger time scales $t/T_{\rm U}>1$ the
amplitude of the wave fluctuations becomes comparable to the
amplitude of condensate fluctuations so the considering quasilinear
approximation breaks down there. Let us study sensitivity of our
solutions with respect to higher-order corrections in detail.

\subsection{Higher-order corrections}
\label{Sec:YM-Ham-analysis:HO}

So far we have considered the $SU(2)$ YM wave dynamics in the
classical YM condensate in the first (leading or linear)
approximation, while the YM condensate dynamics -- in the linear and
next-to-linear (or quasi-linear) approximations. The range of
applicability of our quasi-classical analysis is limited to small
quantum-wave fluctuations $\widetilde{A}_{ik}$ in the condensate $U$
considered as a classical background, i.e. in the
$\widetilde{A}_{ik}\ll U$ asymptotics.

\begin{figure*}[!h]
 \centerline{\includegraphics[width=0.4\textwidth]{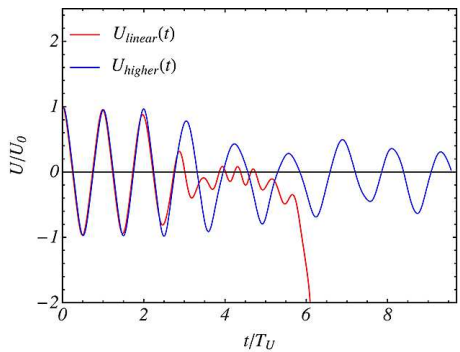}}
   \caption{Time evolution of the YM condensate in the complete next-to-linear problem given
   by a numerical solution of Eqs.~(\ref{eq1}), (\ref{eq2}) and
   (\ref{BackReact}) (red line) and in the problem with extra higher
   order terms included (\ref{effmass3}) (blue line).}
 \label{fig:YMC-higher}
\end{figure*}

It has been demonstrated above that the interactions between the
wave modes and the condensate lead to an increase of energy
accumulated by the wave modes at expense of a corresponding decrease
of YM condensate energy. This means that as some point in evolution of the
system the amplitude of wave modes becomes too large so that the
initial approximation $\widetilde{A}_{ik}\ll U$ breaks down. Such a
breakdown can be noticed in numerical results in the next-to-linear
approximation, e.g. in YM condensate dynamics given by a numerical solution of
Eqs.~(\ref{eq1}), (\ref{eq2}) and (\ref{BackReact}) illustrated in
Fig.~\ref{fig:YMC-higher} by red line. It is clearly seen from this
figure that at relatively large time scales increasing the time
domain of effective energy swap from the condensate to waves, the
YM condensate energy becomes singular and unbounded. The latter anomaly is due
to breakdown of the next-to-linear approximation. In what follows, we
show that inclusion of the principal part of the higher-order terms
into the linear equations (\ref{eq1}) and (\ref{eq2}) for the wave
modes allows to eliminate such anomalies and reveals qualitative
stability of the energy swap effect under consideration.

The equations of motion for the wave $\widetilde{A}_{ik}$ modes in a
given order are normally constructed after elimination of the
equation of motion for the YM condensate from generic Eq.~(\ref{main}). Denote
the left side of Eq.~(\ref{main}) as $Y_{lk}$ such that equations
for the $\widetilde{A}_{lk}$ modes take the following form
\begin{equation}
\widetilde{Y}_{lk} = Y_{lk} - \langle Y_{lk} \rangle = 0\,,
\label{Ylk}
\end{equation}
where $\langle \dots \rangle$ denotes averaging over the Heisenberg
state vector as usual. For simplicity, we take into account the last
two terms in Eq.~(\ref{main}) only. Omitting all other higher-order
terms in Eq.~(\ref{Ylk}) we have
\begin{eqnarray}
 && \partial_0\partial_0\widetilde{A}_{lk} - \partial_i\partial_i\widetilde{A}_{lk} +
    \partial_i\partial_k\widetilde{A}_{li} \nonumber \\
 && +\; g\, e_{lmk}\partial_i\widetilde{A}_{mi}U + 2g\,e_{lip}\partial_i\widetilde{A}_{pk}U +
    g\,e_{lmi}\partial_k\widetilde{A}_{mi}U \label{Alk3} \\
 && -\; g^2\,\widetilde{A}_{kl}U^2 + g^2\,\widetilde{A}_{lk}U^2 +
    2g^2\,\delta_{lk}\widetilde{A}_{ii}U^2 +
    g^2\,(\widetilde{A}_{pi}\widetilde{A}_{pi}\widetilde{A}_{lk} -
    \widetilde{A}_{li}\widetilde{A}_{pk}\widetilde{A}_{pi} ) + \dots = 0\,. \nonumber
\end{eqnarray}

Now let us multiply $\widetilde{Y}_{lk}(x,t)$ by
$\widetilde{A}_{lk}$ in a different space-time point $(x',t')$ in
order to make the terms of the fourth order such that
\begin{equation}
 \langle \widetilde{A}_{lk}(x',t')\widetilde{Y}_{lk}(x,t) \rangle = 0\,,
\end{equation}
due to Eq.~(\ref{Ylk}). Here, each fourth-order term can then be
transformed as follows
\begin{eqnarray}
\langle A_1A_2A_3A_4 \rangle =\frac13\big[\langle A_1A_2\rangle \langle A_3A_4\rangle +
\langle A_1A_3\rangle \langle A_2A_4\rangle +\langle A_1A_4\rangle \langle A_2A_3\rangle] \,.
\label{psi1234}
\end{eqnarray}
Dropping $\widetilde{A}_{lk}(x',t')$ function and performing Fourier
transform of the remaining terms we obtain
\begin{eqnarray}
 && \partial_0\partial_0\widetilde{A}_{lk} + p^2\,\widetilde{A}_{lk} -
 p^2\,n_in_k\,\widetilde{A}_{li} + igp\,e_{lmk}n_i\widetilde{A}_{mi}\,U \nonumber \\
 && +\; 2ig\,p\,e_{lip}n_i\partial_i\widetilde{A}_{pk}\,U + ig\,p\,e_{lmi}n_k\widetilde{A}_{mi}\,U -
 g^2\,\widetilde{A}_{kl}\,U^2  \nonumber \\
 && +\; g^2\,\widetilde{A}_{lk}\,U^2 + 2g^2\,\delta_{lk}\widetilde{A}_{ii}\,U^2 - \frac{g^2}{12}\,
 \Big[ \widetilde{A}_{li}\sum_{\vec{p}}\langle\widetilde{A}_{pi}\widetilde{A}^\dagger_{pk} +
 \widetilde{A}^\dagger_{pi}\widetilde{A}_{pk}\rangle \nonumber \\
 && +\; \widetilde{A}_{pk}\sum_{\vec{p}}\langle\widetilde{A}_{li}\widetilde{A}^\dagger_{pi} +
 \widetilde{A}^\dagger_{li}\widetilde{A}_{pi}\rangle + \widetilde{A}_{pi}\sum_{\vec{p}}
 \langle\widetilde{A}_{li}\widetilde{A}^\dagger_{pk} + \widetilde{A}^\dagger_{li}\widetilde{A}_{pk}\rangle  \nonumber \\
 && -\; 2\widetilde{A}_{pi}\sum_{\vec{p}}\langle\widetilde{A}_{pi}\widetilde{A}^\dagger_{lk} +
 \widetilde{A}^\dagger_{pi}\widetilde{A}_{lk}\rangle - \widetilde{A}_{lk}\sum_{\vec{p}}
 \langle\widetilde{A}_{pi}\widetilde{A}^\dagger_{pi} + \widetilde{A}^\dagger_{pi}\widetilde{A}_{pi}\rangle \Big] +
 \dots = 0 \,. \label{effmass3}
 \end{eqnarray}
We notice here that the third-order terms have transformed to
effective mass terms given by averages of various two operator
products. In particular, Fig.~\ref{fig:YMC-higher} (blue line)
illustrates that such higher-order effective mass terms eliminate
formal singularities in the YM condensate such that the numerical solution
stabilises and the energy swap effect discussed above remains at the
qualitative level.

Certainly, this simplified analysis is not complete and is aimed
only at illustrating that inclusion of the major part of
higher-order terms significantly improves and stabilises the results
of the next-to-linear model. In a fully consistent model one has to
take into account contributions from all the higher-order terms both
in the wave equations of motion and in the YM condensate equation
simultaneously, which will be done elsewhere.

\section{Dynamics of the Yang-Mills plasma}
\label{Sec:YM-plasma}

So far we have considered either pure condensate limit or interacting system 
``condensate + waves'' in the small wave approximation. Now, it is worth to analyse
another important limiting case of pure Yang-Mills (gluon) plasma composed of wave excitations 
only which corresponds to the case of vanishingly small condensate and large waves, i.e. 
$|U(t)|\ll |\widetilde A_{ik}|$.

The basic equation which describes dynamics of the pure gluon plasma in $SU(3)$ theory 
has the following form
\begin{eqnarray} \nonumber
&& \partial_0\partial_0\widetilde{A}_{ai}
-\partial_k\partial_k\widetilde{A}_{ai}
+\partial_i\partial_k\widetilde{A}_{ak} \\ \nonumber
&& +\, 2gf_{abc}\partial_k\widetilde{A}_{ci}\widetilde{A}_{bk}
+gf_{abc}\partial_i\widetilde{A}_{bk}\widetilde{A}_{ck}
+gf_{abc}\partial_k\widetilde{A}_{bk}\widetilde{A}_{ci} \\
&& +\, g^2f_{cba}f_{cb'c'}\,\widetilde{A}_{bk}\widetilde{A}_{b'k}\widetilde{A}_{c'i}=0\,,
\label{non-transformed equation}
\end{eqnarray}
where $a=1,\,\dots\,8$. Multiplying this equation by $\widetilde{A}_{a'i'}(\vec{x'},t')$ and averaging the result 
over the state vector we obtain
\begin{eqnarray} \nonumber
&& \langle\widetilde{A}_{a'i'}(\vec{x'},t')(\partial_0\partial_0\widetilde{A}_{ai}(\vec{x},t)
-\partial_k\partial_k\widetilde{A}_{ai}(\vec{x},t)
+\partial_i\partial_k\widetilde{A}_{ak}(\vec{x},t) \\ \nonumber
&& +\, 2gf_{abc}\partial_k\widetilde{A}_{ci}(\vec{x},t)\widetilde{A}_{bk}(\vec{x},t)
+gf_{abc}\partial_i\widetilde{A}_{bk}(\vec{x},t)\widetilde{A}_{ck}(\vec{x},t)
+gf_{abc}\partial_k\widetilde{A}_{bk}(\vec{x},t)\widetilde{A}_{ci}(\vec{x},t) \\
&& +\, g^2f_{cba}f_{cb'c'}(\widetilde{A}_{bk}(\vec{x},t)m^{b'k}_{c'i}(t)
+\widetilde{A}_{b'k}(\vec{x},t)m^{bk}_{c'i}(t)
+\widetilde{A}_{c'i}(\vec{x},t)m^{bk}_{b'k}(t))\rangle =0\,,
\label{<eq eff mass 3-terms>} 
\end{eqnarray}
where the effective mass matricies
\begin{eqnarray}
m^{bk}_{b'k'}\equiv \frac{1}{3}\langle\widetilde{A}_{bk}\widetilde{A}_{b'k'}\rangle 
=\frac{1}{12}\sum_{\vec{p}}\langle\widetilde{A}^{r\vec{p}}_{bk}\widetilde{A}^{\dagger r\vec{p}}_{b'k'}+
\widetilde{A}^{\dagger r\vec{p}}_{bk}\widetilde{A}^{r\vec{p}}_{b'k'}\rangle  \,.
\label{3-terms mass} 
\end{eqnarray}
In the derivation of effective mass terms we implied that the fourth-order terms resulted from third-order ones 
in Eq.~(\ref{non-transformed equation}) are transformed according to Eq.~(\ref{psi1234}).
Dropping $\widetilde{A}_{a'i'}(\vec{x'},t')$ in Eq.~(\ref{<eq eff mass 3-terms>}) we obtain
\begin{eqnarray} \nonumber
&& \partial_0\partial_0\widetilde{A}_{ai}
-\partial_k\partial_k\widetilde{A}_{ai}
+\partial_i\partial_k\widetilde{A}_{ak} \\ \nonumber
&& +\, 2gf_{abc}\partial_k\widetilde{A}_{ci}\widetilde{A}_{bk}
+gf_{abc}\partial_i\widetilde{A}_{bk}\widetilde{A}_{ck}
+gf_{abc}\partial_k\widetilde{A}_{bk}\widetilde{A}_{ci} \\
&& +\, g^2f_{cba}f_{cb'c'}(\widetilde{A}_{bk}m^{b'k}_{c'i}
+\widetilde{A}_{b'k}m^{bk}_{c'i}
+\widetilde{A}_{c'i}m^{bk}_{b'k})=0 \,.
\label{eq eff mass 3-terms} 
\end{eqnarray}

\subsection{Effective gluon mass from wave self-interactions}
\label{Sec:YM-plasma:toy}

For simplicity, at this very first step here we would like to investigate the influence of the odd-power terms in waves, i.e. the first- and
third-order terms in Eq.~(\ref{eq eff mass 3-terms}) only. Let us divide Eq.~(\ref{eq eff mass 3-terms}) 
into two equations for transverse and longitudinal modes using $\widetilde{A}_{ai}=\widetilde{A}_{ai\perp}+\widetilde{A}_{ai\|}$. 
For this purpose, one multiplies Eq.~(\ref{eq eff mass 3-terms}) by $\widetilde{A}_{a'i'\perp}$ and $\widetilde{A}_{a'i'\|}$ 
and then averages over the state vector. For simplicity, the contributions from second order terms in Eq.~(\ref{non-transformed equation}) 
can be omitted since being multiplied by $\widetilde{A}_{a'i'\perp}$ they are expected to be small upon averaging over the state vector 
in the ``soft'' limit $p\to0$ in the beginning of time evolution. While this toy-model is very useful for understanding of generic features 
of the gluon plasma, in the realistic plasma case one should go beyond this approximation (see below). 

Finally, one notices that the equation of motion for the transversely-polarised gluon plasma 
modes takes the form of the Klein-Gordon-Fock equation
\begin{equation}
\partial_0\partial_0\widetilde{A}_{ai\perp}-\partial_k\partial_k\widetilde{A}_{ai\perp}+
m_{\perp}^2\widetilde{A}_{ai\perp}=0
\label{eqmtr} 
\end{equation}
with effective mass
\begin{equation}
m_{\perp}^2=\frac{1}{12}g^2(\langle \widetilde{A}_{ai\|}\widetilde{A}_{ai\|}\rangle +
\langle \widetilde{A}_{ai\perp}\widetilde{A}_{ai\perp}\rangle ) \,.
\label{mtr} 
\end{equation}
Note, self-interactions of the plasma modes described by the cubic terms in Eq.~(\ref{non-transformed equation})
naturally generate constituent mass of transversely polarised gluons (\ref{mtr}).

In analogy to the transverse modes, dynamics of the longitudinal modes in the pure gluon plasma 
is described by
\begin{equation}
\partial_0\partial_0\widetilde{A}_{ai\|}+m_{\|}^2\widetilde{A}_{ai\|}=0\,
\label{longitudinal} 
\end{equation}
where the effective mass of longitudinal modes
\begin{equation}
m_{\|}^2=\frac{1}{12}g^2(\langle \widetilde{A}_{ai\|}\widetilde{A}_{ai\|}\rangle +
\langle \widetilde{A}_{ai\perp}\widetilde{A}_{ai\perp}\rangle ) \,,
\end{equation}
turns out to be equal to the mass of transverse modes,
\begin{equation}
m^2\equiv m_{\|}^2=m_{\perp}^2 \,.
\label{mtrml} 
\end{equation}
Note, both transverse and longitudinal modes contribute to the effective mass (\ref{mtrml}) 
due to their mutual interactions. The dispersion relations for transverse and longitudinal modes 
follow directly from Eqs.~(\ref{eqmtr}) and (\ref{longitudinal}),
\begin{equation}
\varepsilon_{p,\lambda}^2=p^2+m^2 \,, \qquad \varepsilon_{p,0}^2=m^2 \,.
\end{equation}
The solutions of Eqs.~(\ref{eqmtr}) and (\ref{longitudinal}) can then be found in the following form
\begin{equation}
\widetilde{A}_{i}^{a}=
\displaystyle\sum_{\vec{p},\eta=-1,0,1}
\frac{1}{\sqrt{2V\varepsilon_{p,\eta}}}
e^i_{\vec{p},\eta}\,
(c_{\vec{p},\eta}^{a}e^{-i(\varepsilon_{p,\eta} t-\vec{p}\vec{x})}+
c_{\vec{p},\eta}^{a\dagger}e^{i(\varepsilon_{p,\eta} t-\vec{p}\vec{x})}) \,,
\label{wavesol} 
\end{equation}
where $e^i_{\vec{p},\eta}$ is the plane wave polarisation vector.
Substituting this solution into Eq.~(\ref{mtr}) one arrives at
\begin{equation}
m^2=\frac{1}{12V}g^2\Big(\displaystyle\sum_{\vec{p},\lambda=\pm1}\frac{N_{\vec{p},\lambda}}{\varepsilon_{p,\lambda}}
+\displaystyle\sum_{\vec{p}}\frac{N_{\vec{p},0}}{\varepsilon_{p,0}}\Big)+m_0^2 \,, \label{m2expr}
\end{equation}
where
\begin{equation}
m_0^2=\frac{2}{3V}g^2\displaystyle\sum_{\vec{p}}\Big(\frac{1}{\varepsilon_{p,\lambda}}+\frac{1}{2\varepsilon_{p,0}}\Big) \,. \label{m02expr}
\end{equation}
Here, $N_{\vec{p},\lambda}$ and $N_{\vec{p},0}$ should be treated as the occupation numbers 
of longitudinal and transverse gluons, respectively,
\begin{equation}
[c_{\vec{p},\eta}^{\dagger\, a}c_{\vec{p}\,',\eta'}^{a'}]=-\delta_{\vec{p}\vec{p}\,'}\delta_{\eta\eta'}\delta_{aa'}\,, 
\quad \sum_a\langle c_{\vec{p},\eta}^{\dagger\, a}c_{\vec{p}\,',\eta'}^{a} \rangle=
\delta_{\vec{p}\vec{p}\,'}\delta_{\eta\eta'}N_{\vec{p},\eta}\,, \quad \eta,\eta'=-1,0,+1\,,
\end{equation}

Applying the dimensional regularisation technique, one can represent the divergent sums over 
$\vec p$ in Refs.~(\ref{m2expr}) and (\ref{m02expr}) in $1+n$ dimensions where 
$n=3-2\epsilon$ according to the following example
\begin{eqnarray} \nonumber
\frac{g_n^2n}{V_n}\displaystyle\sum_{|\vec{p}|}\frac{1}{\varepsilon_{p}}=
\frac{n\,g_n^2 2\pi^{n/2}l^{n-3}}{(2\pi)^n\Gamma(n/2)}\int^\infty_0 \frac{p^{n-1}dp}{\varepsilon_p}=
-\frac{g_n^2m^2n}{4\pi(n-1)}\Big(\frac{m^2l^2}{4\pi}\Big)^{\frac{n-3}{2}}\Gamma\Big(\frac{3-n}{2}\Big)\,,
\end{eqnarray}
where the dispersion relation $\varepsilon_p^2=p^2+m^2$ is employed and the parameter $l$ has been introduced to preserve 
the overall dimension. Repeating the same procedure for other relevant sums and then performing $\epsilon\to0$ 
expansion one obtains the following final expressions
\begin{eqnarray}
&& \frac{g_n^2n}{V_n}\displaystyle\sum_{|\vec{p}|}\frac{1}{\varepsilon_{p}}=
-\frac{3g^2}{8\pi^2}m^2\Big(\frac{1}{\epsilon}-\ln\frac{m^2}{\mu^2}+O(\epsilon)\Big)\,, \label{ren1} \\
&& \frac{g_n^2n}{V_n}\displaystyle\sum_{|\vec{p}|}\varepsilon_{p}=
-\frac{3g^2}{32\pi^2}m^4\Big(\frac{1}{\epsilon}-\ln\frac{m^2}{\sqrt{e}\mu^2}+O(\epsilon)\Big)\,, \label{ren2} \\
&& \frac{g_n^2n}{V_n}\displaystyle\sum_{|\vec{p}|}\frac{|\vec{p}|^2}{\varepsilon_{p}}=
\frac{9g^2}{32\pi^2}m^4\Big(\frac{1}{\epsilon}-\ln\frac{e^{\frac{1}{6}}m^2}{\mu^2}+O(\epsilon)\Big) \,, \label{ren3}
\end{eqnarray}
where the renormalization scale is defined as
\begin{equation}
\mu^2=\frac{4\pi e^{\frac{1}{3}-c}}{l^2} \,, \qquad c=0.5772157 \,.
\end{equation}
In what follows we assume that all the terms proportional to $1/\varepsilon$ in Eqs.~(\ref{ren1}) -- (\ref{ren3}) 
are cancelled in a complete remormalizable theory and thus can be discarded.

The form of dispersion relations does not enable us to renormalized the contributions from longitudinal modes, so let us leave them 
in the initial form and transform the vacuum terms induced by the transverse modes only. In this way, we end up with 
the following expression for renormalized mass
\begin{equation}
m^2=\frac13\Big[\frac{1}{4V}g^2\displaystyle\sum_{\vec{p},\lambda=\pm1}\frac{n_{\vec{p},\lambda}}{\varepsilon_{p,\lambda}}
+\frac{1}{4\pi^2}g^2m^2\ln\frac{m^2}{\mu^2}
+\frac{1}{4V}g^2\displaystyle\sum_{\vec{p}}\frac{n_{\vec{p},0}}{\varepsilon_{p,0}}+
\frac{1}{V}g^2\displaystyle\sum_{\vec{p}}\frac{1}{\varepsilon_{p,0}} \Big] \,.
\label{eqmass} 
\end{equation}

A validation criterion for the effective mass result (\ref{eqmass}) is based on a consistency of the considered 
quantum field theory and thermodynamic approaches. Since we considered the gluon plasma in thermodynamic
equilibrium, the same expression for effective mass should be derived from the minimality condition for 
thermodynamic potentials. For example, such a condition for the plasma energy reads
\begin{equation}
\Big(\frac{\partial\langle H\rangle }{\partial m}\Big)_S=0 \,.
\label{minH} 
\end{equation}
Then, the time independence of entropy means that the occupation numbers are constant. 
Under this condition the expression for the plasma energy accounting for contributions from 
quadratic and quartic terms only takes the final form
\begin{equation}
\langle H\rangle =\displaystyle\sum_{\vec{p},\lambda=\pm1}\varepsilon_{p,\lambda}n_{\vec{p},\lambda}+
8\displaystyle\sum_{\vec{p}}\varepsilon_{p,\lambda}+
\displaystyle\sum_{\vec{p}}\varepsilon_{p,0}n_{\vec{p},0}+
4\displaystyle\sum_{\vec{p}}\varepsilon_{p,0}-\frac{3m^4V}{g^2}\,.
\label{energy} 
\end{equation}
It is straightforward to check that by substitution of Eq.~(\ref{energy}) into Eq.~(\ref{minH}) one obtains 
the same expression for the effective mass as the one in Eq.~(\ref{eqmass}) thus validating the above calculations.

\subsection{Plasma equations with second- and third-order wave self-interactions}
\label{Sec:YM-plasma:exact}

Our goal now is to analyse the influence of the second order terms in Eq.~(\ref{eq eff mass 3-terms}) 
to the gluon plasma dynamics. It will be explicitly demonstrated that these terms also contribute 
to the constituent gluon mass which then appears to be non-local in time.

Consider the YM wave field $\widetilde{A}_{ai}$ as the sum of ``real'' and ``virtual'' parts
\begin{equation}
\widetilde{A}_{ai}=\widetilde{A}^{r}_{ai}+\widetilde{A}^{v}_{ai}
\label{def rv}
\end{equation}
such that the ``virtual'' part $|\widetilde{A}^{v}_{ai}|\ll |\widetilde{A}^{r}_{ai}|$ and 
the coupling constant $g\ll 1$ are assumed to be small to the same order, i.e. 
$|\widetilde{A}^{v}_{ai}|/|\widetilde{A}^{r}_{ai}|\sim g$, for simplicity.

In the first-order approximation, the equation for the ``real'' field $\widetilde{A}^{r}_{ai}$ 
has the following form
\begin{equation}
\partial_0\partial_0\widetilde{A}^{r}_{ai}
-\partial_k\partial_k\widetilde{A}^{r}_{ai}
+\partial_i\partial_k\widetilde{A}^{r}_{ak}=0 \,.
\label{r1} 
\end{equation}
In the second-order approximation, the ``real'' and ``virtual'' parts satisfy the following 
equation up to the terms proportional to $g^2$
\begin{eqnarray} \nonumber
&& \partial_0\partial_0
\widetilde{A}^{r}_{ai}
+\partial_0\partial_0\widetilde{A}^{v}_{ai}
-\partial_k\partial_k
\widetilde{A}^{r}_{ai}-
\partial_k\partial_k\widetilde{A}^{v}_{ai}
+\partial_i\partial_k
\widetilde{A}^{r}_{ak}+
\partial_i\partial_k\widetilde{A}^{v}_{ak} \\  \nonumber
&& +\,2gf_{abc}\partial_k
\widetilde{A}^{r}_{ci}\widetilde{A}^{r}_{bk}+
2gf_{abc}\partial_k
\widetilde{A}^{r}_{ci}\widetilde{A}^{v}_{bk}+
2gf_{abc}\partial_k
\widetilde{A}^{v}_{ci}\widetilde{A}^{r}_{bk}
+gf_{abc}\partial_i
\widetilde{A}^{r}_{bk}\widetilde{A}^{r}_{ck} \\  \nonumber
&& +\, gf_{abc}\partial_i
\widetilde{A}^{r}_{bk}\widetilde{A}^{v}_{ck} 
+ gf_{abc}\partial_i
\widetilde{A}^{v}_{bk}\widetilde{A}^{r}_{ck}
+gf_{abc}\partial_k
\widetilde{A}^{r}_{bk}\widetilde{A}^{r}_{ci}
+gf_{abc}\partial_k
\widetilde{A}^{r}_{bk}\widetilde{A}^{v}_{ci} \\
&& +\, gf_{abc}\partial_k
\widetilde{A}^{v}_{bk}\widetilde{A}^{r}_{ci}
+g^2f_{cba}f_{cb'c'}(\widetilde{A}^{r}_{bk}m^{b'k}_{c'i}
+\widetilde{A}^{r}_{b'k}m^{bk}_{c'i}
+\widetilde{A}^{r}_{c'i}m^{bk}_{b'k}) \,,
\label{r2}
\end{eqnarray}
where the effective mass terms $m^{bk}_{b'k'}$ 
are found in Eq.~(\ref{3-terms mass}).

The equation for the ``virtual'' part is obtained by subtracting the first-order equation for the ``real'' one 
(\ref{r1}) and keeping the terms linear in $g$ only such that
\begin{eqnarray}  \nonumber
&& \partial_0\partial_0\widetilde{A}^{v}_{ai}
-\partial_k\partial_k\widetilde{A}^{v}_{ai}
+\partial_i\partial_k\widetilde{A}^{v}_{ak} \\
&& +\, 2gf_{abc}\partial_k
\widetilde{A}^{r}_{ci}\widetilde{A}^{r}_{bk}
+gf_{abc}\partial_i
\widetilde{A}^{r}_{bk}\widetilde{A}^{r}_{ck}
+gf_{abc}\partial_k
\widetilde{A}^{r}_{bk}\widetilde{A}^{r}_{ci}=0 \,.
\label{v} 
\end{eqnarray}
The equation for the ``real'' part in the second (in $g$) approximation 
is then obtained by subtracting Eq.~(\ref{v}) from Eq.~(\ref{r2}).

One immediately notices that the ``virtual'' part can be expressed 
via square of the ``real'' part from Eq.~(\ref{v}). Performing Fourier 
transform of Eq.~(\ref{v}) it appears to be exactly solvable by 
variational methods. Corresponding solutions for the ``virtual'' part
expressed in terms of the ``real'' one read
\begin{eqnarray} \nonumber
\widetilde{A}^{v0}_{ai}(t)&=&
\int^t_{t_0} t' dt' \frac{1}{V}\int d^3 x' gf_{abc} 
\Big(2\partial_k\widetilde{A}^{r}_{ci}(x')\widetilde{A}^{r}_{bk}(x') \\ \nonumber
&+& \partial_i\widetilde{A}^{r}_{bk}(x')\widetilde{A}^{r}_{ck}(x')+
\partial_k\widetilde{A}^{r}_{bk}(x')\widetilde{A}^{r}_{ci}(x')\Big) \\  \nonumber
&-& \int^t_{t_0} dt' \frac{1}{V}\int d^3 x' gf_{abc} 
\Big(2\partial_k\widetilde{A}^{r}_{ci}(x')\widetilde{A}^{r}_{bk}(x') \\
&+& \partial_i\widetilde{A}^{r}_{bk}(x')\widetilde{A}^{r}_{ck}(x')+
\partial_k\widetilde{A}^{r}_{bk}(x')\widetilde{A}^{r}_{ci}(x')\Big)\,t \,,
\label{0 sol} 
\end{eqnarray}
\begin{eqnarray} \nonumber
\widetilde{A}^{v\vec{p}}_{a\|}(t)&=&
\int^t_{t_0} t' dt' \frac{1}{V}\int d^3 x' gf_{abc} e^{-i\vec{p}\vec{x'}} n_i 
\Big(2\partial_k\widetilde{A}^{r}_{ci}(x')\widetilde{A}^{r}_{bk}(x')  \\ \nonumber
&+& \partial_i\widetilde{A}^{r}_{bk}(x')\widetilde{A}^{r}_{ck}(x')+
\partial_k\widetilde{A}^{r}_{bk}(x')\widetilde{A}^{r}_{ci}(x')\Big) \\ \nonumber
&-& \int^t_{t_0} dt' \frac{1}{V}\int d^3 x' gf_{abc} e^{-i\vec{p}\vec{x'}} 
n_i \Big(2\partial_k\widetilde{A}^{r}_{ci}(x')\widetilde{A}^{r}_{bk}(x') \\
&+& \partial_i\widetilde{A}^{r}_{bk}(x')\widetilde{A}^{r}_{ck}(x')+
\partial_k\widetilde{A}^{r}_{bk}(x')\widetilde{A}^{r}_{ci}(x')\Big)\,t \,,
\label{long sol} 
\end{eqnarray}
\begin{eqnarray} \nonumber
\widetilde{A}^{v\vec{p}}_{a\alpha}(t)&=&-\int^t_{t_0} dt'
\frac{1}{V}\int d^3 x' \frac{g}{2ip} f_{abc} e^{-i\vec{p}\vec{x'}}e^{ip(t-t')}s^{\alpha}_i
\Big(2\partial_k\widetilde{A}^{r}_{ci}(x')\widetilde{A}^{r}_{bk}(x') \\  \nonumber
&+& 
\partial_i\widetilde{A}^{r}_{bk}(x')\widetilde{A}^{r}_{ck}(x')+
\partial_k\widetilde{A}^{r}_{bk}(x')\widetilde{A}^{r}_{ci}(x')\Big) \\  \nonumber
&+&
\int^t_{t_0} dt' \frac{1}{V}\int d^3 x' \frac{g}{2ip} f_{abc} 
e^{-i\vec{p}\vec{x'}}e^{ip(t'-t)} s^{\alpha}_i
\Big(2\partial_k\widetilde{A}^{r}_{ci}(x')\widetilde{A}^{r}_{bk}(x') \\
&+&
\partial_i\widetilde{A}^{r}_{bk}(x')\widetilde{A}^{r}_{ck}(x')+
\partial_k\widetilde{A}^{r}_{bk}(x')\widetilde{A}^{r}_{ci}(x')\Big) \,.
\label{tr sol}
\end{eqnarray}
Above, $\widetilde{A}^{v\vec{p}}_{a\|}(t)$ and $\widetilde{A}^{v\vec{p}}_{a\alpha}(t)$ 
are the longitudinal and transverse parts of $\widetilde{A}^{v\vec{p}}_{ai}$, 
$\widetilde{A}^{v0}_{ai}(t)$ is the mode with zero momentum, and 
$\widetilde{A}^{r}_{ck}(x')=\widetilde{A}^{r}_{ck}(t',\vec{x'})$. 
For the initial conditions at $t=t_0$, we adopted that 
$\widetilde{A}^{v\vec{p}}_{a\|}(t_0)=\widetilde{A}^{v\vec{p}}_{a\alpha}(t_0)=
\widetilde{A}^{v0}_{ai}(t_0)=0$ for all the polarisations and momenta. Indeed,
this condition agrees well will the initial procedure of splitting a single wave mode into 
two parts, the large ``real'' and small ``virtual'' ones, whose initial values must be consistent.

By a substitution of Eqs.~(\ref{long sol}) and (\ref{tr sol}) into Eq.~(\ref{r2}), the second order terms 
can be transformed to the third-order ones over the ``real'' field which then evaluated by means of 
the same algorithm as used above. Finally, Eq.~(\ref{r2}) gets transformed to
\begin{eqnarray} \nonumber
&& \partial_0\partial_0
\widetilde{A}^{r\vec{q}}_{ai}
+q^2\widetilde{A}^{r\vec{q}}_{ai} - q_iq_k\widetilde{A}^{r\vec{q}}_{ak}+
\int_{t0}^t\widetilde{A}^{r\vec{q}}_{c'i'}(t') M^{\vec{q}}_{aic'i'}(t,t')dt' \\
&& +\, g^2f_{cba}f_{cb'c'}(\widetilde{A}^{r\vec{q}}_{bk}m^{b'k}_{c'i}
+\widetilde{A}^{r\vec{q}}_{b'k}m^{bk}_{c'i}+
\widetilde{A}^{r\vec{q}}_{c'i}m^{bk}_{b'k})=0 \,,
\label{v to r2}
\end{eqnarray}
where $m^{ai}_{bk}$ are the contributions from third-order terms in 
wave modes given by Eq.~(\ref{3-terms mass}), and the second-order contribution reads
\begin{eqnarray} \nonumber
&& M^{\vec{q}}_{aic'i'}(t,t')=
-\frac18\sum_{\vec{p}}\biggl[\Big(\hat{a}^{acb'c'i'k}_{\vec{q}-\vec{p}}(2q_{k'}-p_{k'})(q_k+p_k) \\ \nonumber
&&+\,  \hat{a}^{acc'b'k'k}_{\vec{q}-\vec{p}}(q_{i'}-2p_{i'})(q_k+p_k)\Big)\langle 
\widetilde{A}^{r\vec{p}}_{ci}(t)\widetilde{A}^{\dagger r\vec{p}}_{b'k'}(t')\rangle \\  \nonumber
&&+\, \Big(\hat{a}^{acb'c'i'k}_{\vec{q}+\vec{p}}(2q_{k'}+p_{k'})(q_k-p_k)+
\hat{a}^{acc'b'k'k}_{\vec{q}+\vec{p}}(q_{i'}+2p_{i'})(q_k-p_k)\Big)\langle 
\widetilde{A}^{\dagger r\vec{p}}_{ci}(t)\widetilde{A}^{r\vec{p}}_{b'k'}(t')\rangle \\ \nonumber
&&+\,  \hat{a}^{acc'b'k'k}_{\vec{q}-\vec{p}}(q_{k'}-p_{k'})(q_k+p_k)\langle 
\widetilde{A}^{r\vec{p}}_{ci}(t)\widetilde{A}^{\dagger r\vec{p}}_{b'i'}(t')\rangle \\ \nonumber
&&+\,  \hat{a}^{acc'b'k'k}_{\vec{q}+\vec{p}}(q_{k'}+p_{k'})(q_k-p_k)\langle 
\widetilde{A}^{\dagger r\vec{p}}_{ci}(t)\widetilde{A}^{r\vec{p}}_{b'i'}(t')\rangle \\ \nonumber
&&+\,  \Big(\hat{a}^{abb'c'i'k}_{\vec{q}-\vec{p}}(q_{k'}-p_{k'})(q_i-2p_i)+
\hat{a}^{abc'b'k'i}_{\vec{q}-\vec{p}}(q_{i'}-p_{i'})(2q_k-3p_k) 
\end{eqnarray}
\begin{eqnarray}
&&+\,  \hat{a}^{abc'b'k'k}_{\vec{q}-\vec{p}}(q_{i'}-2p_{i'})(q_i-2p_i)+
\hat{a}^{abc'b'i'i}_{\vec{q}-\vec{p}}(2q_{k'}-p_{k'})(2q_k-3p_k)\Big)
\langle \widetilde{A}^{r\vec{p}}_{bk}(t)\widetilde{A}^{\dagger r\vec{p}}_{b'k'}(t')\rangle \\ \nonumber
&&+\,  \Big(\hat{a}^{abb'c'i'k}_{\vec{q}+\vec{p}}(q_{k'}+p_{k'})(q_i+2p_i)+
\hat{a}^{abc'b'k'i}_{\vec{q}+\vec{p}}(q_{i'}+p_{i'})(2q_k+3p_k) \\ \nonumber
&&+\,  \hat{a}^{abc'b'k'k}_{\vec{q}+\vec{p}}(q_{i'}+2p_{i'})(q_i+2p_i)+
\hat{a}^{abc'b'i'i}_{\vec{q}+\vec{p}}(2q_{k'}+p_{k'})(2q_k+3p_k)\Big)
\langle \widetilde{A}^{\dagger r\vec{p}}_{bk}(t)\widetilde{A}^{r\vec{p}}_{b'k'}(t')\rangle \\ \nonumber
&&+\,  \Big(\hat{a}^{abc'b'k'i}_{\vec{q}-\vec{p}}(q_{k'}-p_{k'})(2q_k-3p_k)+
\hat{a}^{abc'b'k'k}_{\vec{q}-\vec{p}}(q_{k'}-p_{k'})(q_i-2p_i)\Big)
\langle \widetilde{A}^{r\vec{p}}_{bk}(t)\widetilde{A}^{\dagger r\vec{p}}_{b'i'}(t')\rangle \\ \nonumber
&&+\,  \Big(\hat{a}^{abc'b'k'i}_{\vec{q}+\vec{p}}(q_{k'}+p_{k'})(2q_k+3p_k) \\
&&+\,  \hat{a}^{abc'b'k'k}_{\vec{q}+\vec{p}}(q_{k'}+p_{k'})(q_i+2p_i)\Big)
\langle \widetilde{A}^{\dagger r\vec{p}}_{bk}(t)\widetilde{A}^{r\vec{p}}_{b'i'}(t')\rangle \biggr] \,.
\label{MatrixSO} 
\end{eqnarray}
Here, the Fourier coefficients are given by
\begin{eqnarray}
\hat{a}^{acb'c'i'k}_{\vec{p}}&=&\frac{1}{3}
f_{abc}f_{bb'c'}\Big(\frac{-g^2}{ip}e^{ip(t-t')}s^{\alpha}_{i'}s^{\alpha}_{k}
+ \frac{g^2}{ip}e^{ip(t'-t)}s^{\alpha}_{i'}s^{\alpha}_{k}+2g^2n_{i'}n_{k}(t'-t)\Big) \,, \label{hat a}  \\
\hat{a}^{acb'c'i'k}_{0}&=&\frac{2}{3}f_{abc}f_{bb'c'}g^2 \delta_{i'k}(t'-t) \,. \label{hat a0}
\end{eqnarray}
Equation (\ref{v to r2}) describes dynamics of the gluon plasma including the contribution of 
all second- and third-order terms in Eq.~(\ref{non-transformed equation}). One observes
that the second-order self-interactions induce a time non-local contribution to the effective 
gluon mass compared to the local ones induced by the third-order self-interactions.

\subsection{Equations for condensate and waves with all-order wave self-interactions}
\label{Sec:YM-plasma:exact-YMC}

All the relevant higher-order interaction terms can be incorporated into the equations of motion 
for the YM ``condensate + waves'' system in an analogical way as was done above in 
Sect.~\ref{Sec:YM-plasma:exact} for the pure gluon plasma case. Before separation
of the YM condensate from YM fields, the YM equations of motion can be written in terms
of ``real'' and ``virtual'' fields  in the form of Eqs.~(\ref{r2}) and (\ref{v}) where
wave modes should replaced by complete fields everywhere, i.e. ${\widetilde A}\to A$.
Correspondingly, the solutions for $A^v$ in terms of $A^r$ have exactly the same form
as Eqs.~(\ref{0 sol}) -- (\ref{tr sol}) but with replacements ${\widetilde A}\to A$. Using the
latter expressions in the equation for the ``real'' part one completely excludes the ``virtual''
components from consideration and analyzes dynamics of the largest ``real'' part only.

Let us now consider for simplicity $SU(2)$ theory and separate of the YM condensate 
and wave components in the ``real'' part in usual manner
\begin{eqnarray}
A^r_{ik}(t,\vec x)=\delta_{ik}U(t)+\widetilde{A}^r_{ik}(t,\vec x)\,.
\end{eqnarray}
Then one ends up with an integro-differential equation for 
the corresponding condensate
\begin{eqnarray} \nonumber
&& \partial_0\partial_0 U + 2g^2 U^3+
\frac{1}{3}\int_{t_0}^t dt' U(t') M^{0}_{aabb}(t,t')+
\frac{g^2}{36}U\sum_{\vec{p}}\langle \widetilde{A}^{r\vec{p}\dagger}_{ck}\widetilde{A}^{r\vec{p}}_{ck}+
\widetilde{A}^{r\vec{p}}_{ck}\widetilde{A}^{r\vec{p}\dagger}_{ck} \\ 
&& +\, 2\widetilde{A}^{r\vec{p}\dagger}_{aa}\widetilde{A}^{r\vec{p}}_{ii}+
2\widetilde{A}^{r\vec{p}}_{aa}\widetilde{A}^{r\vec{p}\dagger}_{ii}-
\widetilde{A}^{r\vec{p}\dagger}_{ck}\widetilde{A}^{r\vec{p}}_{kc}-
\widetilde{A}^{r\vec{p}}_{ck}\widetilde{A}^{r\vec{p} \dagger}_{kc}\rangle =0 \,,
\end{eqnarray}
where $M^{0}_{aabb}(t,t')\equiv M^{\vec{q}=0}_{aabb}(t,t')$ matrix should be 
found from Eq.~(\ref{MatrixSO}) for $\vec{q}=0$. 

The equation for the YM waves takes the form
\begin{eqnarray}   \nonumber
&& \partial_0\partial_0\widetilde{A}^{r\vec{q}}_{ai}
+q^2\widetilde{A}^{r\vec{q}}_{ai}-q_iq_k\widetilde{A}^{r\vec{q}}_{ak}+
\int_{t0}^t\widetilde{A}^{r\vec{q}}_{c'i'}(t') M^{\vec{q}}_{aic'i'}(t,t')dt' \\   \nonumber
&& +\, \widetilde{A}^{r\vec{q}}_{cl}(t)\int_{t_0}^tdt'\sum_{\vec{p}}N^{\vec{p}\vec{q}}_{aicl}(t,t')+
U(t)\int_{t_0}^t dt' U(t')P^{\vec{q}}_{aib'k'}\widetilde{A}^{r\vec{q}}_{b'k'}(t') \\   \nonumber
&& +\, \frac{g^2}{3}U^2(\widetilde{A}^{r\vec{q}}_{ai}-\widetilde{A}^{r\vec{q}}_{ia}+2\delta_{ai}
\widetilde{A}^{r\vec{q}}_{kk})+g^2(2\widetilde{A}^{r\vec{q}}_{bk}m^{bk}_{ai} \\
&& +\, \widetilde{A}^{r\vec{q}}_{ai}m^{bk}_{bk}-
\widetilde{A}^{r\vec{q}}_{bk}m^{ak}_{bi}-\widetilde{A}^{r\vec{q}}_{ak}m^{bk}_{bi}-
\widetilde{A}^{r\vec{q}}_{bi}m^{bk}_{ak})=0 \,,
\label{waves exact} 
\end{eqnarray}
where tensors $N^{\vec{p}\vec{q}}_{aicl}(t,t')$ and $P^{\vec{q}}_{aib'k'}$ 
(with coefficients $\hat{a}_{\vec{q}}$, $\hat{a}_{0}$ given by Eq.~(\ref{hat a}), (\ref{hat a0}), respectively) 
are defined as follows
\begin{eqnarray*}
N^{\vec{p}\vec{q}}_{aicl}(t,t')&=&\frac{1}{8}(2q_i \hat{a}^{acb'c'i'l}_0-2\delta_{li}q_k 
\hat{a}^{acb'c'i'k}_0+q_l \hat{a}^{acb'c'i'i}_0)
(2p_{k'}\langle \widetilde{A}^{r\vec{p}}_{c'i'}\widetilde{A}^{r\vec{p}\dagger}_{b'k'}\rangle \\ 
&-& 2p_{k'}\langle \widetilde{A}^{r\vec{p}\dagger}_{c'i'}\widetilde{A}^{r\vec{p}}_{b'k'}\rangle +
p_{i'}\langle \widetilde{A}^{r\vec{p}}_{b'k'}\widetilde{A}^{r\vec{p}\dagger}_{c'k'}\rangle \\ 
&-& p_{i'}\langle \widetilde{A}^{r\vec{p}\dagger}_{b'k'}\widetilde{A}^{r\vec{p}}_{c'k'}\rangle +
p_{k'}\langle \widetilde{A}^{r\vec{p}}_{b'k'}\widetilde{A}^{r\vec{p}\dagger}_{c'i'}\rangle -
p_{k'}\langle \widetilde{A}^{r\vec{p}\dagger}_{b'k'}\widetilde{A}^{r\vec{p}}_{c'i'}\rangle ) \,, \\
P^{\vec{q}}_{aib'k'}(t,t')&=&2q_l\delta_{k'i}(q_k \hat{a}^{aklb'i'i}_{\vec{q}}-
\frac{1}{2} q_i \hat{a}^{aklb'i'k}_{\vec{q}}-
\frac{1}{2} q_k \hat{a}^{ailb'i'k}_{\vec{q}}) \\ 
&+& q_{i'}(q_k \hat{a}^{akb'k'i'i}_{\vec{q}}-
\frac{1}{2} q_i \hat{a}^{akb'k'i'k}_{\vec{q}}-
\frac{1}{2} q_k \hat{a}^{aib'k'i'k}_{\vec{q}}) \\ 
&+& q_{k'}(q_k \hat{a}^{akb'i'i'i}_{\vec{q}}-
\frac{1}{2} q_i \hat{a}^{akb'i'i'k}_{\vec{q}}-
\frac{1}{2} q_k \hat{a}^{aib'i'i'k}_{\vec{q}}) \,.
\end{eqnarray*}

\section{Condensate and waves in a supersymmetric Yang-Mills theory}
\label{Sec:SYM}

Now we would like to extend our analysis to the ${\cal N}=4$
supersymmetric Yang-Mills theory with $SU(2)$ which contains
additional spinor, scalar and pseudo-scalar d.o.f. (see e.g.
Ref.~\cite{Gliozzi}). The corresponding Lagrangian reads
\begin{eqnarray} \nonumber
{\cal L}_{\rm SUSY}&=&-\frac{1}{4}F^a_{\mu\nu}F_a^{\mu\nu}
+\frac{i}{2}\bar{\lambda}^a_{k'}\gamma^\mu D_\mu\lambda^a_{k'}
+\frac{1}{2}(D_\mu C^a_{(i)})^2 +\frac{1}{2}(D_\mu B^a_{(i)})^2  \\
&-& \frac{g}{2}\epsilon^{abc}\bar{\lambda}^a_{k'}(\alpha_{k'l'}^{(i)}
C^b_{(i)}+\beta_{k'l'}^{(i)}\gamma_5 B^b_{(i)})\lambda^c_{l'} \nonumber \\
&+& \frac{g^2}{4}\, \Big[(\epsilon^{abc}C^b_{(i)}C^c_{(j)})^2 +
(\epsilon^{abc}B^b_{(i)}B^c_{(j)})^2 +
(\epsilon^{abc}C^b_{(i)}B^c_{(j)})^2\Big]\,. \label{Lsuper}
\end{eqnarray}
Here $a,b,c=1\,...\,3$ are the isotopic $SU(2)$ indices,
$i,j=1\,...\,3$ numerate different types of scalar $C$ and
pseudoscalar $B$ fields, $k',l'=1\,...\,4$ numerate different
flavors of fermions $\lambda$. Next, we decompose additional
supersymmetric modes into transverse and longitudinal components in
momentum space as follows
\begin{eqnarray*}
C_k^{(i)\vec{p}}&=&n_k K_{(i)} + s^{\alpha}_k M_{(i)\alpha}\,, \\
B_k^{(i)\vec{p}}&=&n_k K'_{(i)} + s^{\alpha}_k M'_{(i)\alpha}\,, \\
\lambda_k^{k'\vec{p}}&=&n_k \vartheta^{k'}+ s^{\alpha}_k
\zeta^{k'}_{\alpha}\,.
\end{eqnarray*}
Also, we perform the corresponding tensor decompositions for the YM
field $A_\mu^a$ which enters the covariant derivative $D_\mu$ and
stress tensor $F_{\mu\nu}$ in the same way as is done above.

Next, let us rewrite the supersymmetric part of the Lagrangian
density (\ref{Lsuper}) in terms of new Fourier modes
$K,\,M,\,\vartheta$ and $\zeta$. The ${\cal N}=4$ supersymmetric YM
Lagrangian density (\ref{Lmain}) is given in terms of wave modes and
condensate by
\begin{eqnarray} \nonumber
 {\mathcal L}_{\rm SUSY}
 &=&\frac12\Big\{\partial_0\psi_\lambda\,\partial_0\psi_\lambda^\dagger +
 \partial_0\phi_\sigma\,\partial_0\phi_\sigma^\dagger +
 \partial_0\Phi\,\partial_0\Phi^\dagger +
 \frac{1}{2}\partial_0\Lambda\,\partial_0\Lambda^\dagger +
 \partial_0\eta_\sigma\,\partial_0\eta_\sigma^\dagger \\  \nonumber
 &+& \partial_0\lambda\,\partial_0\lambda^\dagger -
  p^2\,\psi_\lambda\,\psi_\lambda^\dagger -
 \frac{p^2}{2}\,\phi_\sigma\,\phi_\sigma^\dagger - p^2\,\Phi\,\Phi^\dagger -
 \frac{p^2}{2}\,\eta_\sigma\,\eta_\sigma^\dagger - p^2\lambda\,\lambda^\dagger
 \\  \nonumber
 &+& \frac{p^2}{2}\,e^{\gamma\sigma}(\eta_\sigma\phi_\gamma^\dagger
 + \phi_\gamma\eta_\sigma^\dagger) -
 igp\,U\,e^{\sigma\gamma}\eta_\sigma\eta_\gamma^\dagger +
 igp\,U\,Q^{\lambda\gamma}\psi_\lambda\psi_\gamma^\dagger \\  \nonumber
 &+& igp\,U\,e^{\sigma\gamma}\phi_\sigma\phi_\gamma^\dagger +
 igp\,U\,(2\Phi\lambda^\dagger - 2\lambda\Phi^\dagger + \Lambda\lambda^\dagger -
 \lambda\Lambda^\dagger)  \\  \nonumber
 &-& 2g^2\,U^2\,\eta_\sigma\eta_\sigma^\dagger - 2g^2\,U^2\,\lambda\lambda^\dagger -
 g^2\,U^2\,(4\Phi\Phi^\dagger + 2\Phi\Lambda^\dagger + 2\Lambda\Phi^\dagger + \Lambda\Lambda^\dagger) 
 \\  \nonumber
 &+& \frac{1}{2}\partial_0 K_{(i)}\partial_0 K_{(i)}^\dagger +
 \frac{1}{2}\partial_0M_{(i)\alpha}\partial_0M_{(i)\alpha}^\dagger -
 \frac{1}{2}p^2K_{(i)}K_{(i)}^\dagger - \frac{1}{2}p^2M_{(i)\alpha}M_{(i)\alpha}^\dagger -
 \\  \nonumber
 &-& igp\,U\,e^{\beta\alpha}M_{(i)\alpha}M_{(i)\beta}^\dagger - g^2\,U^2\,K_{(i)}K_{(i)}^\dagger 
 g^2\,U^2\,M_{(i)\alpha}M_{(i)\alpha}^\dagger  \\  \nonumber
 &+& \frac{1}{2}\,\partial_0 K^\prime_{(i)}\,\partial_0 K_{(i)}^{\prime \dagger} +
 \frac{1}{2}\,\partial_0M^\prime_{(i)\alpha}\,\partial_0M_{(i)\alpha}^{\prime \dagger} -
 \frac{1}{2}\,p^2K^\prime_{(i)}\,K_{(i)}^{\prime \dagger} -
 \frac{1}{2}\,p^2M^\prime_{(i)\alpha}\,M_{(i)\alpha}^{\prime \dagger}  \\
  \nonumber
 &-& igp\,Ue^{\beta\alpha}M^{\prime}_{(i)\alpha}\,M_{(i)\beta}^{\prime \dagger} -
 g^2\,U^2\,K^{\prime}_{(i)}K_{(i)}^{\prime \dagger} 
 g^2\,U^2\,M^\prime_{(i)\alpha}M_{(i)\alpha}^{\prime \dagger} \\  \nonumber
 &+& \frac{i}{2}\,\bar{\vartheta}^{k'}\gamma_0\partial_0\vartheta^{k'} +
 \frac{i}{2}\,\bar{\zeta}^{k'}_\alpha\gamma_0\partial_0\zeta^{k'}_\alpha -
 \frac{p}{2}\,n_k\bar{\vartheta}^{k'}\gamma_k\vartheta^{k'} -
 \frac{p}{2}\,n_k\bar{\zeta}^{k'}_\alpha\gamma_k\zeta^{k'}_\alpha  \\
 &-& \frac{i}{2}\,gUe_{ilm}n_is_m^\alpha\bar{\vartheta}^{k'}\gamma_l\zeta^{k'}_\alpha -
 \frac{i}{2}\,gUe_{ilm}n_ms_i^\alpha\bar{\zeta}^{k'}_\alpha\gamma_l\vartheta^{k'} -
 \frac{i}{2}\,gUe_{ilm}s_i^\alpha
 s_m^\beta\bar{\zeta}^{k'}_\alpha\gamma_l\zeta^{k'}_\beta\Big\}\,,
 \label{LSUSY}
\end{eqnarray}
and the corresponding Hamiltonian density (\ref{H}) has a form
\begin{eqnarray}  \nonumber
 \mathcal{H}_{\rm SUSY}&=&\,\frac12\Big\{\partial_0\psi_\lambda\,\partial_0\psi_\lambda^\dagger +
 \partial_0\phi_\sigma\,\partial_0\phi_\sigma^\dagger + \partial_0\Phi\,\partial_0\Phi^\dagger +
 \frac{1}{2}\,\partial_0\Lambda\,\partial_0\Lambda^\dagger +
 \partial_0\eta_\sigma\,\partial_0\eta_\sigma^\dagger \\  \nonumber
 &+& \partial_0\lambda\,\partial_0\lambda^\dagger + p^2\,\psi_\lambda\psi_\lambda^\dagger +
 \frac{p^2}{2}\,\phi_\sigma\phi_\sigma^\dagger + p^2\,\Phi\Phi^\dagger +
 \frac{p^2}{2}\,\eta_\sigma\eta_\sigma^\dagger + p^2\,\lambda\lambda^\dagger
 \\  \nonumber
 &+& \frac{p^2}{2}\,e^{\gamma\sigma}(\eta_\sigma\phi_\gamma^\dagger +
 \phi_\gamma\eta_\sigma^\dagger) + igp\,U\,e^{\sigma\gamma}\eta_\sigma\eta_\gamma^\dagger -
 igp\,U\,Q^{\lambda\gamma}\psi_\lambda\psi_\gamma^\dagger \\ \nonumber
 &-& igp\,U\,e^{\sigma\gamma}\phi_\sigma\phi_\gamma^\dagger -
 igp\,U\,(2\Phi\lambda^\dagger - 2\lambda\Phi^\dagger + \Lambda\lambda^\dagger - \lambda\Lambda^\dagger) \\ \nonumber
 &+& 2g^2\,U^2\,\eta_\sigma\eta_\sigma^\dagger + 2g^2\,U^2\,\lambda\lambda^\dagger +
 g^2\,U^2\,(4\Phi\Phi^\dagger + 2\Phi\Lambda^\dagger + 2\Lambda\Phi^\dagger + \Lambda\Lambda^\dagger) \\ \nonumber
 &+& \frac{1}{2}\,\partial_0 K_{(i)}\,\partial_0 K_{(i)}^\dagger +
 \frac{1}{2}\,\partial_0M_{(i)\alpha}\,\partial_0M_{(i)\alpha}^\dagger +
 \frac{1}{2}\,p^2\,K_{(i)}K_{(i)}^\dagger + \frac{1}{2}\,p^2\,M_{(i)\alpha}M_{(i)\alpha}^\dagger \\ \nonumber
 &+& igp\,U\,e^{\beta\alpha}M_{(i)\alpha}M_{(i)\beta}^\dagger + g^2\,U^2\,K_{(i)}K_{(i)}^\dagger +
 g^2\,U^2\,M_{(i)\alpha}M_{(i)\alpha}^\dagger \\  \nonumber
 &+& \frac{1}{2}\,\partial_0 K^\prime_{(i)}\,\partial_0 K_{(i)}^{\prime \dagger} +
 \frac{1}{2}\,\partial_0M^\prime_{(i)\alpha}\,\partial_0M_{(i)\alpha}^{\prime \dagger} +
 \frac{1}{2}\,p^2\,K^\prime_{(i)}K_{(i)}^{\prime \dagger} +
 \frac{1}{2}\,p^2\,M^\prime_{(i)\alpha}M_{(i)\alpha}^{\prime \dagger} \\ \nonumber
 &+& igp\,U\,e^{\beta\alpha}M^{\prime}_{(i)\alpha}M_{(i)\beta}^{\prime \dagger} +
 g^2\,U^2\,K^{\prime}_{(i)}K_{(i)}^{\prime \dagger}
 g^2\,U^2\,M^\prime_{(i)\alpha}M_{(i)\alpha}^{\prime \dagger} + \\ \nonumber
 &+& \frac{p}{2}\,n_k\bar{\vartheta}^{k'}\gamma_k\vartheta^{k'} +
 \frac{p}{2}\,n_k\bar{\zeta}^{k'}_\alpha\gamma_k\zeta^{k'}_\alpha +
 \frac{i}{2}\,g\,U\,e_{ilm}n_is_m^\alpha\bar{\vartheta}^{k'}\gamma_l\zeta^{k'}_\alpha \\
 &+& \frac{i}{2}\,g\,U\,e_{ilm}n_ms_i^\alpha\bar{\zeta}^{k'}_\alpha\gamma_l\vartheta^{k'} +
 \frac{i}{2}\,g\,U\,e_{ilm}s_i^\alpha
 s_m^\beta\bar{\zeta}^{k'}_\alpha\gamma_l\zeta^{k'}_\beta\Big\}\,.
 \label{HSUSY}
\end{eqnarray}

The equations of motion for the extra supersymmetric d.o.f.
$K,\,M,\,\vartheta^{k'},\,\zeta_{\alpha}^{k'}$ can be constructed in
the standard way as Lagrange (or Hamilton) equations based upon
Eq.~(\ref{LSUSY}) (or Eq.~(\ref{HSUSY}))
\begin{eqnarray}
 && \partial_0\partial_0 K_{(i)} + p^2\,K_{(i)} + 2g^2\,U^2\,K_{(i)}=0\,,
 \label{Keq} \\
 && \partial_0\partial_0 M_{(i)\alpha} + p^2\,M_{(i)\alpha} +
 2igp\,U\,e^{\alpha\beta}M_{\beta} + 2g^2\,U^2\,M_{(i)\alpha} = 0\,,
 \label{Meq} \\
 && \gamma_0\partial_0\vartheta^{k'} + ip\,n_j\gamma_j\vartheta^{k'} -
 g\,U\,e_{ikl}\gamma_kn_is^\alpha_l\zeta^{k'}_\alpha = 0\,,
 \label{varthetaeq} \\
 && \gamma_0\partial_0\zeta_{\alpha}^{k'} + ipn_j\gamma_j\zeta_{\alpha}^{k'} -
 g\,U\,e_{ikl}\gamma_ks^\alpha_is^\beta_l\zeta^{k'}_\beta -
 g\,U\,e_{ikl}\gamma_kn_ls^\alpha_i\vartheta^{k'} = 0\,. \label{zetaeq}
\end{eqnarray}
The equations for pseudoscalar modes $K'$ and $M'$ are the same as
equations for $K$ (\ref{Keq}) and $M$ (\ref{Meq}), respectively. A
numerical analysis of these equations in the linear approximation
(with free YM condensate) has shown that qualitative behavior of the
(pseudo)scalar modes is analogical to that of the YM wave modes
discussed above. Finally, the equation of motion for the YM condensate in the
next-to-linear approximation accounting for ``back reaction'' effects
of the wave modes (including supersymmetric ones) to the condensate
reads
\begin{eqnarray*}
 &&\partial_0\partial_0 U + 2g^2\,U^3 +
 \frac{g^2}{6}\,U\,\sum_{\vec{p}}\Big\langle 2\eta_\sigma\eta_\sigma^\dagger +
 2\lambda\lambda^\dagger + 4\Phi\Phi^\dagger + \Lambda\Lambda^\dagger + 2\Phi\Lambda^\dagger \\
 && +\; 2\Lambda\Phi^\dagger + 2\eta^\dagger_\sigma\eta_\sigma + 2\lambda^\dagger\lambda +
 4\Phi^\dagger\Phi + \Lambda^\dagger\Lambda + 2\Phi^\dagger\Lambda + 2\Lambda^\dagger\Phi \Big\rangle \\
 && +\; \frac{ig}{12}\,\sum_{\vec{p}} p\, \Big\langle 2\Lambda^\dagger\lambda - 2\Lambda\lambda^\dagger +
 2\Phi^\dagger\lambda - 2\Phi\lambda^\dagger + 2\lambda\Phi^\dagger - 2\lambda^\dagger\Phi \\
 && -\, Q^{\lambda\sigma} (\psi_\lambda\psi^\dagger_\sigma - \psi^\dagger_\lambda\psi_\sigma) -
 e^{\sigma\gamma}(\phi_\sigma\phi^\dagger_\gamma - \phi^\dagger_\sigma\phi_\gamma) \\
 && +\; \phi^\dagger_\sigma\eta_\sigma -\phi_\sigma\eta^\dagger_\sigma + \eta_\sigma^\dagger\phi_\sigma -
 \eta_\sigma\phi^\dagger_\sigma + e^{\sigma\gamma}(\eta_\sigma\eta^\dagger_\gamma -
 \eta^\dagger_\sigma\eta_\gamma)\Big\rangle \\ 
 && +\; \frac{g^2}{6}\,U\,\sum_{\vec{p}} \Big\langle K_{(i)}K_{(i)}^\dagger + K_{(i)}^\dagger K_{(i)} +
 M_{(i)\alpha}^\dagger M_{(i)\alpha} + M_{(i)\alpha}M_{(i)\alpha}^\dagger \Big\rangle  \\
 && +\; \frac{ig}{12}\, \sum_{\vec{p}}p\,\Big\langle e^{\alpha\beta}(M_{(i)\alpha}^\dagger M_{(i)\beta} -
 M_{(i)\alpha}M_{(i)\beta}^\dagger)\Big\rangle + \frac{g^2}{6}\,U\,\sum_{\vec{p}}\Big\langle
 K_{(i)}^{\prime}K _{(i)}^{\prime \dagger} \\
 && +\; K_{(i)}^{\prime \dagger}K_{(i)}^{\prime} + M_{(i)\alpha}^{\prime \dagger}M_{(i)\alpha}^{\prime} +
 M_{(i)\alpha}^{\prime}M_{(i)\alpha}^{\prime \dagger} \Big\rangle + \frac{ig}{12}\,\sum_{\vec{p}}p\,
 \Big\langle e^{\alpha\beta}(M_{(i)\alpha}^{\prime \dagger}M_{(i)\beta}^{\prime} \\
 && -\; M_{(i)\alpha}^{\prime} M_{(i)\beta}^{\prime \dagger})\Big\rangle -
 \frac{ig}{6}\,\sum_{\vec{p}}\Big\langle e_{lmp} n_m s_p^\alpha \bar{\vartheta}^{k'}\gamma_l\zeta^{k'}_\alpha \\
 && +\; e_{lmp} n_p s_m^\alpha \bar{\zeta_\alpha}^{k'}\gamma_l\vartheta^{k'} +
 e_{lmp} s_m^\alpha s_p^\beta \bar{\zeta_\alpha}^{k'}\gamma_l\zeta^{k'}_\beta \Big\rangle = 0\,.
\end{eqnarray*}
Taking into consideration only additional scalar and pseudoscalar
fields in numerical analysis we notice that the qualitative picture
of YM condensate dynamics shown in Figs.~\ref{fig:YMC-int} is not changed.
Also, energy of the extra d.o.f.'s grows effectively due to the energy
swap effect in the parametric resonance-like instability region
similarly to the other YM wave modes. 

As a validation test of our quasi-classical approach the canonical quantization of the 
YM wave modes in the classical YM condensate has been performed in Appendix \ref{A2} where
the quantum Hamilton equations in commutators have been constructed and are shown 
to be consistent with the YM equations of motion (\ref{eq1}), (\ref{eq2}) and (\ref{Keq}) -- (\ref{zetaeq}).
Note, a consistent analysis of equations (\ref{varthetaeq}) and (\ref{zetaeq}) for the spinor modes
$\vartheta^{k'},\,\zeta_{\alpha}^{k'}$ can only be performed in the
framework of quantum field theory approach, which is planned for
further studies.

\section{Cosmological evolution of the Yang-Mills condensate}
\label{Sec:YM-cosm}

Until now, we have discussed the properties of the YM condensates 
in trivial Minkowski background. It is instructive to extend this consideration 
to a realistic case of flat Friedmann universe since these findings can be useful e.g. 
in studies of the cosmological QCD phase transition, QCD vacuum compensation 
to the Dark Energy, particle production mechanisms after Inflation etc.

\subsection{Free condensate case}
\label{Sec:YM-cosm:freeYMC}

To start with, let us neglect the wave modes and study evolution of 
the homogeneous YM condensate $U=U(t)$ in the $SU(2)$ theory in the early 
Universe disregarding all other forms of matter. The 4-interval in the 
flat Friedmann model is
\begin{equation}
ds^2=a(\eta)^2(d\eta^2-dx^2-dy^2-dz^2)\,,
\label{ds2} 
\end{equation}
where $\eta$ is the conformal time related to the cosmological time as
\begin{equation}
dt=ad\eta \,.
\label{dtdeta} 
\end{equation}
The corresponding metric is
\begin{equation}
g_{\mu\nu}=a^2g_{\mu\nu}^M \,,
\label{fridg} 
\end{equation}
where $g_{\mu\nu}^M$ is the Minkowski metric. 
The corresponding Einstein equations with conformal metric (\ref{fridg}) 
and energy-momentum tensor of the classical Yang-Mills fields read \cite{YM-eqs}
\begin{eqnarray}
&&\frac{1}{\varkappa}\left(R_\mu^\nu-\frac12\delta_\mu^\nu
R\right)=\frac{1}{g_{\rm YM}^2}\frac{1}{\sqrt{-g}}\left(
-F_{\mu\lambda}^aF^{\nu\lambda}_a+ \frac14\delta_\mu^\nu
F_{\sigma\lambda}^aF^{\sigma\lambda}_a\right)\,,\quad \sqrt{-g}=a^4(\eta)\,, \nonumber\\
&&\left(\frac{\delta^{ab}}{\sqrt{-g}}\partial_\nu\sqrt{-g}-f^{abc}A_\nu^c\right)
\frac{F_b^{\mu\nu}}{\sqrt{-g}}=0\,,\quad F_{\mu\nu}^a=\partial_\mu
A_\nu^a-\partial_\nu A_\mu^a+f^{abc}A_\mu^bA_\nu^c\,,
\label{classicalYM}
\end{eqnarray}
where the gluon vacuum polarisation effects are not included in 
the considered free YM condensate case. As usual,
raising and lowering the Lorenz indices are done by the Minkowski metric
$g_{\mu\nu(\rm M)}$.

Neglecting the wave modes, the classical Einstein-Yang-Mills equations 
(\ref{classicalYM}) for the condensate $U=U(\eta)$ and 
the scale factor $a=a(\eta)$ are reduced to
\begin{eqnarray}
\frac{3}{\varkappa}\frac{a'^2}{a^4}&=&\frac{3}{2g_{\rm
YM}^2a^4}\left(U'^2+U^4\right)\,, \qquad U''+2U^3=0\,, \label{class}
\end{eqnarray}
The exact general solution of this system for the conformal evolution of 
the YM condensate corresponds to the non-linear oscillation whose 
frequency depends on initial value of the condensate $U_0=U(0)$, namely
\begin{eqnarray}
&& U'^2+U^4=C^{4},\quad \int_{U_0}^U\frac{dU}{\sqrt{C^4-U^4}}=\eta\,, \\
&& U(\eta)\simeq U_0\cos\biggl(\frac65\,U_0\eta\biggr) \,, \qquad
\qquad C=U_0\,, \qquad U'(0)=0\,,
\label{classol}
\end{eqnarray}
where $U_0,\,C$ are the integration constants.
We notice that the YM condensate in the Friedmann Universe 
behaves as an ultra-relativistic medium with energy density
$\varepsilon_{\rm YM}\sim 1/a^4$ and the equation of state $p_{\rm
YM}=\varepsilon_{\rm YM}/3$ \cite{Cembranos:2012ng}.

\subsection{Quasi-free condensate and vacuum polarisation effects}
\label{Sec:YM-cosm:YMC-vacpol}

Let us discuss the conformal time evolution of the quasi-free YM condensate 
$U=U(\eta)$ in the $SU(2)$ theory in the Friedmann Universe incorporating 
the vacuum polarisation effects. 
The corresponding YM Lagrangian accounting the one-loop vacuum 
polarisation reads \cite{Savvidy}:
\begin{eqnarray}\nonumber
L_{\rm YM}=
-\frac{11}{128\pi^2}\frac{F^a_{\mu\nu}F_a^{\mu\nu}}{\sqrt{-g}}\ln\biggl(\frac{J}{\Lambda_{\rm
QCD}^4}\biggr)\,,\qquad
J=\frac{1}{\xi^4}\frac{|F_{\alpha\beta}^aF^{\alpha\beta}_a|}{\sqrt{-g}}\,,
\label{Lagr}
\end{eqnarray}
where parameter $\xi$ reflects an ambiguity in normalisation of the invariant $J$. 
The corresponding Einstein-Yang-Mills equations now read \cite{Pasechnik}
\begin{eqnarray}
&&\frac{1}{\varkappa}\left(R_\mu^\nu-\frac12\delta_\mu^\nu R\right)=
{T_\mu^\nu}^{,\,{\rm mat}} + \bar{\Lambda}\delta_\mu^\nu +
\frac{11}{32\pi^2}\frac{1}{\sqrt{-g}}\biggl[\biggl(-F_{\mu\lambda}^aF^{\nu\lambda}_a
\nonumber
\\
&& \qquad\qquad +\,\frac14\delta_\mu^\nu
F_{\sigma\lambda}^aF^{\sigma\lambda}_a\biggr)
\ln\frac{e|F_{\alpha\beta}^aF^{\alpha\beta}_a|}{\sqrt{-g}\,(\xi\Lambda_{\rm
QCD})^4}-\frac14 \delta_\mu^\nu \,
F_{\sigma\lambda}^aF^{\sigma\lambda}_a\biggr]\,,\label{modYM} \\
&&\left(\frac{\delta^{ab}}{\sqrt{-g}}\partial_\nu\sqrt{-g}-f^{abc}A_\nu^c\right)
\left(\frac{F_b^{\mu\nu}}{\sqrt{-g}}
\ln\frac{e|F_{\alpha\beta}^aF^{\alpha\beta}_a|}{\sqrt{-g}\,(\xi\Lambda_{\rm
QCD})^4}\right)=0\,, \nonumber
\end{eqnarray}
where $e\simeq 2.71$ is the base of the natural logarithm;
$\Lambda_{\rm QCD}$ is the QCD energy scale;
${T_{\mu}^{\nu}}^{,\,{\rm mat}}=(\varepsilon+p)u_\mu
u^\nu-\delta_{\mu}^{\nu} p$ is the energy-momentum tensor of usual 
matter and radiation components; $\bar{\Lambda}$ is the constant vacuum 
energy density.

In the considered limit of vanishing YM waves, the components of the total 
energy-momentum tensor read
\begin{eqnarray}\nonumber
{T_0^0}^{,\,{\rm tot}}&=&{T_0^0}^{,\,{\rm mat}}+\bar{\Lambda}+
\frac{33}{64\pi^2}\frac{1}{a^4}\biggl[(U'^2+U^4)
\ln\frac{6e|U'^2-U^4|}{a^4(\xi\Lambda_{\rm
QCD})^4}+U'^2-U^4\biggr]\,,
\quad {T_{0}^{\beta}}^{,\,{\rm tot}}={T_{0}^{\beta}}^{,\,{\rm mat}}\,,\\
{T_{\alpha}^{\beta}}^{,\,{\rm tot}}&=&{T_{\alpha}^{\beta}}^{,\,{\rm
mat}}+ \bar{\Lambda}\delta_{\alpha}^{\beta}+
\frac{11}{32\pi^2}\frac{1}{a^4}\delta_{\alpha}^{\beta}
\biggl[-\frac12(U'^2+U^4)\ln\frac{6e|U'^2-U^4|} {a^4(\xi\Lambda_{\rm
QCD})^4}+\frac32(U'^2-U^4)\biggr] \,. \label{TeI}
\end{eqnarray}
The trace of the Einstein equations and the equation of motion of 
the YM condensate are written as follows
\begin{eqnarray}
&&\frac{6}{\varkappa}\frac{a''}{a^3}=\varepsilon-3p + 4\bar{\Lambda}
+ {T_{\mu}^{\mu}}^{,\,{\rm YM}}\,,\quad {T_{\mu}^{\mu}}^{,\,{\rm
YM}}=\frac{33}{16\pi^2}
\frac{1}{a^4}\left(U'^2-U^4\right)\,,\label{sysYM-1} \\
&&\frac{\partial}{\partial\eta}\left(U'\ln\frac{6e|U'^2-U^4|}
{a^4(\xi\Lambda_{\rm QCD})^4}\right)+2U^3
\ln\frac{6e|U'^2-U^4|}{a^4(\xi\Lambda_{\rm QCD})^4}=0 \,,
\label{sysYM-2}
\end{eqnarray}
respectively. It turns out that the $(0,0)$ Einstein equation
\begin{eqnarray}
\frac{3}{\varkappa}\frac{a'^2}{a^4}=\varepsilon + \bar{\Lambda} +
\frac{33}{64\pi^2}\frac{1}{a^4}
\left[\left(U'^2+U^4\right)\ln\frac{6e|U'^2-U^4|}{a^4(\xi\Lambda_{\rm
QCD})^4}+ U'^2-U^4\right] \label{int00}
\end{eqnarray}
is the exact first integral of the system of equations
(\ref{sysYM-1}) and (\ref{sysYM-2}), while the exact first integral
of second equation (\ref{sysYM-2}) is
\begin{eqnarray}
U'^2-U^4=a^4\frac{(\xi\Lambda_{\rm QCD})^4}{6e}\,.
\label{sysYM-2-int}
\end{eqnarray}
Applying this relation in Eq.~(\ref{TeI}), one arrives to
\begin{eqnarray}
&&{T_0^0}^{,\,{\rm tot}}={T_0^0}^{,\,{\rm mat}}+\bar{\Lambda}+
\frac{33}{64\pi^2}\frac{(\xi\Lambda_{\rm QCD})^4}{6e}\,,\nonumber\\
&&{T_{\alpha}^{\beta}}^{,\,{\rm tot}}={T_{\alpha}^{\beta}}^{,\,{\rm
mat}}+\left(\bar{\Lambda}+ \frac{33}{64\pi^2}\frac{(\xi\Lambda_{\rm
QCD})^4}{6e}\right)\delta_{\alpha}^{\beta}\,. \label{TeI-sol}
\end{eqnarray}
Therefore, the quasi-free YM condensate incorporating the 
vacuum polarisation effects already as one loop exhibits drastically 
different behaviour w.r.t. to the free condensate limit. Firstly, its
contribution to the total energy density of the Universe appears to
be constant and independent on the scale factor like it is expected
for a vacuum-like medium with the equation of state $p_{\rm
YM}=-\varepsilon_{\rm YM}$ compared to free condensate equation of state 
$p_{\rm YM}=\varepsilon_{\rm YM}/3$. Secondly, it provides a positive 
contribution to the cosmological constant which would potentially allow
to compensate large negative contributions to the vacuum energy density
potentially coming from quantum-topological excitations contributing to $\bar{\Lambda}$, e.g. 
from the vacuum averaged bilinear terms in YM waves $\langle \widetilde A^2 \rangle$ \cite{Pasechnik}.
The latter argument is consedered to be critical for observed smallness of the $\Lambda$-term 
density $\Lambda_{obs}$ which should be provided by the condition
\begin{eqnarray}
\bar{\Lambda}+\frac{33}{64\pi^2}\frac{(\xi\Lambda_{\rm QCD})^4}{6e}\equiv\Lambda_{\rm obs}\,,
\end{eqnarray}
such that the macroscopic expansion of the Universe is driven by 
the ordinary Friedmann equation
\begin{eqnarray}
\frac{3}{\varkappa}\frac{a'^2}{a^4}=\varepsilon + \Lambda_{\rm obs} \,.  \label{sys-fin-1}
\end{eqnarray}
A thorough analysis of cosmological evolution of the complete YM system 
``condensate + waves'' incorporating vacuum polarisation effects 
is planned for future studies.

\subsection{Super-Yang-Mills condensate decay in quasi-linear approximation}
\label{Sec:YM-cosm:SYMC}

Let us discuss now the effect of wave-condensate interactions in quasi-linear approximation in 
the expanding Universe. For this purpose, consider the Lagrangian of ${\cal N}=4$ supersymmetric 
$SU(2)$ model in arbitrary metric $g_{\mu\nu}$ in the quadratic approximation in waves
(in what follows, we denote determinant of spacetime metric as $g_{det}$ in order to not 
confuse it up with the coupling constant)
\begin{eqnarray} \nonumber
&& \mathcal{L}\sqrt{-g_{det}}=\sqrt{-g_{det}}\Big(-\frac{1}{4}F^a_{\mu\nu}F^a_{\lambda\theta}g^{\lambda\mu}g^{\theta\nu}
+\frac{i}{2}\bar{\lambda}^a_{k'}\gamma_\lambda g^{\lambda\mu}D_\mu\lambda^a_{k'}
+\frac{1}{2}(D_\lambda C^a_{(i)})(D_\mu C^a_{(i)})g^{\lambda\mu} \\
&& +\, \frac{1}{2}(D_\lambda B^a_{(i)})(D_\mu B^a_{(i)})g^{\lambda\mu}\Big) \,.
\label{Lsuperg} 
\end{eqnarray}
This Lagrangian is not conformally invariant due to the presence of scalar and pseudo-scalar fields. 
Note, the conformal invariance can be easily restored by adding terms proportional to the scalar curvature $R$, i.e. 
\begin{eqnarray} \nonumber
&& \mathcal{L}\sqrt{-g_{det}}=\sqrt{-g_{det}}\Big(-\frac{1}{4}F^a_{\mu\nu}F^a_{\lambda\theta}g^{\lambda\mu}g^{\theta\nu}
+\frac{i}{2}\bar{\lambda}^a_{k'}\gamma_\lambda g^{\lambda\mu}D_\mu\lambda^a_{k'}
+\frac{1}{2}(D_\lambda C^a_{(i)})(D_\mu C^a_{(i)})g^{\lambda\mu} \\
&& +\,\frac{1}{2}(D_\lambda B^a_{(i)})(D_\mu B^a_{(i)})g^{\lambda\mu}+
\frac{1}{12}R C^a_{(i)}C^a_{(i)}
+\frac{1}{12}R B^a_{(i)}B^a_{(i)}\Big) \,.
\label{Lsupergmod} 
\end{eqnarray}
The corresponding theory is free of radiative corrections and hence vacuum polarisation effects considered above 
due to conformal symmetry. So this model can be convenient for studies of conformal time evolution of the 
YM condensate induced purely by its interactions with the wave modes, at least, at the first step.

It is instructive to introduce the following conventions
\begin{eqnarray}
A_{\mu}=A^{M}_{\mu}\,,\quad A^{\mu}=\frac{1}{a^2} A^{\mu}_M \,, \quad
C^a_{(i)}=\frac{1}{a}C^{Ma}_{(i)}\,, \quad B^a_{(i)}=\frac{1}{a}B^{Ma}_{(i)}\,, \quad
\lambda^a_{k'}=\frac{1}{a}\lambda^{Ma}_{k'} \,,
\label{masshtab} 
\end{eqnarray}
Then performing transition to Friedman metric (\ref{fridg}) we arrive at the same expression 
for Lagrangian as for Minkowski metric (with $a=1$) due to the conformal invariance 
of Eq.~(\ref{Lsupergmod}). Indeed,
\begin{eqnarray} \nonumber
&& \mathcal{L}\sqrt{-g_{det}}=-\frac{1}{4}F^{Ma}_{\mu\nu}F_{Ma}^{\mu\nu}
+\frac{i}{2}\bar{\lambda}^{Ma}_{k'}\gamma^\mu D_\mu\lambda^{Ma}_{k'}
+\frac{1}{2}(D^\mu C^{Ma}_{(i)})(D_\mu C^{Ma}_{(i)}) \\
&& +\, \frac{1}{2}(D^\mu B^{Ma}_{(i)})(D_\mu B^{Ma}_{(i)})\,,
\label{LsuperM} 
\end{eqnarray}
where time derivatives are taken over the conformal time $\eta$. So due to conformal invariance dynamics of 
the YM-fields in Friedman space corresponds to that in Minkowski metric, but in conformal time.

The Friedman equation allows to find a relation between conformal and cosmological times as
\begin{equation}
\frac{3}{a^4}\Big(\frac{\partial a}{\partial\eta}\Big)^2=\frac{\kappa}{a^4}\langle {T^0_0}^M \rangle \,, \qquad 
\langle {T^0_0}^M \rangle\equiv c={\rm const} \,,
\label{lawa} 
\end{equation}
where ${T^0_0}^M=a^4T^{0}_0$. The corresponding solution is
\begin{equation}
a=\sqrt{\frac{\kappa c}{3}}\eta\,, \qquad t=\frac{1}{2}\sqrt{\frac{\kappa c}{3}}\eta^2\,.
\label{tetasol} 
\end{equation}

Our analysis of quasilinear YM theory shows that the YM condensate decay time is inversely proportional 
to the initial amplitude of YM condensate oscillations $U_0$, i.e.
\begin{equation}
\Delta\eta=\frac{\alpha}{gU_0}\,,
\label{timeU} 
\end{equation}
where the numerical coefficient $\alpha\approx10$ for the wave modes $\Phi,\Lambda,\lambda$. 
Inclusion of other modes is problematic due to the limitations of quasi-linear theory (see above) but can 
be consistently performed by accounting for higher order effects in waves.

Staying within the quasi-linear approximation and assuming that all the energy was concentrated in 
the YM condensate in the initial moment of time then one has
\begin{equation}
c=\frac{3}{2}g^2U_0^4\,.
\label{T00U} 
\end{equation}
By substitution of Eqs.~(\ref{T00U}) and (\ref{timeU}) into Eq.~(\ref{tetasol}) one 
finds the cosmological time of YM condensate decay, approximately,
\begin{equation}
\Delta t\simeq\sqrt{\frac{\kappa}{8}}\frac{\alpha^2}{g}\,, \qquad \alpha\sim 10\,.
\label{timeUsol} 
\end{equation}
This provides us with the typical time scale for the YM condensate decay into particles 
which can be considered as an additional mechanism for particles production in the early Universe.
An extension of cosmological dynamics of the YM system ``condensate + waves'' to 
an arbitrary YM theory accounting for higher order effects in wave modes will be done elsewhere.

\section{An overview of the two-condensate $SU(4)$ model}
\label{Sec:Two-YMC}

Let us consider a more complicated YM theory based on a large gauge group 
containing a few $SU(2)$ subgroups. The simplest group of this type is $SU(4)$. 
It contains two $SU(2)$ gauge subgroups both isomorphic 
to the $SO(3)$ spatial symmetry group which means that the YM field, 
described by local $SU(4)$ gauge theory, contains two condensates. 
The main focus of this Section is to discuss dynamical features 
of such a heterogenic vacuum system.

The generators of $SU(4)$ gauge group can be written as
\begin{eqnarray*}
\lambda^{1} = \left( \begin{array}{cccc}
0 & 1 & 0 & 0\\
1 & 0 & 0 & 0\\
0 & 0 & 0 & 0\\
0 & 0 & 0 & 0\\
\end{array} \right)\,, \;
\lambda^{2} = \left( \begin{array}{cccc}
0 & -i & 0 & 0\\
i & 0 & 0 & 0\\
0 & 0 & 0 & 0\\
0 & 0 & 0 & 0\\
\end{array} \right)\,, \;
\lambda^{3} = \left( \begin{array}{cccc}
1 & 0 & 0 & 0\\
0 & -1 & 0 & 0\\
0 & 0 & 0 & 0\\
0 & 0 & 0 & 0\\
\end{array} \right)\,, \;
\lambda^{4} = \left( \begin{array}{cccc}
0 & 0 & 0 & 0\\
0 & 0 & 0 & 0\\
0 & 0 & 0 & 1\\
0 & 0 & 1 & 0\\
\end{array} \right)
\end{eqnarray*}

\begin{eqnarray*}
\hspace{-0.55cm}
\lambda^{5} = \left( \begin{array}{cccc}
0 & 0 & 0 & 0\\
0 & 0 & 0 & 0\\
0 & 0 & 0 & -i\\
0 & 0 & i & 0\\
\end{array} \right)\,, \;
\lambda^{6} = \left( \begin{array}{cccc}
0 & 0 & 0 & 0\\
0 & 0 & 0 & 0\\
0 & 0 & 1 & 0\\
0 & 0 & 0 & -1\\
\end{array} \right)\,, \;
\lambda^{7} = \left( \begin{array}{cccc}
0 & 0 & 1 & 0\\
0 & 0 & 0 & 0\\
1 & 0 & 0 & 0\\
0 & 0 & 0 & 0\\
\end{array} \right)\,, \;
\lambda^{8} = \left( \begin{array}{cccc}
0 & 0 & -i & 0\\
0 & 0 & 0 & 0\\
i & 0 & 0 & 0\\
0 & 0 & 0 & 0\\
\end{array} \right)
\end{eqnarray*}

\begin{eqnarray*}
\hspace{-0.55cm}
\lambda^{9} = \left( \begin{array}{cccc}
0 & 0 & 0 & 0\\
0 & 0 & 1 & 0\\
0 & 1 & 0 & 0\\
0 & 0 & 0 & 0\\
\end{array} \right)\,, \;
\lambda^{10} = \left( \begin{array}{cccc}
0 & 0 & 0 & 0\\
0 & 0 & -i & 0\\
0 & i & 0 & 0\\
0 & 0 & 0 & 0\\
\end{array} \right)\,, \;
\lambda^{11} =\frac{1}{\sqrt{2}} \left( \begin{array}{cccc}
1 & 0 & 0 & 0\\
0 & 1 & 0 & 0\\
0 & 0 & -1 & 0\\
0 & 0 & 0 & -1\\
\end{array} \right)\,, \;
\lambda^{12} = \left( \begin{array}{cccc}
0 & 0 & 0 & 1\\
0 & 0 & 0 & 0\\
0 & 0 & 0 & 0\\
1 & 0 & 0 & 0\\
\end{array} \right)
\end{eqnarray*}

\begin{eqnarray}
\hspace{-0.55cm}
\lambda^{13} = \left( \begin{array}{cccc}
0 & 0 & 0 & -i\\
0 & 0 & 0 & 0\\
0 & 0 & 0 & 0\\
i & 0 & 0 & 0\\
\end{array} \right)\,, \;
\lambda^{14} = \left( \begin{array}{cccc}
0 & 0 & 0 & 0\\
0 & 0 & 0 & 1\\
0 & 0 & 0 & 0\\
0 & 1 & 0 & 0\\
\end{array} \right)\,, \;
\lambda^{15} = \left( \begin{array}{cccc}
0 & 0 & 0 & 0\\
0 & 0 & 0 & -i\\
0 & 0 & 0 & 0\\
0 & i & 0 & 0\\
\end{array} \right)
\label{su4generators}
\end{eqnarray}
The generators $\lambda^1,\,\lambda^2,\,\lambda^3$ and
$\lambda^4,\,\lambda^5,\,\lambda^6$ correspond to $SU(2)$ subgroups,
and structure constants are given by
\begin{equation}
f_{abc}=\frac{1}{4i}\,{\rm Tr}([\lambda_a,\lambda_b]\lambda_c)\,.
\end{equation}

In the Hamilton gauge, two different YM condensates $U_i=U_i(t),\,i=1,2$
corresponding to each $SU(2)$ subgroup can be introduced in analogy
with Eq.~(\ref{TheFirst}), i.e.
\begin{equation}
 A^a_i=\delta_{1\,i}^a\, U_1 + \delta_{2\,i}^a\, U_2 +
 \widetilde{A}^a_i\,, \qquad a=1\,\dots\,15\,,
\end{equation}
where
\begin{equation}
\delta_{1\,i}^a = \left( \begin{array}{ccc}
1 & 0 & 0 \\
0 & 1 & 0 \\
0 & 0 & 1 \\
0 & 0 & 0 \\
0 & 0 & 0 \\
0 & 0 & 0 \\
. & . & . \\
0 & 0 & 0 \\
\end{array} \right)\,, \qquad
\delta_{2\,i}^a = \left( \begin{array}{ccc}
0 & 0 & 0 \\
0 & 0 & 0 \\
0 & 0 & 0 \\
1 & 0 & 0 \\
0 & 1 & 0 \\
0 & 0 & 1 \\
. & . & . \\
0 & 0 & 0 \\
\end{array} \right) \,.
\end{equation}

The equations of motion are given by general formula from the
classical YM theory (\ref{eq}). The equations for free
(non-interacting) condensates $U_1$ and $U_2$ can be easily
extracted from Eq.~(\ref{eq})
\begin{eqnarray*}
 \partial_0\partial_0 U_1 + 2g^2\,U_1^3 = 0\,, \qquad 
 \partial_0\partial_0 U_2 + 2g^2\,U_2^3 = 0\,.    
\end{eqnarray*}
Note, these equations do not contain mixed terms like $U_1^n\,U_2^m$
and coincide with Eq.~(\ref{eqU}). This means that the condensates
$U_1(t)$ and $U_2(t)$ in the $SU(4)$ theory do not interact with
each other directly. As we will demonstrate later, they can interact
only by means of particle exchanges.

The linear equations of motion for the Fourier transformed wave
modes $\widetilde{A}^{\vec{p}\,a}_i$ (we omit index $\vec{p}$ below)
are
\begin{eqnarray} \nonumber
 && \partial_0\partial_0\widetilde{A}^{a}_k + p^2\,\widetilde{A}^{a}_k -
    p^2\,n_kn_i\widetilde{A}^{a}_i + 2igp\,U_1\,f_{aic}n_i\widetilde{A}^{c}_k +
    2igp\,U_2\,f_{a(i+3)c}n_i\widetilde{A}^{c}_k + \\
 && igp\,U_1\,f_{abk}n_i\widetilde{A}^{b}_i +
    igp\,U_2\,f_{ab(k+3)}n_i\widetilde{A}^{b}_i +
    igp\,U_1\,f_{abi}n_k\widetilde{A}^{b}_i +
    igp\,U_2\,f_{ab(i+3)}n_k\widetilde{A}^{b}_i = 0\,.
 \label{su4lineq}
\end{eqnarray}
One can show by a direct calculation that equations for the wave
modes corresponding to each of the $SU(2)$ subgroups (with $a=1,2,3$
and $a=4,5,6$, respectively) coincide with analogical equations in
the one-condensate model (\ref{main}) (or with Eqs.~(\ref{eq1}) and
(\ref{eq2}) in terms of the tensor basis modes).

Now let us investigate the quasi-linear ``back reaction'' effect of
the wave modes to YM condensates. Including next (second) order in waves, the
equation for the $U_1$ condensate reads
\begin{eqnarray} \nonumber
 && \partial_0\partial_0 U_1 + 2g^2\,U_1^3 + \frac{ig}{12}\,\sum_{\vec{p}}
    p\,f_{kbc} \Big\langle 2n_i\widetilde{A}^{b \dagger}_i\widetilde{A}^{c}_k -
    2n_i\widetilde{A}^{b}_i\widetilde{A}^{c \dagger}_k +
    n_i\widetilde{A}^{c \dagger}_k\widetilde{A}^{b}_i - \\
 && n_i\widetilde{A}^{c}_k\widetilde{A}^{b \dagger}_i +
    n_k\widetilde{A}^{b \dagger}_i\widetilde{A}^{c}_i -
    n_k\widetilde{A}^{b}_i\widetilde{A}^{c \dagger}_i \Big\rangle +
    \frac{g^2}{12}\,U_1\,\sum_{\vec{p}} \Big\langle
    f_{kbi}f_{bde}(\widetilde{A}^{d}_i\widetilde{A}^{e \dagger}_k +
    \widetilde{A}^{d \dagger}_i\widetilde{A}^{e}_k) + \nonumber \\
 && f_{kbc}f_{bdk}(\widetilde{A}^{d}_i\widetilde{A}^{c \dagger}_i +
    \widetilde{A}^{d \dagger}_i\widetilde{A}^{c}_i) +
    f_{kbc}f_{bie}(\widetilde{A}^{e}_k\widetilde{A}^{c \dagger}_i +
    \widetilde{A}^{e \dagger}_k\widetilde{A}^{c}_i) \Big\rangle + \nonumber \\
 && \frac{g^2}{12}\,U_2\,\sum_{\vec{p}}
    \Big\langle f_{kb(i+3)}f_{bde}(\widetilde{A}^{d}_i\widetilde{A}^{e \dagger}_k +
    \widetilde{A}^{d \dagger}_i\widetilde{A}^{e}_k) +
    f_{kbc}f_{bd(k+3)}(\widetilde{A}^{d}_i\widetilde{A}^{c \dagger}_i +
    \widetilde{A}^{d \dagger}_i\widetilde{A}^{c}_i) + \nonumber \\
 && f_{kbc}f_{b(i+3)e}(\widetilde{A}^{e}_k\widetilde{A}^{c \dagger}_i +
    \widetilde{A}^{e \dagger}_k\widetilde{A}^{c}_i) \Big\rangle = 0 \,.
\label{su4U1backreaction}
\end{eqnarray}
The corresponding equation for the second condensate $U_2$ has an
analogical form. By a direct calculation it can be shown from
Eq.~(\ref{su4U1backreaction}), analogical equation for $U_2$ and
Eq.~(\ref{su4lineq}) that the wave modes corresponding to the first
and second $SU(2)$ subgroup interact only with its own condensate
$U_1$ and $U_2$, respectively, in the same way as they do in the
one-condensate model considered above. Remarkably enough, other 27
modes corresponding $a=7\,\dots\,15$ generators of $SU(4)$ interact
with both condensates at the same time. This means that interaction
between $U_1$ and $U_2$ condensates is realised via particle
exchanges only related to these remaining 27 wave modes. This effect
is explicitly confirmed by a numerical analysis.

In general, time evolution of wave modes of the $SU(4)$ theory and
two YM condensates is analogical to the case of one-condensate system
illustrated in Fig.~\ref{fig:YMC-int}, i.e. energy of the both
condensates is transferred into the ultra-relativistic YM plasma.
This study suggests that the observed effect of energy swap is a
generic feature of YM dynamics.

\section{Summary}
\label{Sec:concl}

Starting from the basic idea about an important dynamical role of
the YM condensate (\ref{TheFirst}), we have constructed a consistent
quasi-classical approach based on Hamilton formulation and canonical
quantisation of the wave modes in the classical YM condensate. This approach
has been applied in numerical analysis of the system of YM wave modes (or
particles after quantisation) in the ultra-relativistic gluon plasma interacting 
with the YM condensate (in the limit of small interactions between 
the wave modes) as well as in analytic investigation of generic 
YM ``condensate + waves'' dynamics with all the interaction terms included.

First, the quantum energy spectrum of free (no waves) YM condensate has been 
found from the stationary Schr\"odinger equation. It turned out that this 
spectrum corresponds to a potential well of the fourth power, and a
convenient analytical approximation to the discrete numerical
solution has been proposed.

Next, we have derived the YM equations of motion for 
the ``condensate + waves'' system in 
linear approximation (\ref{eqU}), (\ref{eq1}) and (\ref{eq2})
in the $SU(2)$ gauge theory and numerically investigated
their solutions. As an important effect which removes degeneracy 
in the underlined gauge theory the interactions of wave modes with 
the YM condensate lead to dynamical generation of the longitudinal d.o.f.'s 
in the gluon plasma. 

Also, in order to understand how condensate-waves interactions affect 
the energy balance between these two YM subsystems, we have investigated 
their dynamics in quasi-linear and next-to-linear 
approximations accounting for the ``back reaction'' effect of particles 
to the condensate (\ref{BackReact}). The higher order terms have also 
been studied and found to be important for late-time stability of the 
YM solutions. The results of numerical analysis 
are presented in Figs.~\ref{fig:YMC-int} and \ref{fig:Phi-energy} and 
demonstrate a characteristic energy swap effect, namely, an effective 
energy transfer from initially large condensate fluctuations to
the YM wave modes. This effect of condensate decay can thus be 
considered as a possible source for heating up the ultra-relativistic 
gluon plasma as well as potentially gives rise to particle production 
mechanisms in the hot cosmological plasma in the early Universe. 

At the next step, the equations for pure gluon plasma (without the 
YM condensate) in the $SU(3)$ gauge theory have been derived in different 
approximations. The third-order interaction terms in waves dynamically 
induce a local constituent gluon mass. The corresponding renormalized 
expression appears to be consistent with thermodynamic predictions. 
The second-order interactions initiate the dynamical gluon mass terms 
in a time non-local form. In addition, the exact integro-differential equations 
for the condensate and waves evolution incorporating all the higher-order 
interaction terms have been derived for the $SU(2)$ theory. 

It has been shown that dynamics of waves and condensate in the
extended ${\cal N}=4$ supersymmetric YM theory is analogical to the
one of pure YM theory discussed above. In particular, it has been
indicated that interaction of the supersymmetric (pseudo)scalar wave
modes with the YM condensate leads to a similar energy swap effect between
them. A similar energy swap effect has also been found in the heterogenic 
system of two interacting YM condensates in the $SU(4)$ gauge theory. 
These findings strongly suggest that the observed dynamics in
energy balance of the interacting YM ``wave + condensate'' system is
a general phenomenon and a specific property inherent to a variety of 
YM theories with condensate.

We also considered conformal time dynamics of free non-interacting YM condensate 
in the Friedmann Universe. The non-linear oscillations of the condensate behaving 
as radiation matter in the course of Universe expansion appears to be unstable w.r.t. 
quantum corrections. In order to address the latter issue 
the leading vacuum polarisation effects at one-loop have been incorporated. 
Such a quasi-free condensate contribution to the energy density dramatically 
changes its behavior to a positive constant in time. This behaviour is consistent 
with the vacuum equation of state and thus can be used to compensate 
large negative topological vacua terms \cite{Pasechnik}. The latter would potentially 
offer a consistent way to address an observable smallness of the cosmological 
constant but a self-consistent analysis of the waves with the vacuum polarisation 
effects will be critical for that. As a toy-model void of radiative corrections, 
the Super-Yang-Mills theory with condensate interacting with small waves has 
been considered in the Friedmann Universe and the condensate decay time 
has been estimated in quasi-linear approximation. The latter approach once 
applied to a realistic interacting gauge theory could provide an access to 
typical time scale for particle production, in particular, during the QCD transition 
epoch and/or in the preheating stage of a gauge-field driven inflation 
in Cosmology. These important aspects will be discussed elsewhere.

Finally, we have investigated the gauge dependence of the YM condensate and 
waves. It turned out that it is possible to establish a connection between 
the results in the Hamilton gauge and those in arbitrary gauge by the 
method of absorbing the gauge dependence into a coordinate 
transformation. In this framework the YM condensate 
in arbitrary gauge can be considered as a functional of the complete 
YM solution in the Hamilton gauge and the gauge-fixing function. 
Our thorough analysis of non-degenerate YM dynamics performed 
in the Hamilton gauge has appeared to be justified as the most 
straightforward and convenient since these results can be then, 
in principle, transformed into those in any given gauge.

To summarize, as one of the main results of this paper, the energy redistribution effect
from the YM condensate to the YM wave modes in the case of initially 
small wave amplitudes has been found and investigated
from the first principles of canonical quasi-classical YM theory 
in different settings and approximations. As was mentioned above, 
this effect can be of major importance for cosmological processes 
in the early Universe, in particular, in the processes of particle production 
during the preheating period after cosmic inflation which is planned 
for further studies. In addition, an extension of the quasi-classical 
approach to a full quantum field theory formalism (including fermion modes 
and vacuum polarisation effects) could become one of the next important 
steps in further theoretical understanding of interacting 
YM dynamics of ``condensate + waves'' systems.

\appendix

\section{The method of infinitesimal parameter}
\label{A1}

The major difficulties of the canonical quantization of a free YM
field (without the YM condensate) arise due to the presence of its
time-retarded zero component in the Lagrangian (\ref{L}). One of the
ways to resolve this issue is based upon the method of asymptotic expansion 
in the configuration space elaborated in Refs.~\cite{BVF1,BVF2}. Here, we
advise an alternative {\it method of infinitesimal parameter} which is also 
applicable in the Hamilton gauge (\ref{gauge}).

Let us construct the YM propagator in the form of chronologically
ordered operator product averaged over the state vector. Consider a YM system 
containing a YM field $A^{a}_\mu$ and a charged multiplet of fermions interacting 
with $A^{a}_\mu$. The Lagrangian and Hamiltonian densities of such a system
are
\begin{eqnarray}
\mathcal{L}&=&-\frac{1}{4}F_{\mu\nu}^{a}F_{a}^{\mu\nu}+i\overline{q}_f\gamma^\mu
\Big(\partial_\mu q_f - \frac{i}{2} g A_{\mu}^{a}\lambda^a q_f\Big)
- m_f\overline{q}_fq_f \,, \\
\mathcal{H}&=&\frac{1}{2}\Big((F^{a}_{0k})^2 + (F^{a}_{ik})^2\Big)
-\frac{1}{2}g\,\bar{q}_f\gamma^iA^{a}_i\lambda^{a}q_f
-i\bar{q}_f\gamma^i\partial_iq_f+m_f\bar{q}_fq_f \,,
\end{eqnarray}
respectively. Then, the corresponding Lagrange equations of motion
read
\begin{eqnarray*}
 && \partial^\mu F_{\mu\nu}^{a} - gF_{\mu\nu}^{b}f_{abc}A^{\mu}_{c} +
\frac{1}{2}g\overline{q}_f\gamma_\nu\lambda^{a}q_f = 0\,, \\
 && i\gamma^\mu\partial_\mu q_f + \frac{1}{2}g\gamma^\mu A^{a}_\mu
\lambda^a q_f - m_fq_f=0\,, \\
 && -i\partial_\mu\overline{q}_f\gamma^\mu +
 \frac{1}{2}g\overline{q}_f\gamma^\mu A^{a}_\mu \lambda^a -
 m_f\overline{q}_f = 0 \,.
\end{eqnarray*}
The canonical quantization procedure is based upon the
(anti)commutation relations between the field operators
\begin{eqnarray*}
 && [E_{i}^{a}(t,\vec{x}),\,A_{k}^{b}(t,\vec{x}\,')]_- =
   - i\delta_{ik}\delta_{ab}\delta(\vec{x}-\vec{x}\,') \,, \\
 && [q^{\alpha}_f(t,\vec{x}),\,P^{\beta}_{f'}(t,\vec{x}\,')]_+ =
   i\delta_{\alpha\beta}\delta_{ff'}\delta(\vec{x}-\vec{x}\,') \,,
\end{eqnarray*}
where generalized momenta conjugated to the fields $A^{a}_k$ and
$q^{\alpha}_f$ are found as
\begin{eqnarray*}
 && E^a_k = \frac{\delta}{\delta(\partial_0A^{a}_k)}
 \int\mathcal{L}d^{\,3}x = F^{a}_{0k}\,, \\
 && P^{\alpha}_f = \frac{\delta}{\delta(\partial_0q_f^\alpha)}
 \int\mathcal{L}d^{\,3}x = i\bar{q}^{\alpha}_f\gamma^0 \,,
\end{eqnarray*}
respectively. Here, we kept color index $\alpha$ of a quark flavor
$f$ for transparency, while it is often omitted in other places. The
system of quantum equations of motion can be written in the Heisenberg
representation
\begin{eqnarray*}
 && \partial_0 F_{0k}^{a} = \partial_i F_{ik}^{a} - gF_{ik}^{b}f_{abc}A_{i}^{c} -
 \frac{1}{2}g\,\bar{q}_f\gamma_k\lambda^{a}q_f \,, \\
 && i\gamma^\mu\partial_\mu q_f +
 \frac{1}{2}g\,\gamma^i A^{a}_i \lambda^a q_f - m_fq_f = 0 \,, \\
 && -i\partial_\mu\bar{q}_f\gamma^\mu +
 \frac{1}{2}g\,\bar{q}_f\gamma^i A^{a}_i \lambda^a -
 m_f\bar{q}_f = 0 \,, \\
 && \partial^0\Big(\partial^i F_{i0}^{a} - gf_{abc}F_{i0}^{b}A^{i}_{c} +
 \frac{1}{2}g\bar{q}_f\gamma_0\lambda^{a}q_f\Big) = 0\,,
\end{eqnarray*}
as well as in the interaction representation
\begin{eqnarray}
 \partial_0\partial_0 A^{a}_{i} - \partial_k\partial_k A_{i}^{a}
  + \partial_i \partial_k A_{k}^{a} = 0 \,,  \quad
 i\gamma^\mu\partial_\mu q_f - m_fq_f = 0 \,, \quad
 i\partial_\mu\bar{q}_f\gamma^\mu + m_f\bar{q}_f = 0 \,. \label{eqs-int}
\end{eqnarray}
Performing the Fourier transformation, one arrives at the system of 
equations for the longitudinal $A_{\vec{p}\,0}^{a}$ and transverse components
$A_{\vec{p}\,\lambda}^{a},\,\lambda=1,2$ of the YM field in the following form
\begin{eqnarray}
 \partial_0\partial_0A_{\vec{p}\,0}^{a} = 0\,, \qquad
 \partial_0\partial_0A_{\vec{p}\,\lambda}^{a}
   + |\vec{p}|^2\,A_{\vec{p}\,\lambda}^{a} = 0 \,, \label{long-trans}
\end{eqnarray}
respectively. The equation for the longitudinal mode does not have a
wave solution, so it is impossible to take into account its
contribution in the YM propagator constructed as vacuum average of
the chronologically ordered operator product. 

In order to resolve this problem we can modify the Hamiltonian density by means of
adding an extra small ``virtual'' term depending on an infinitesimal
parameter $\zeta\ll 1$ and vanishing at $\zeta\to0$, i.e.
\begin{eqnarray*}
 \mathcal{H}_{mod}&=&\int\Big[\frac{1}{2}\big(\partial_0 A_{k}^{a}\,\partial_0
 A_{k}^{a} + \partial_l A_{k}^{a}\,\partial_l A_{k}^{a} - (1-\zeta^2)
 \partial_l A_{k}^{a}\,\partial_k A_{l}^{a}\big) -
 i\bar{q}_{f}\gamma^i\partial_iq_{f} +
 m_f\bar{q}_{f}q_{f}\Big] d^3x\,.
\end{eqnarray*}
Due to such a modification, the equation of motion for the longitudinal
component becomes
\begin{equation}
\partial_0\partial_0A_{\vec{p}\,0}^{a} +
\zeta^2\,|\vec{p}|^2\,A^a_{\vec{p}\,0}=0\,,
\end{equation}
such that it acquires an infinitesimal frequency. This modified
equation enables us to incorporate the longitudinal mode into the YM
propagator which is given by (in the limit $\zeta\to0$)
\begin{equation}
D^{ab}_{ik}(x-x') = \frac{\delta_{ab}}{(2\pi)^4}\,\int
\Big(\delta_{ik}-\frac{p_ip_k}{p_0^2+i\epsilon}\Big)
\frac{d^{\,4}p}{p^2-p_0^2-i\epsilon}\,e^{-ip(x-x')} \,,
\label{prop-A}
\end{equation}
where $p_0$ is the zeroth component of the YM field momentum $p$. Then the
fermion propagator takes the standard form, i.e.
\begin{eqnarray}
 D_{ff'}(x-x') = \frac{\delta_{ff'}}{(2\pi)^4}\, \int (\gamma_\mu p^\mu+m_f)\,
 \frac{d^{\,4}p}{p^2-m_f^2+i\epsilon}\,e^{-ip(x-x')}\,. \label{prop-q}
\end{eqnarray}
Note, the formulae (\ref{prop-A}) and (\ref{prop-q}) coincide with 
the corresponding Green functions constructed for the initial (non-modified) 
YM equations of motion (\ref{eqs-int}), thus, validating the proposed method.

\section{Canonical quantization of the Yang-Mills wave modes}
\label{A2}

Let us now perform canonical quantisation of the YM wave modes in
the classical YM condensate and therefore construct the quasi-classical YM
theory. For this purpose, as the matter of the Bohr's correspondence
principle we introduce operators instead of field functions in the
Hamiltonian density of, for example, the ${\cal N}=4$ supersymmetric
YM theory (\ref{HSUSY}). Then we impose (anti)commutation relations
to the field operators for each wave mode as follows
\begin{eqnarray*}
 P_\Phi=\frac12\partial_0 \Phi^\dagger\,, &\;& [P_\Phi,\Phi]_-=-i \,, \\
 P_\Lambda=\frac{1}{4}\,\partial_0 \Lambda^\dagger\,, &\;&
    [P_\Lambda,\Lambda]_-=-i \,, \\
 P_\lambda=\frac12\partial_0 \lambda^\dagger\,, &\;&
    [P_\lambda,\lambda]_-=-i \,, \\
 P^\sigma_\phi=\frac12\partial_0 \phi_\sigma^\dagger\,, &\;&
    [P^\sigma_\phi,\phi_\gamma]_-=-i\delta_{\sigma\gamma} \,, \\
 P^\sigma_\eta=\frac12\partial_0 \eta_\sigma^\dagger\,, &\;&
    [P^\sigma_\eta,\eta_\gamma]_-=-i\delta_{\sigma\gamma} \,, \\
 P^\sigma_\psi=\frac12\partial_0 \psi_\sigma^\dagger\,, &\;&
    [P^\sigma_\psi,\psi_\gamma]_-=-i\delta_{\sigma\gamma} \,, \\
 P_K^{(i)}=\frac{1}{4}\partial_0K_{(i)}^\dagger\,, &\;&
    [P^{(i)}_K,K_{(j)}]_- = -i\delta_{ij} \,, \\
 P_M^{(i)\alpha}=\frac{1}{4}\partial_0M_{(i)\alpha}^\dagger\,, &\;&
    [P^{(i)\alpha}_M,M_{(j)\beta}]_-=-i\delta_{ij}\delta_{\alpha\beta} \,, \\
 P_{\vartheta}^{k'}=\frac{i}{4}\bar{\vartheta}^{k'}\gamma_0\,,
 &\;& 
 [\vartheta^{l'},P^{k'}_{\vartheta}]_+=i\delta_{l'k'} \,,\\ 
 P_{\zeta}^{k'\alpha}=\frac{i}{4}\bar{\zeta}^{k'}_\alpha\gamma_0\,,
 &\;&
    [\zeta^{l'}_\alpha,P^{k'\beta}_{\zeta}]_+=
    i\delta_{l'k'}\delta_{\alpha\beta} \,.
    \label{comm-rel}
\end{eqnarray*}
Commutation relations for pseudoscalar modes $K'$ and $M'$ are the
same as for scalar ones $K$ and $M$, respectively. In addition,
analogical formulas for Hermitian conjugate modes
$\Phi^\dagger,\,\Lambda^\dagger,$ etc should be added.

Finally, quantum Hamilton equations in commutators can be
constructed in the standard way. For example, for the $\Phi$ mode we
have
\begin{equation}
\partial_0 \Phi=i[\mathcal{H},\Phi]_-\,, \qquad \partial_0 P_\Phi=i[\mathcal{H},P_\Phi]_-\,.
\end{equation}
Such equations written for all wave modes coincide with the
corresponding equations of motion which were constructed previously
(\ref{eq1}), (\ref{eq2}) and (\ref{Keq}) -- (\ref{zetaeq}). The
latter is an important validation test of our calculations.\\

\acknowledgments

This work was supported in part by the Crafoord Foundation (Grant
No. 20120520). R. P. is grateful to the ``Beyond the LHC'' Program
at Nordita (Stockholm) for support and hospitality during completion
of this work.



\end{document}